\documentclass[fleqn,usenatbib]{aa}  
\usepackage{hyperref}
\hypersetup{colorlinks,citecolor=blue}
\usepackage{graphicx}
\usepackage{natbib}
\usepackage{lscape}
\usepackage{longtable}
\usepackage{soul}

\usepackage{txfonts}
\usepackage{float}

\def\mstar  {$M_{\star}$}

\def\macc   {$\dot{M}_{\rm acc}$}
\def\lacc   {$L_{\rm acc}$}
\def\mdust {$M_{\rm dust}$}
\def\mdisk {$M_{\rm disk}$}

\def\msun {$M_{\odot}$}
\def\lsun {$L_{\odot}$}
\def\lstar {$L_\star$}
\def\teff {$T_{\rm eff}$}
\def\sori {$\sigma$-Ori}
\def\sorionis {$\sigma$-Orionis}

\newcommand{\Lacc}{{$L_{\rm acc}$}}

\newcommand{\mT}[1]{\,\mathrm{#1 M}_\oplus} 

\usepackage{xcolor}

\begin{document}

   \title{Testing external photoevaporation in the $\sigma$-Orionis cluster \\with spectroscopy and disk mass measurements}


   \author{K. Maucó\inst{\ref{instESO}} \thanks{Based on observations collected at the European Southern Observatory under ESO programmes 0104.C-0454(A) and 108.22CB.001.}
          \and
          C.~F. Manara\inst{\ref{instESO}}
          \and
          M. Ansdell\inst{\ref{instNASA}}
          \and
          G. Bettoni\inst{\ref{instUniMI},\ref{instMPE}}
          \and
          R. Claes
          \inst{\ref{instESO}}
          \and 
          J. Alcala\inst{\ref{instNA}} 
          \and \\
        A. Miotello\inst{\ref{instESO}}
          \and 
          S. Facchini\inst{\ref{instUniMI}}
          \and 
          T. J. Haworth \inst{\ref{instQM}}
          \and
          G. Lodato\inst{\ref{instUniMI}}
          \and
        J. P. Williams\inst{\ref{instHW}}
          }
   \institute{ European Southern Observatory, Karl-Schwarzschild-Strasse 2, 85748 Garching bei M\"unchen, Germany\label{instESO}\\
              \email{kmaucoco@eso.org}
         \and
             NASA Headquarters, 300 E Street SW, Washington, DC 20546, USA\label{instNASA}
        \and
            Dipartimento di Fisica, Universit\'a degli Studi di Milano, Via Giovanni Celoria 16, 20133 Milano, Italy\label{instUniMI}
        \and 
            Max-Planck Institute for Extraterrestrial Physics, Gießenbachstraße 1, 85748 Garching, Germany\label{instMPE}
        \and 
            INAF -- Osservatorio Astronomico di Capodimonte, via Moiariello 16, 80131 Napoli, Italy\label{instNA}
        \and
            Astronomy Unit, School of Physics and Astronomy, Queen Mary University of London, London E1 4NS, UK\label{instQM}
        \and 
            Institute for Astronomy, University of Hawai‘i at Mānoa, Honolulu, HI, USA\label{instHW}
             }


 
  \abstract
   {The evolution of protoplanetary disks is regulated by an interplay of several processes, either internal to the system or related to the environment. As most of the stars and planets, including our own Solar System, have formed in massive stellar clusters that contain OB-type stars, studying the effects of UV radiation on disk evolution is of paramount importance.}
   {Here we test the impact of external photoevaporation on the evolution of disks in the mid-age ($\sim$3--5 Myr) \sorionis\,cluster by conducting the first combined large-scale UV to IR spectroscopic and mm-continuum survey of this region.}
   {We study a sample of 50 targets located at increasing distances from the central, massive OB system \sori. We combine new spectra obtained with VLT/X-Shooter, used to measure mass accretion rates and stellar masses, with new and previously published ALMA measurements of disk dust and gas fluxes and masses.}
   {We confirm the previously found decrease of \mdust\, in the inner $\sim$0.5 pc of the cluster. This is particularly evident when considering the disks around the more massive stars ($\ge$ 0.4 \msun), where those located in the inner part ($<$ 0.5 pc) of the cluster have \mdust\, about an order of magnitude lower than the more distant ones. 
   About half of the sample is located in the region of the \macc\,vs \mdisk\, expected by models of external photoevaporation, namely showing shorter disk lifetimes than expected for their ages. The shorter disk lifetimes is observed for all targets with projected separation from \sori\,$<$ 0.5 pc, proving that the presence of a massive stellar system affects disk evolution.
 }
   {External photoevaporation is a viable mechanism to explain the observed shorter disk lifetimes and lower \mdust\,
   in the inner $\sim$0.5 pc of the \sorionis\,cluster, where the effects of this process are more pronounced. 
   Follow-up observations of the low stellar mass targets are crucial to constrain disk dispersion time scales in the cluster and to confirm the dependence of the external photoevaporation process with stellar host mass.
   This work confirms that the effects of external photoevaporation are significant down to at least impinging radiation as low as $\sim 10^{4}$ G$_0$.
   }


   \maketitle
%

\section{Introduction}

Protoplanetary disks, made of gas and dust, are the byproduct of the star formation process and are the places where planets form. Their evolution is mediated by the interplay of several physical processes most likely acting simultaneously, which makes understanding disk evolution challenging \citep[][for a review]{Manara2023}. The standard theory is framed in the steady-state viscous paradigm, where the transfer of angular momentum in the disk drives its evolution, and results in accretion onto the central star \citep[e.g.,][]{Hartmann2016}. Dispersal mechanisms, such as winds and outflows, also contribute to the evolution through the depletion of disk material \citep[e.g.,][]{frank2014,Ercolano2017,Winter2022,pascucci2022}. Mass loss processes can have an internal origin, such as inside-out clearing produced by the ionizing radiation of the host star, or come from an external source, for example, the local environment. Dynamical interactions between stars and external photoevaporation, driven by high-energy radiation fields from OB massive stars, are among the most commonly discussed processes affecting disk evolution in clustered environments \citep[e.g.,][]{winter2018,reiter2022,cuello2023}.

Given the variety of properties found in planetary systems in our Galaxy, the way forward for understanding disk evolution must include the analysis of general disk and host star properties measured in a large statistical sample of systems at different evolutionary stages and environments. This makes it possible to identify correlations between the parameters (e.g., disk mass, disk radii, mass accretion rates) and their possible connection with the age of the region or its environment. Thanks to the availability of sensitive, wide-band optical spectrographs, such as the X-Shooter instrument on the Very Large Telescope (VLT), and radio interferometers, like the Atacama Large Millimeter and sub-millimeter Array (ALMA), it is now possible to measure some of these general properties \citep[][for a review]{Miotello2023}. In particular, the mass accreted onto the central star per unit time (\macc), drawn from UV-optical spectra, and the disk mass (\mdisk), from ALMA observations, have proven to be very useful for this task \citep{Manara2023}. 
For instance, surveys of young stars in different star-forming regions (SFRs) have found a tentative trend of decreasing \macc\, with age \citep[e.g.,][]{Sicilia2010,Antoniucci2014,briceno2019,Manzo2020}, predicted by viscous evolution models \citep[e.g.,][]{LyndenBell1974,Hartmann1998}. This observational trend, however, has large uncertainties, mainly due to unreliable age estimates for individual stars \citep[e.g.,][]{Soderblom2014} and correlated uncertainties between stellar properties and estimated individual ages \citep{DaRio2014}. Finally, an unexpectedly large fraction of high accretors are found in old ($>$5Myr) regions \citep{Ingleby2014,Manara2020,Manara2021,Testi2022}. 

Furthermore, measurements of \mdisk\,(estimated from dust emission and assuming a gas-to-dust ratio of 100) are now available for large samples of disks \citep[e.g.,][]{Ansdell2017,Ansdell2016,Pascucci2016,Barenfeld2016,Grant2021,vanTerwisga2023}, which in combination with \macc, has allowed us to connect what is happening in the innermost regions ($\lesssim$1 au) with outer disk properties and thus test disk evolution models. According to viscous evolution, \macc\,should positively correlate with \mdisk\,(predicted from the gas mass) in such a way as to expect a tighter correlation at older ages \citep{Rosotti2017,Lodato2017,Somigliana22}. The \mdisk\,--\macc\,relation has now been empirically established for nearby SFRs, although with a (puzzling) large spread regardless of the age of the region \citep[e.g.,][]{Manara2016b,Manara2020,Mulders2017}, pointing to a deviation from the purely viscous evolution theory, possibly toward a further importance of MHD winds in driving accretion in the disk \citep[][Somigliana et al. subm.]{Manara2023,Tabone22}. 

However, these studies have mainly focused on nearby ($<$ 300 pc) low-mass SFRs that distinctly lack OB stars \citep[e.g., Taurus,][]{Andrews2013}, and do not represent the environment where most planets have formed or the birth environment of our Solar System \citep[e.g., ][]{lada_lada2003,fatuzzo2008,Adams2010,Winter2020}.
Given the increasing relevance attributed to environmental factors in modulating disk evolution and planet formation, several authors have now included the effects of external photoevaporation by massive stars in models of viscous disk evolution \citep[e.g., ][]{Clarke2007, Anderson2013, Facchini2016, Haworth2018, Sellek2020b, coleman2022}. The ratio \mdisk/\macc\,has gained particular attention as a proxy of disk evolution, and as a possible discriminant between external effects and other internal disk evolution mechanisms. \citet{Rosotti2017} showed that externally irradiated disks show a \mdisk/\macc\,significantly lower than the expected value for a given system age, due to the radical disk mass depletion characteristic of this scenario. Similarly, external truncation in multiple stellar systems leads to a similar decrease of \mdisk/\macc\,\citep{Zagaria2022}.

An ideal region to test these aforementioned predictions is the \sorionis\,cluster. Its intermediate age \citep[$\sim$3-5 Myr,][]{Oliveira2004,Hernandez2014}, makes it young enough to remain bound, yet old enough for its central OB system \citep[\sori,][]{caballero2007} to have left its imprint. In contrast to the more extreme examples of externally irradiated disks we have, the Orion proplyds \citep{Odell1993}, where EUV photons drive mass loss and shape the proplyds in close proximity ($<$0.03 pc) to $\theta^1$ Ori C \citep{Johnstone1998}, the dispersal of disks in \sorionis\,is controlled by far-UV (FUV) radiation \citep[e.g.,][]{Adams2004, Facchini2016, Haworth2018} as a result of the lower mass of its OB system (compared to $\theta^1$ Ori C) and the larger separation of the stars to the center, depleting the disks close to \sori. 
This was shown in the ALMA survey of \sorionis\,\citep{Ansdell2017}, which found a dearth of massive (\mdust $> 3 M_{\oplus}$) disks close ($<$ 0.5 pc) to the central OB stars, and a smooth distance-dependent trend in the disk dust mass distribution, in line with previous results in the NGC2024 and the Orion Nebula Clusters \citep{Mann2014,Mann2015}, and in other less massive regions in Orion \citep{vanTerwisga2023}. This observed depletion of disk masses in \sorionis\,was later reproduced using external photoevaporative models \citep{Winter2020}. However, several other effects are at play, including dynamics in the clusters, and this trend could be coincidental \citep{Parker2021}.

Just measuring disk dust masses is not enough to firmly assess the effects of external photoevaporation on disk evolution in massive star-forming regions. Two additional observational probes can be used. The ratio of forbidden emission lines is also a way to detect signs of externally photoevaporated disks. \citet{Rigliaco2009} used this probe to claim that the SO587 disk in \sorionis\,is currently being externally photoevaporated, and has also been supported by photoevaporative models recently \citep{ballabio2023}. Additional forbidden emission line data analyzed by \citet{Gangi2023} for 3 targets in the \sorionis\, cluster are however still not conclusive tell-tale tests of external photoevaporation, due both to the strong nebular contamination and the small sample. The other observational proxy of external photoevaporation, the correlation between \mdisk\,and \macc, has not yet been well established due to the lack of accurate mass accretion rates for sources with detected sub-mm fluxes. Previous estimates of accretion rates for \sorionis\,members were obtained either for a small sub-sample of very low-mass stars \citep{Rigliaco2012} or using indirect tracers such as U-band photometry \citep{Rigliaco2011} or the H$\alpha$ line from low-resolution spectroscopy \citep{Mauco2016}. Therefore, this latter proxy is for the first time used in this work for the \sorionis\, cluster.

Here we present the results of the first large-scale spectroscopic survey of disk-bearing stars in the \sorionis\, cluster in which mass accretion rates are analyzed together with - new and previously published - disk masses.  Our main objective is to study, for the first time,
the relationship between \macc\,and \mdisk, and to further constrain the dependence of \mdisk\, with the distance from the massive system \sori. After describing the sample in Sect.~\ref{sec:sample} and the observations and data reduction in Sect.\ref{sec:obs}, we present our results on stellar parameters, and disk mass estimates in Sect.~\ref{sec:results}. We discussed the implications of our findings in the context of external photoevaporation in Sect.~\ref{sec:discussion}. Finally, we summarize our conclusions in Sect.~\ref{sec:conclusions}.
 
\section{Sample} \label{sec:sample}

The \sorionis\, cluster is located in the Orion OB1 association, which is one of the largest and nearest OB associations spanning over 200 deg$^2$ on the sky \citep[see the review in][]{Reipurth2008}. Their OB stars were first recognized by \citet{Garrison1967} as a group of 15 B-type stars around the massive hierarchical triple system \sori, whose most massive component is an O9.5V star \citep{caballero2007, Simon2015}, shaping the photodisociation region known as the Horsehead Nebula \citep[e.g.,][]{abergel2003,Pety2005} and setting the UV field strength in the cluster (see Fig.~\ref{fig:Go_massive_stars}). In the last decades, several hundred low-mass stars and brown dwarfs have been already identified as part of the cluster \citep[e.g.,][]{Reipurth2008}. The disks around the low-mass stars were first identified using \textit{Spitzer} photometry \citep{Hernandez2007,Luhman2008} and then followed with \textit{Herschel} \citep{Mauco2016} and, more recently, imaged with ALMA at 1.3 mm \citep{Ansdell2017} and followed down to the brown dwarf limit \citep{damian2023a,damian2023b}. The low reddening
toward its center \citep[E(B-V) $\lesssim$ 0.1 mag, e.g.,][]{Brown1994,Bejar1999,Sherry2008} makes it an excellent natural laboratory to study protoplanetary disk evolution in the entire range of stellar masses and in the context of externally irradiated disks in moderate-to-high UV environments.

\begin{figure}[t]
\centering \includegraphics[width = 0.5\textwidth]{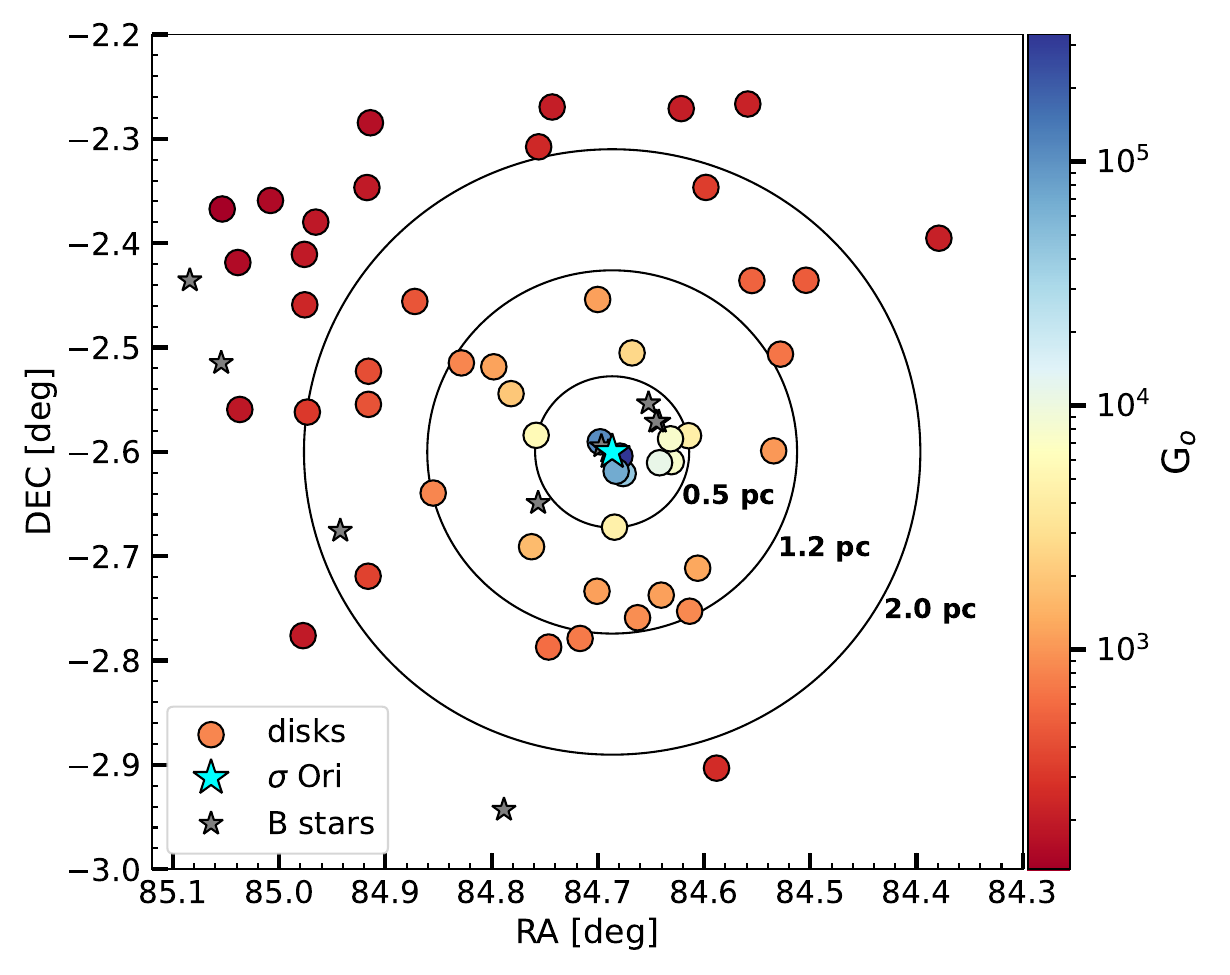}
\caption{Spatial distribution of \sorionis\,sources (points), and massive O-B stars (star symbols) in the cluster. The massive, multiple system \sori\,is indicated in cyan while the rest of the B-type stars in gray. The color bar shows the incident FUV field strength (in terms of the dimensionless parameter $G_o$) due to the massive stars. Black circles show projected distances of 0.5, 1.2, and 2.0 pc.}\label{fig:spatial_distribution_stars}
\end{figure}

Our X-Shooter sample consists of 50 disk-bearing stars in the \sorionis\, cluster with ALMA observations \citep{Ansdell2017} and located at different projected distances from \sori\,(see Fig.~\ref{fig:spatial_distribution_stars}). Of the 50 stars observed with X-Shooter, 43 have been detected by ALMA. 
The sample includes the objects studied in \citet{Rigliaco2012, Rigliaco2009},
and mainly consists of late-K and M spectral type (SpT) stars at different evolutionary stages based on the classification of their spectral energy distribution \citep[SED,][]{Luhman2003}, as reported by \citet{Hernandez2007,Rigliaco2011,Mauco2016}. 
Our sample includes five disks with central cavities or transition disks (TD), one class I star (SO1153), which in the \citet{Luhman2003} classification points to a strong IR excess rather than to an embedded object (this source is visible at UV-optical wavelengths), and the rest are class II stars. The list of the observed targets is reported in Table~\ref{tab:sample}.  

The $\textit{Gaia}$  EDR3 astrometric solutions for the sample are generally good, with low renormalized unit weight errors (RUWE). Only 8 targets (SO397, SO490, SO563, SO583, SO587, SO736, SO823, SO897) have RUWE values $>$ 1.4, considered an appropriate nominal limit for $\textit{Gaia}$  EDR3 \citep{Gaia2021}. 
For all targets, we assumed the individual distances inverting the parallaxes from Gaia EDR3 \citep[arithmetic distances,][]{Gaia2021}. We then estimated the average distance to the cluster, considering only sources with RUWE $<$ 1.4, and found a median distance of 401 pc. This is compatible with the values reported by \citet{damian2023b}. Therefore, for all our targets we assumed their arithmetic distances unless the values were unreliable – RUWE $>$ 1.4 and/or distance differing more than 60 pc from the mean distance to the region (target SO936) – or not available (targets SO435, SO562, and SO1155), in which case we assumed the median distance to the members of the region. 
Distances for the sample are also listed in Table~\ref{tab:sample}.

Through this analysis, we found four targets, namely SO73, SO299, SO411, and SO848, whose distances are lower than the median by $\sim$40 pc and yet have RUWE values $<$1.4. These can be possible members of the more sparse Orion OB1a sub-association in front of \sorionis\, \citep{briceno2019}. For SO411 this seems to be the case based on its proper motions  \citep{Perez2018}, however, for the rest of these stars we cannot know for certain. Therefore, we have included them in the analysis assuming their arithmetic distances from \textit{Gaia}, and we have pointed them out whenever they appear as outliers from the main population. 
Similarly, the star SO828 with a distance of 449.5 pc (i.e., $\sim$50 pc away from the median distance to the members of the region) is treated in the same way.

\section{Observations, and data reduction} \label{sec:obs}

\subsection{Spectroscopy with VLT/X-Shooter}
Observations were carried out between October 2019 and February 2020 (Pr.Id. 0104.C-0454(A), PI Ansdell) and between November 2021 and January 2022 (Pr.Id. 108.22CB.001, PI Ansdell) in Service Mode at the ESO Very Large Telescope (VLT). The X-Shooter instrument \citep{Vernet2011} was used for all observations. This instrument acquires spectra simultaneously in three arms: UVB ($\lambda\sim 300 - 550$ nm), VIS ($\lambda\sim 500 - 1050$ nm), and NIR ($\lambda\sim 1000 - 2500$ nm). All the stars were observed with a nodding pattern using a set of narrow slits (1.0"–0.4"–0.4" in the UVB–VIS–NIR arms, respectively), yielding the highest spectral resolution ($\sim$ 5400, 18400, 11600, respectively). For flux calibrating the spectra, a short ($\sim$1 min to 10 min depending on target brightness) exposure in stare with a set of wide slits (5.0") prior to the science exposure was taken. 

Data reduction was done using the X-Shooter pipeline v.3.2.0 (P104 data) and v.3.5.0 (P108 data)  \citep{Modigliani2010} run within the ESO Reflex environment \citep{Freuding2013} using the same procedure as in previous similar analyzes \citep[e.g.,][]{Alcala2017,Manara2020,venuti2019}. The pipeline runs the classical reduction steps, including bias subtraction, flat-fielding, wavelength calibration, flexure and atmospheric dispersion correction, background removal (in stare mode) or combination of spectra obtained in a nodding cycle, and the extraction of the 1D spectrum. Telluric correction was then performed on the high-resolution spectra with the molecfit tool \citep{Smette2015}, which models the telluric absorption lines on the observed spectra using information on the atmospheric conditions in the night. Finally, the high-resolution spectra were rescaled to those obtained with the wider slit in order to account for slit losses and obtain absolute flux calibration. 
This methodology leads to accurate flux calibration of the spectra \citep[e.g.,][]{Manara2021}.

Particular care was taken in the case of the resolved binary system SO1267, where the two traces of the two targets, separated by 1.4", were manually extracted using the IRAF software. Throughout this paper, the source indicated as SO1267 refers to SO1267A. For the targets observed on nights with humidity higher than $\sim$40\% or with PWV$\sim$9.5 mm, we did use the flux standard observed in the closest night with optimal conditions, to avoid introducing incorrect shapes in the NIR arm of the spectra. Finally, for SO844 and SO1154 we did rescale the narrow slit spectra to non-simultaneous photometric data, since the wide slit spectra had non-reliable fluxes lower than the narrow slit ones, possibly due to the presence of thin cirrus at the time of the observations. 

\subsection{ALMA cycle 4 data}\label{sec:obs_ALMA}

In this paper we use new, higher sensitivity Band 6 Cycle~4 ALMA observations obtained with eight Execution Blocks (EBs) on 29, 30 October 2016, 2, 3 November 2016, 14 May 2017, 2, and 4 July 2017 (Project ID: 2016.1.00447.S; PI: Williams). The array configuration used between 40 and 44 12m antennas, with baselines of $\sim$20–2650 m in July 2017, leading to a spatial resolution of $\sim$0.18", and shorter baselines of $\sim$15 - 1125 m in May 2017 and in 2016, with corresponding spatial resolution $\sim$0.26". The correlator setup included two broadband continuum windows centered on 234.3 and 216.5 GHz with bandwidths of 1.875 GHz and channel widths of 31.25 and 1.129 MHz, respectively. The bandwidth-weighted mean continuum frequency was 225.77 GHz (1.33 mm). The spectral windows covered the $^{12}$CO (230.538 GHz), $^{13}$CO (220.399 GHz), and C$^{18}$O (219.560 GHz) $J = 2-1$ transitions at velocity resolutions of 0.079 - 0.096 km/s. These spectral windows had bandwidths of 58.59 MHz and channel widths of 60.6 kHz - 0.071 MHz.

The raw data were pipeline calibrated at NRAO using the CASA package (version 4.7.2). The pipeline calibration included: absolute flux calibration with observations of J0522-3627 or J0423-0120; bandpass calibration with observations of J0510+1800 or J0522-3627; and gain calibration with observations of J0532-0307. We estimate an absolute flux calibration error of $\sim$10\% based on the amplitude variations of gain calibrators over time.
The imaging of the continuum and line data was performed similarly to what was done by \cite{Ansdell2017}, cleaning with a Briggs robust weighting parameter of 0.5. We find a median 1.33 mm continuum RMS of 50 $\mu$Jy and the median $^{12}$CO RMS is 11 mJy in 0.5 km s$^{-1}$ channels. The achieved RMS for the Representative Window centered on $^{13}$CO $(J=2-1)$ (220.399 GHz) is of 9.5 mJy Beam$^{-1}$ with a bandwidth of 0.096km/s and a 0.30$\times$0.22 arcsec beam, while the requested sensitivity was of 3.3 mJy $^{-1}$over 1.0 km s$^{-1}$ and a beam size of 0.22 arcsec. The achieved continuum RMS is of 4.5 $10^{-2}$ mJy Beam$^{-1}$ with a bandwidth of 3.4 GHz and a 0.27$\times$0.19 arcsec beam. Continuum and $^{12}$CO images are shown in Fig.~\ref{fig:cont_images}, and ~\ref{fig:12COmom0_images}, respectively, in Appendix \ref{sec:appendix_ALMA}.

\begin{figure}[]
\includegraphics[width=0.48\textwidth]{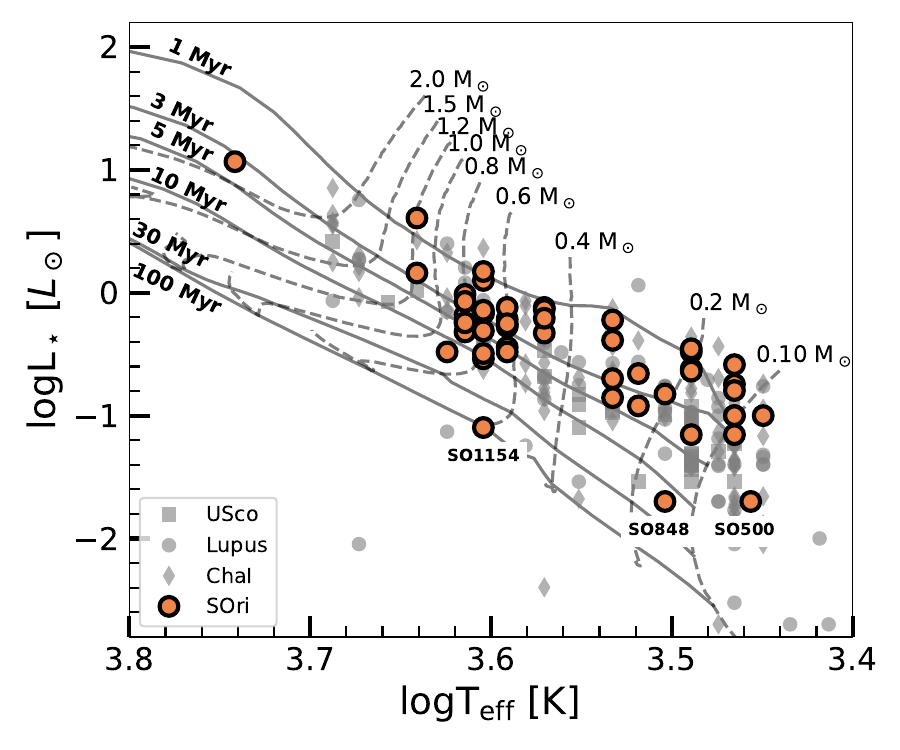}
\caption{Hertzsprung-Russell diagram for $\sigma$-Orionis disks (orange circles) including those from R12. Sources from other SFRs are shown by gray symbols. Isochrones for 1, 3, 5, 10, 30, and 100 Myr from \citet{Siess2000} are overplotted. Evolutionary tracks are from \citet{Baraffe2015}.
     \label{fig:HRD}}
\end{figure}

\begin{figure}
\includegraphics[width=0.48\textwidth]{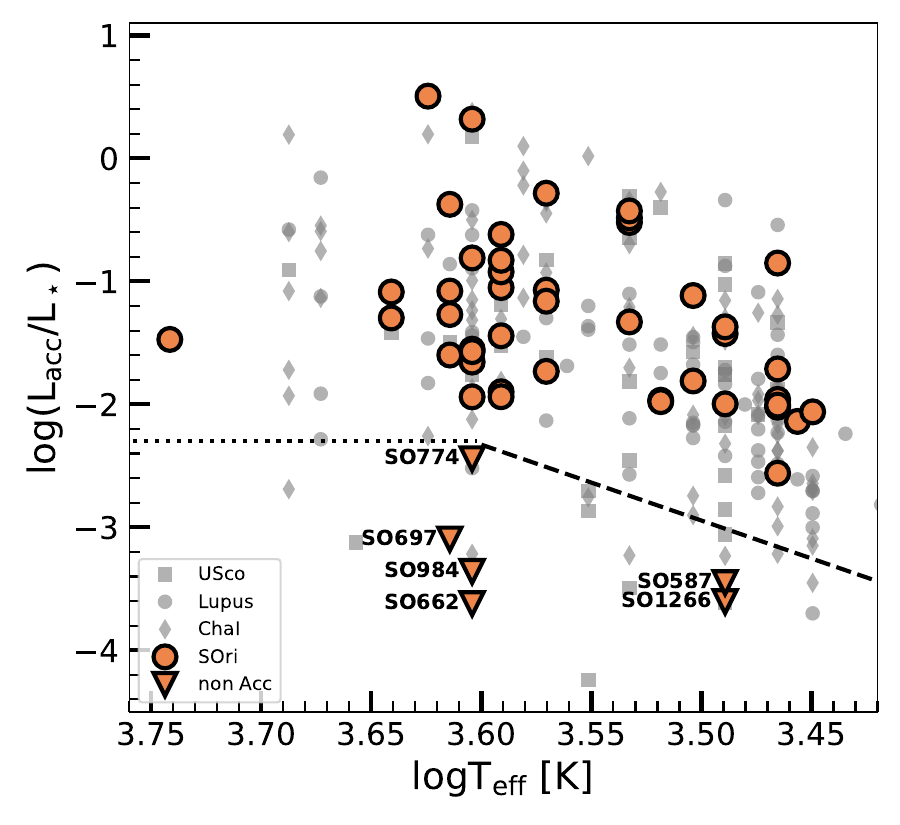}
\caption{Ratio between the accretion and stellar luminosities vs effective temperature. \sorionis\,sources are indicated by orange circles while stars in other young SFRs by gray symbols. The dotted and dashed lines represent the locus of the chromospheric emission defined by \citet{Manara2013a,Manara2017b}. Downward triangles indicate the non-accretors identified in this work. 
     \label{fig:lacc_Teff}}
\end{figure}

\section{Results} \label{sec:results}

\subsection{Stellar and accretion properties} \label{sec:stellar_param}

X-Shooter provides absolute flux calibrated spectra with sufficient spectral resolution and wavelength coverage to simultaneously characterize stellar, accretion, wind, jet, and ionization properties of young stellar objects \citep[e.g,][]{Bacciotti2011,Rigliaco2012,Alcala2014,Frasca2017,Manara2016a,Manara2021}. The continuum regions needed to determine stellar and accretion parameters range from $\lambda \, \sim$ 300-364 nm (the Balmer continuum) to $\lambda \, \sim$ 700 nm (where several molecular bands are present). Various absorption lines along the spectrum are required to constrain stellar spectral type and photospheric parameters \citep[e.g.,][]{Manara2013b}.

In order to derive the stellar and accretion properties of the targets, we follow the same fitting procedure as \citet{Manara2013b}. 
In short, we model the spectra by adding a photospheric template spectrum plus a slab model to match the observed, dereddened spectrum. 
The grid of Class III photospheric templates includes targets with SpT from G- to late M taken from \citet{Manara2013a,Manara2017b}, different slab models, and extinction values ($A_{V}$), assuming the \citet{Cardelli1989} reddening law ($R_V$ = 3.1). The output from the models is the excess luminosity due to accretion (\lacc), given by the integrated flux of the best-fit
slab models, and the stellar luminosity (\lstar), which is estimated by measuring the normalization of the Class III templates that best match the observations. Distances were estimated as described in Sect.~\ref{sec:sample}. In Fig.~\ref{fig:best_fit_plots} in the appendix, we show the best-fit spectrum of each of our targets. We note that, as expected, $A_V$ is typically low, reaching values above or equal to 1.0 mag only in 7 targets.

For the sake of comparison with other star-forming regions, we considered the same assumptions as \citet[][]{Manara2023} and derived all the stellar and accretion parameters in a similar way. Therefore, we measure all luminosities (\lstar, \Lacc) using the new $\textit{Gaia}$ distances and obtain $T_{\rm eff}$ from SpT using the calibration by \citet{Herczeg2014}. In Table~\ref{tab:bestfit} we list the stellar and accretion parameters estimated for our sample, including those from the \citet{Rigliaco2012} sample, which are recalculated with the same assumptions that we just stated, including rescaling the distance from 360 pc to the \textit{Gaia}-based ones.

Using the $T_\mathrm{eff}$ and $L_\star$ from the best-fit we were able to locate each target on the Hertzsprung-Russell diagram (HRD), as shown in Fig.~\ref{fig:HRD}.
The targets in the \sorionis\, cluster are located in the region of the HRD consistent with their expected age (3-5 Myrs). Three targets are located at lower \lstar\, with respect to the bulk of the population at the same \teff, namely SO500, SO848, and, SO1154. SO500 is a known brown dwarf \citep{Rigliaco2011} and its location on the HRD is in line with other sub-stellar objects. For SO1154, partial obscuration of the star by a highly inclined disk could explain their positions on the HRD. A highly inclined disk can add gray extinction and make the star under-luminous, resulting in more uncertain estimates of \lstar\, and of the mass accretion rate \citep{Alcala2014}. This target is the one with the highest measured $A_V$=1.8 mag, supporting the hypothesis of (partial) obscuration by a disk. Finally, SO848 could either be also a highly inclined disk, or a foreground object, as discussed in Sect.~\ref{sec:sample}.

In order to check the estimates from the fit, we compared the values of \lacc\, obtained with the fitting procedure described above, with those derived from the luminosity of 10 emission lines, namely CaK, H$\delta$, H$\gamma$, H$\beta$, $\mathrm{HeI}587$ nm, H$\alpha$, $\mathrm{HeI}667$ nm, Pa$\gamma$, Pa$\beta$, Br$\gamma$, using the relations between line and accretion luminosity by \citet{Alcala2017}. 
The mean value of \lacc\,derived from the emission lines is generally in agreement with the one obtained by fitting the continuum in the X-Shooter spectrum with no dependence on the wavelengths of the lines, pointing toward correctly estimated $A_V$.

Figure~\ref{fig:lacc_Teff} shows the ratio between accretion and stellar luminosities as a function of the effective temperature, which is a diagram used to check whether the measured accretion luminosity is larger than typical chromospheric emission \citep{Manara2013b}. 
Assuming the locus of chromospheric emission defined by \citet{Manara2017b}, we found 6 non-accreting targets in our sample (downward triangles). As shown in Fig.~\ref{fig:best_fit_plots} these sources exhibit negligible UV excess, in line with their non-accreting nature. The rest of accreting targets have similar \lacc/\lstar\, values at any given \teff\, as those found in other star-forming regions, in line with previous results.

After locating the targets in the HRD, we derive \mstar\,using the non-magnetic models of \citet{Baraffe2015} for colder stars ($T_{\rm eff} \leq$3900 K), and of \citet{Feiden2016} for hotter stars ($T_{\rm eff} >$3900 K). For targets having stellar properties outside of the range of values sampled by these models, we used the \citet{Siess2000} models instead. 
Finally, the \macc\, is obtained from the classic relation $\dot{M}_{\rm acc} = 1.25 \times L_{\rm acc} R_{\star} / (G M_{\star})$ from \citet{Hartmann1998}, using \lacc\, from the fit.
The stellar and accretion parameters of the sample are found in Table~\ref{tab:bestfit}.

\begin{figure}
\includegraphics[width=0.49\textwidth]{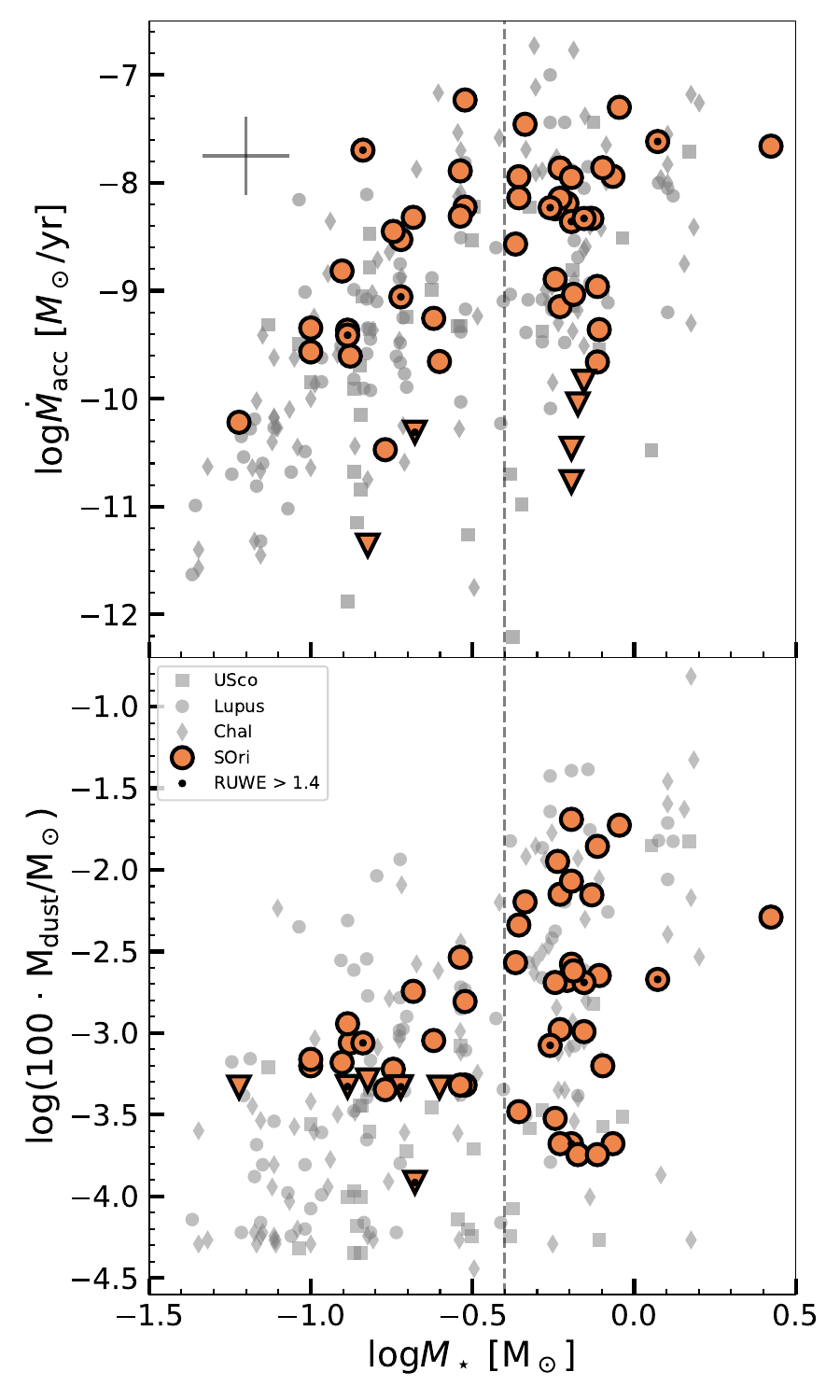}
\caption{\textit{Top:} Mass accretion rates vs stellar mass. The expected uncertainties are indicated by the error bars at the top left. \textit{Bottom:} Disk masses vs stellar mass. All the targets from our sample and from \citet{Rigliaco2012} are plotted. Downward triangles indicate upper limits. The vertical dashed line indicates a \mstar= 0.4 \msun.  
     \label{fig:stellar_param_correlation}}
\end{figure}

The relation between \macc\, and \mstar\, is shown in Fig.~\ref{fig:stellar_param_correlation} (top panel). Given the expected uncertainties on both quantities (error bar), the \sorionis\,disks seem to populate the same parameter space as the one covered by other young SFRs like Lupus, and Chameleon I, and even by the older \citep[5-10 Myr;][]{Pecaut2016} Upper-Scorpius (USco). This will be further discussed in Sect.~\ref{sec:mstar_corr}.

\begin{table*}
\begin{center}
\caption{\label{tab:sample} \sorionis\,Disk Sample}	
\begin{tabular}{lcccccc}
\hline\hline 
Name & RA$_{2000}$ & Dec$_{2000}$ & Distance & $d_{\rm p}$ & Log $G_o$ & Disk type \\
     & hh:mm:ss.s  & dd:mm:ss.s   &  [pc]    &  [pc]  &   & \\  
\hline 
  SO73  & 05:37:30.95 & -02:23:42.8 & $359.2_{+4.2}^{-4.4}$ &  2.32 &  2.34 & -- \\[0.5ex]   
 SO299 & 05:38:00.97 & -02:26:07.9 & $355.5_{+4.3}^{-4.4}$ &  1.52 &  2.70 & TD \\[0.5ex]  
 SO341 & 05:38:06.74 & -02:30:22.8 & $409.0_{+4.3}^{-4.4}$ &  1.31 &  2.83 & II \\[0.5ex]  
 SO362 & 05:38:08.27 & -02:35:56.3 & $402.3_{+4.6}^{-4.8}$ &  1.07 &  3.01 & II \\[0.5ex]  
 SO397 & 05:38:13.20 & -02:26:08.8 & $401.0$ &  1.47 &  2.73 & II \\[0.5ex]  
 SO411 & 05:38:14.12 & -02:15:59.8 & $365.5_{+2.2}^{-2.2}$ &  2.28 &  2.35 & TD \\[0.5ex]  
 SO467 & 05:38:21.19 & -02:54:11.1 & $383.3_{+8.6}^{-9.0}$ &  2.13 &  2.41 & -- \\[0.5ex]  
 SO490 & 05:38:23.58 & -02:20:47.6 & $401.0$ &  1.88 &  2.52 & II \\[0.5ex]  
 SO500 & 05:38:25.44 & -02:42:41.3 & $409.2_{+37.2}^{-45.4}$ &  0.98 &  3.09 & II \\[0.5ex]  
 SO518 & 05:38:27.26 & -02:45:09.7 & $399.0_{+3.9}^{-4.0}$ &  1.18 &  2.93 & II \\[0.5ex]  
 SO520 & 05:38:27.51 & -02:35:04.2 & $402.6_{+6.3}^{-6.5}$ &  0.52 &  3.64 & II \\[0.5ex]  
 SO540 & 05:38:29.16 & -02:16:15.7 & $406.0_{+3.5}^{-3.6}$ &  2.38 &  2.32 & II \\[0.5ex]  
 SO562 & 05:38:31.41 & -02:36:33.8 & $401.0$ &  0.39 &  3.88 & II \\[0.5ex]  
 SO563 & 05:38:31.58 & -02:35:14.9 & $401.0$ &  0.39 &  3.88 & II \\[0.5ex]  
 SO583 & 05:38:33.68 & -02:44:14.2 & $401.0$ &  1.01 &  3.06 & II \\[0.5ex]  
 SO587 & 05:38:34.06 & -02:36:37.5 & $401.0$ &  0.32 &  4.06 & II \\[0.5ex]  
 SO646 & 05:38:39.03 & -02:45:32.2 & $404.6_{+6.6}^{-6.8}$ &  1.13 &  2.96 & II \\[0.5ex]  
 SO662 & 05:38:40.27 & -02:30:18.5 & $401.2_{+3.3}^{-3.4}$ &  0.68 &  3.41 & II \\[0.5ex]  
 SO682 & 05:38:42.28 & -02:37:14.8 & $409.8_{+4.7}^{-4.8}$ &  0.17 &  4.63 & II \\[0.5ex]  
 SO687 & 05:38:43.02 & -02:36:14.6 & $412.8_{+4.2}^{-4.3}$ &  0.06 &  5.52 & II \\[0.5ex]  
 SO694 & 05:38:43.87 & -02:37:06.8 & $392.3_{+9.2}^{-9.6}$ &  0.13 &  4.85 & -- \\[0.5ex]  
 SO697 & 05:38:44.23 & -02:40:19.7 & $404.5_{+2.4}^{-2.4}$ &  0.51 &  3.66 & II \\[0.5ex]  
 SO726 & 05:38:47.46 & -02:35:25.2 & $403.9_{+6.8}^{-7.0}$ &  0.10 &  5.03 & II \\[0.5ex]  
 SO736 & 05:38:48.04 & -02:27:14.2 & $401.0$ &  1.03 &  3.05 & II \\[0.5ex]  
 SO739 & 05:38:48.19 & -02:44:00.8 & $433.3_{+20.3}^{-22.3}$ &  1.02 &  3.06 & II\\[0.5ex]  
 SO774 & 05:38:52.01 & -02:46:43.7 & $403.3_{+3.3}^{-3.4}$ &  1.28 &  2.86 & II \\[0.5ex]  
 SO818 & 05:38:58.32 & -02:16:10.1 & $405.4_{+4.1}^{-4.2}$ &  2.37 &  2.32 & TD \\[0.5ex]  
 SO823 & 05:38:59.11 & -02:47:13.3 & $401.0$ &  1.37 &  2.79 & II \\[0.5ex]  
 SO844 & 05:39:01.37 & -02:18:27.5 & $415.5_{+3.7}^{-3.8}$ &  2.18 &  2.39 & II \\[0.5ex]  
 SO848 & 05:39:01.94 & -02:35:02.9 & $356.3_{+16.3}^{-18.0}$ &  0.46 &  3.75 & II\\[0.5ex]  
 SO859 & 05:39:02.98 & -02:41:27.2 & $407.9_{+6.4}^{-6.6}$ &  0.84 &  3.22 & II \\[0.5ex]  
 SO897 & 05:39:07.61 & -02:32:39.1 & $401.0$ &  0.77 &  3.29 & TD \\[0.5ex]  
 SO927 & 05:39:11.51 & -02:31:06.5 & $413.6_{+4.7}^{-4.8}$ &  1.0  &  3.07 & II \\[0.5ex]  
 SO984 & 05:39:18.83 & -02:30:53.1 & $409.6_{+3.1}^{-3.2}$ &  1.18 &  2.92 & II \\[0.5ex]  
SO1036 & 05:39:25.20 & -02:38:22.0 & $395.0_{+3.4}^{-3.5}$ &  1.19 &  2.92 & II \\[0.5ex]  
SO1075 & 05:39:29.35 & -02:27:21.0 & $390.0_{+8.2}^{-8.6}$ &  1.60 &  2.66 & II \\[0.5ex]  
SO1152 & 05:39:39.38 & -02:17:04.5 & $398.6_{+3.8}^{-3.9}$ &  2.71 &  2.21 & -- \\[0.5ex]  
SO1153 & 05:39:39.82 & -02:31:21.8 & $396.6_{+4.2}^{-4.3}$ &  1.68 &  2.62 & I  \\[0.5ex]  
SO1154 & 05:39:39.83 & -02:33:16.0 & $401.0$ &  1.64 &  2.64 & -- \\[0.5ex]  
SO1155 & 05:39:39.90 & -02:43:09.0 & $401.0$ &  1.81 &  2.55 & -- \\[0.5ex]  
SO1156 & 05:39:40.17 & -02:20:48.0 & $403.8_{+2.6}^{-2.6}$ &  2.42 &  2.30 & II \\[0.5ex]  
SO1248 & 05:39:51.73 & -02:22:47.2 & $398.4_{+7.6}^{-7.9}$ &  2.47 &  2.28 & -- \\[0.5ex]  
SO1260 & 05:39:53.63 & -02:33:42.7 & $386.3_{+6.2}^{-6.4}$ &  1.95 &  2.49 & II \\[0.5ex]  
SO1266 & 05:39:54.21 & -02:27:32.6 & $399.1_{+10.4}^{-11.0}$ &  2.24 &  2.37 & II\\[0.5ex]  
SO1267 & 05:39:54.29 & -02:24:38.6 & $400.5_{+5.2}^{-5.3}$ &  2.42 &  2.30 & -- \\[0.5ex]  
SO1274 & 05:39:54.60 & -02:46:34.0 & $407.3_{+2.7}^{-2.7}$ &  2.42 &  2.30 & II \\[0.5ex]  
SO1327 & 05:40:01.96 & -02:21:32.6 & $397.7_{+5.7}^{-5.8}$ &  2.79 &  2.18 & II \\[0.5ex]  
SO1361 & 05:40:08.89 & -02:33:33.7 & $406.0_{+3.9}^{-4.0}$ &  2.50 &  2.27 & II \\[0.5ex]  
SO1362 & 05:40:09.33 & -02:25:06.7 & $399.4_{+10.2}^{-10.7}$ &  2.76 &  2.19 & II\\[0.5ex]  
SO1369 & 05:40:12.87 & -02:22:02.0 & $402.5_{+2.5}^{-2.5}$ &  3.05 &  2.10 & -- \\[0.5ex]  
\hline
\end{tabular}
\end{center}
\end{table*}

\begin{table*}
\begin{center}
\caption{\label{tab:bestfit} Stellar and accretion properties and disk masses}	
\begin{tabular}{lcccccccccc}
\hline\hline 
Name & SpT & $T_{\rm eff}$ & $A_V$ & $L_\star$ & $\log L_{\rm acc}$ & $M_\star$ & $\log \dot{M}_{\rm acc}$ & F$_{\rm mm}$& $M_\mathrm{dust}$ & F$_{^{12}\rm CO}$\\ 
     &     & [K] & [mag] & [\lsun] & [\lsun] & [\msun] & [\msun/yr] & [mJy] & $[\mT{}]$ & [mJy]\\
\hline 
   SO73 &     M3 &   3410 &   1.0 &   0.2 & -1.13 &  0.29 &  -7.89 & 0.53 $\pm$  0.13 &   1.6 $\pm$   0.4 & $<$ 66.0           \\ [0.5ex]
   SO299 &   M3.5 &   3300 &   0.2 &  0.22 & -2.62 &  0.24 &  -9.26 & 1.01 $\pm$  0.14 &   3.0 $\pm$   0.4 & $<$ 66.0           \\ [0.5ex]
   SO341 &     M0 &   3900 &   0.8 &  0.55 & -1.18 &  0.59 &  -8.14 & 1.19 $\pm$  0.13 &   3.5 $\pm$   0.1 & $<$34.35           \\ [0.5ex]
   SO362 &     M3 &   3410 &   1.4 &   0.6 &  -0.7 &   0.3 &  -7.23 & 0.56 $\pm$  0.13 &   1.6 $\pm$   0.1 & $<$34.02           \\ [0.5ex]
   SO397 &   M4.5 &   3085 &   0.0 &  0.24 & -2.62 &  0.19 &  -9.06 & $<$  0.4 & $<$   1.6 & $<$  69.0                          \\ [0.5ex]
   SO411 &     G4 &   5516 &   0.6 & 11.67 &  -0.4 &  2.65 &  -7.66 & 5.16 $\pm$  0.13 &  17.1 $\pm$   0.1 & 130.35 $\pm$ 18.02 \\ [0.5ex]
   SO467 &   M5.5 &   2920 &   0.3 &  0.07 & -3.18 &   0.1 &  -9.57 & 0.61 $\pm$  0.13 &   2.1 $\pm$   0.5 & $<$ 66.0           \\ [0.5ex]
   SO490 &   M5.5 &   2920 &   0.0 &   0.1 & -3.01 &  0.13 &  -9.41 & $<$  0.4 & $<$   1.6 & $<$  72.0                          \\ [0.5ex]
   SO500 &     M6 &   2860 &   0.0 &  0.02 & -3.84 &  0.06 & -10.22 & $<$  0.4 & $<$   1.6 & $<$  63.0                          \\ [0.5ex]
   SO518 &     K6 &   4115 &   1.6 &  0.48 & -0.69 &   0.8 &  -7.86 & 0.52 $\pm$  0.13 &   2.1 $\pm$   0.1 & 96.61 $\pm$ 18.57  \\ [0.5ex]
   SO520 &   M4.5 &   3085 &   0.1 &  0.23 & -2.01 &  0.18 &  -8.45 & 0.52 $\pm$  0.14 &   2.0 $\pm$   0.5 & $<$ 69.0           \\ [0.5ex]
   SO540 &     K6 &   4115 &   0.5 &  0.57 & -1.84 &  0.77 &  -8.96 &10.69 $\pm$  0.29 &  46.4 $\pm$   0.3 & 1306.92 $\pm$ 45.33\\ [0.5ex]
   SO562 &   M5.5 &   2920 &   0.3 &  0.26 & -1.44 &  0.15 &   -7.7 & 0.71 $\pm$  0.13 &   2.9 $\pm$   0.1 & $<$33.66           \\ [0.5ex]
   SO563 &     M0 &   3900 &   0.6 &  0.36 & -1.27 &  0.64 &  -8.36 &  0.18 $\pm$  0.04 &   0.7 $\pm$   0.1 & $<$  33.0         \\ [0.5ex]
   SO583 &     K4 &   4375 &   1.0 &  4.06 & -0.69 &  1.18 &  -7.62 &  1.9 $\pm$  0.13 &   7.1 $\pm$   0.1 & 68.95 $\pm$ 12.52  \\ [0.5ex]
   SO587 &   M4.5 &   3085 &   0.0 &  0.35 & -3.91 &  0.21 & -10.31 & $<$  0.1 & $<$   0.4 & $<$  33.6                          \\ [0.5ex]
   SO646 &   M3.5 &   3300 &   0.0 &  0.12 &  -2.9 &  0.25 &  -9.66 & $<$  0.4 & $<$   1.6 & $<$  69.0                          \\ [0.5ex]
   SO662 &     K7 &   4020 &   0.3 &  0.68 & -3.79 &  0.64 & -10.77 & 1.54 $\pm$  0.14 &   8.8 $\pm$   0.2 & $<$33.99           \\ [0.5ex]
   SO682 &     M0 &   3900 &   0.7 &  0.76 & -2.02 &  0.57 &  -8.89 & 0.41 $\pm$  0.14 &   1.0 $\pm$   0.1 & $<$30.78           \\ [0.5ex]
   SO687 &     M1 &   3720 &   0.8 &  0.73 & -1.21 &  0.44 &  -7.94 &  0.28 $\pm$  0.04 &   1.1 $\pm$   0.1 & $<$  32.1         \\ [0.5ex]
   SO694 &   M5.5 &   2920 &   0.1 &  0.16 & -2.51 &  0.12 &  -8.82 & 0.61 $\pm$  0.14 &   2.2 $\pm$   0.5 & $<$ 69.0           \\ [0.5ex]
   SO697 &     K6 &   4115 &   0.2 &  0.97 & -3.11 &  0.67 & -10.05 &  0.16 $\pm$  0.04 &   0.6 $\pm$   0.1 & $<$  33.9         \\ [0.5ex]
   SO726 &     M0 &   3900 &   0.6 &  0.56 & -2.19 &  0.59 &  -9.15 &  0.18 $\pm$  0.04 &   0.7 $\pm$   0.1 & $<$  33.4         \\ [0.5ex]
   SO736 &     K7 &   4020 &   0.1 &  1.49 & -1.48 &  0.55 &  -8.23 & 0.45 $\pm$  0.14 &   2.8 $\pm$   0.1 & $<$35.88           \\ [0.5ex]
   SO739 &   M6.5 &   2815 &   0.1 &   0.1 & -3.06 &   0.1 &  -9.35 & 0.52 $\pm$  0.14 &   2.3 $\pm$   0.6 & $<$ 69.0           \\ [0.5ex]
   SO774 &     K7 &   4020 &   0.0 &  0.49 & -2.75 &   0.7 &  -9.84 & 0.76 $\pm$  0.14 &   3.4 $\pm$   0.1 & 104.2 $\pm$ 15.91  \\ [0.5ex]
   SO818 &     K7 &   4020 &   0.4 &  0.29 & -2.11 &  0.78 &  -9.36 & 1.97 $\pm$  0.15 &   7.5 $\pm$   0.6 & 514.0 $\pm$  58.0  \\ [0.5ex]
   SO823 &     K7 &   4020 &   1.5 &  0.32 & -2.43 &  0.77 &  -9.66 &  0.17 $\pm$  0.04 &   0.6 $\pm$   0.1 & $<$  32.2         \\ [0.5ex]
   SO844 &     M1 &   3720 &   0.7 &  0.62 & -1.37 &  0.44 &  -8.14 & 2.85 $\pm$  0.14 &  15.3 $\pm$   0.1 & 172.14 $\pm$ 16.73 \\ [0.5ex]
   SO848 &     M4 &   3190 &   0.0 &  0.02 & -3.51 &  0.17 & -10.47 & 0.52 $\pm$  0.14 &   1.5 $\pm$   0.4 & $<$ 66.0           \\ [0.5ex]
   SO859 &     M3 &   3410 &   0.6 &  0.41 & -1.72 &  0.29 &  -8.31 & 2.49 $\pm$  0.14 &   9.7 $\pm$   0.6 & $<$ 69.0           \\ [0.5ex]
   SO897 &     K6 &   4115 &   0.6 &  0.85 & -1.34 &   0.7 &  -8.33 & 1.71 $\pm$  0.14 &   6.8 $\pm$   0.1 & 78.54 $\pm$ 15.28  \\ [0.5ex]
   SO927 &     M0 &   3900 &   0.6 &  0.33 & -1.92 &  0.65 &  -9.03 & 1.41 $\pm$  0.15 &   8.0 $\pm$   0.1 & 75.95 $\pm$ 10.77  \\ [0.5ex]
   SO984 &     K7 &   4020 &   0.1 &  0.72 &  -3.5 &  0.64 & -10.46 & 6.07 $\pm$  0.15 &  28.4 $\pm$   0.1 & 276.62 $\pm$ 30.32 \\ [0.5ex]
  SO1036 &     M0 &   3900 &   0.7 &  0.53 & -0.89 &  0.59 &  -7.86 & 5.94 $\pm$  0.25 &  23.6 $\pm$   0.2 & 233.88 $\pm$ 31.42 \\ [0.5ex]
  SO1075 &     M3 &   3410 &   0.6 &  0.14 & -1.38 &   0.3 &  -8.22 & 1.48 $\pm$  0.15 &   5.2 $\pm$   0.5 & 165.0 $\pm$  33.0  \\ [0.5ex]
  SO1152 &     M0 &   3900 &   0.8 &  0.61 & -1.26 &  0.58 &  -8.19 & 8.57 $\pm$  0.29 &  37.5 $\pm$   0.3 & 748.6 $\pm$  35.8  \\ [0.5ex]
  SO1153 &     K5 &   4210 &   1.5 &  0.33 &  0.02 &   0.9 &   -7.3 &13.62 $\pm$  0.27 &  62.5 $\pm$   0.2 & 746.87 $\pm$ 37.91 \\ [0.5ex]
  SO1154 &     K7 &   4020 &   1.8 &  0.08 & -0.78 &  0.62 &  -8.19 & 1.44 $\pm$  0.15 &   7.0 $\pm$   0.1 & $<$33.96           \\ [0.5ex]
  SO1155 &     K4 &   4375 &   0.6 &  1.45 & -0.92 &  0.86 &  -7.94 &  0.41 $\pm$  0.04 &   0.7 $\pm$   0.1 & $<$  34.4         \\ [0.5ex]
  SO1156 &     K6 &   4115 &   0.4 &  0.66 & -1.26 &  0.74 &  -8.33 & 5.66 $\pm$  0.15 &  23.4 $\pm$   0.2 & 263.26 $\pm$ 22.57 \\ [0.5ex]
  SO1248 &   M5.5 &   2920 &   0.0 &  0.18 &  -3.3 &  0.13 &   -9.6 & 0.79 $\pm$  0.15 &   2.9 $\pm$   0.6 & $<$ 72.0           \\ [0.5ex]
  SO1260 &     M4 &   3190 &   0.0 &  0.15 & -1.94 &  0.19 &  -8.53 & $<$  0.4 & $<$   1.5 & $<$  69.0                          \\ [0.5ex]
  SO1266 &   M4.5 &   3085 &   0.0 &  0.07 & -4.76 &  0.15 & -11.36 & $<$  0.5 & $<$   1.7 & $<$  72.0                          \\ [0.5ex]
  SO1267 &     M1 &   3720 &   0.6 &  0.76 & -1.85 &  0.43 &  -8.57 & 2.27 $\pm$  0.15 &   9.0 $\pm$   0.1 & 170.01 $\pm$ 18.73 \\ [0.5ex]
  SO1274 &     K7 &   4020 &   0.0 &  0.68 & -0.98 &  0.64 &  -7.95 &15.38 $\pm$  0.42 &  67.9 $\pm$   0.4 & 1018.3 $\pm$ 38.63 \\ [0.5ex]
  SO1327 &   M4.5 &   3085 &   0.1 &  0.33 & -1.91 &  0.21 &  -8.32 & 1.63 $\pm$  0.16 &   6.0 $\pm$   0.6 & $<$ 75.0           \\ [0.5ex]
  SO1361 &     M1 &   3720 &   0.5 &  0.47 & -0.61 &  0.46 &  -7.46 & 5.34 $\pm$  0.15 &  21.2 $\pm$   0.1 & 208.19 $\pm$ 25.69 \\ [0.5ex]
  SO1362 &   M5.5 &   2920 &   0.0 &   0.1 & -2.96 &  0.13 &  -9.37 & 1.02 $\pm$  0.15 &   3.8 $\pm$   0.6 & $<$ 72.0           \\ [0.5ex]
  SO1369 &     K7 &   4020 &   0.0 &  1.26 & -1.45 &  0.57 &  -8.24 &  1.4 $\pm$  0.15 &   6.8 $\pm$   0.1 & 55.79 $\pm$ 14.13  \\ [0.5ex]
\hline
\end{tabular}
\end{center}
\end{table*}


\subsection{Disk masses}

The disk masses are estimated through their submm ALMA flux at 1.3 mm (band 6) from cycle 3 \citep[C3,][]{Ansdell2017} and, when available, from our new, deeper ALMA observations taken in cycle 4 (C4) and reported here (Sect.~\ref{sec:obs_ALMA}). ALMA continuum fluxes are estimated as in \citet{Ansdell2017}, that is by fitting point-source models to the visibility data using the \textit{uvmodelfit} routine in \textit{CASA}. More information on the ALMA data is reported in Appendix~\ref{sec:appendix_ALMA}, which includes the comparison between ALMA fluxes from C3 and C4 observations in Fig.~\ref{fig:ALMA_fluxes_comparison}. The measured fluxes are reported in Table~\ref{tab:bestfit}. In total, we have 6 new continuum detections from the C4 observations. 
These continuum fluxes were converted to dust masses taking into account the same assumptions as \citet{Manara2023} namely, following \citet{Ansdell2016}, we used a prescription for the opacity, $\kappa_{\nu} = 2.3(\nu/230 \rm GHz) cm^2/g$, taken from \citet{Beckwith1990}. We used a single dust temperature, $T_{\rm dust}$ = 20 K, which has been empirically demonstrated to be a good disk-average value  
\citep{tazzari2021}. The total disk mass is then obtained by multiplying the \mdust\, by a gas-to-dust ratio of 100. 
We rescaled the dust masses of \cite{Ansdell2017}, which were estimated assuming $d=385$ pc. The rescaled dust masses and their errors are reported in Table~\ref{tab:bestfit}. 

The dependence of \mdust\,with the stellar mass is reported in Fig.~\ref{fig:stellar_param_correlation}, and shows a similar trend of increasing dust mass with stellar mass as in other star-forming regions, although with a large spread at \mstar$>$0.4 \msun\,(vertical dashed line). We do not attempt a fit of the relation as in \citet{Ansdell2017}, as we will describe in Sect.~\ref{sec:mstar_corr} how we think that, in \sorionis, the spread is possibly a consequence of external photoevaporation. 

We do not attempt to derive disk gas masses from the new detections of $^{12}$CO in the C4 data. However, we will use the fluxes of $^{12}$CO, measured as in \citet{Ansdell2017} using a curve-of-growth method on the moment 0 maps for the detected targets. In total, the C4 data lead to 13 new $^{12}$CO detections. More information is provided in Appendix~\ref{sec:appendix_ALMA}.

\begin{figure*}
\centering \includegraphics[width = 1.0\textwidth]{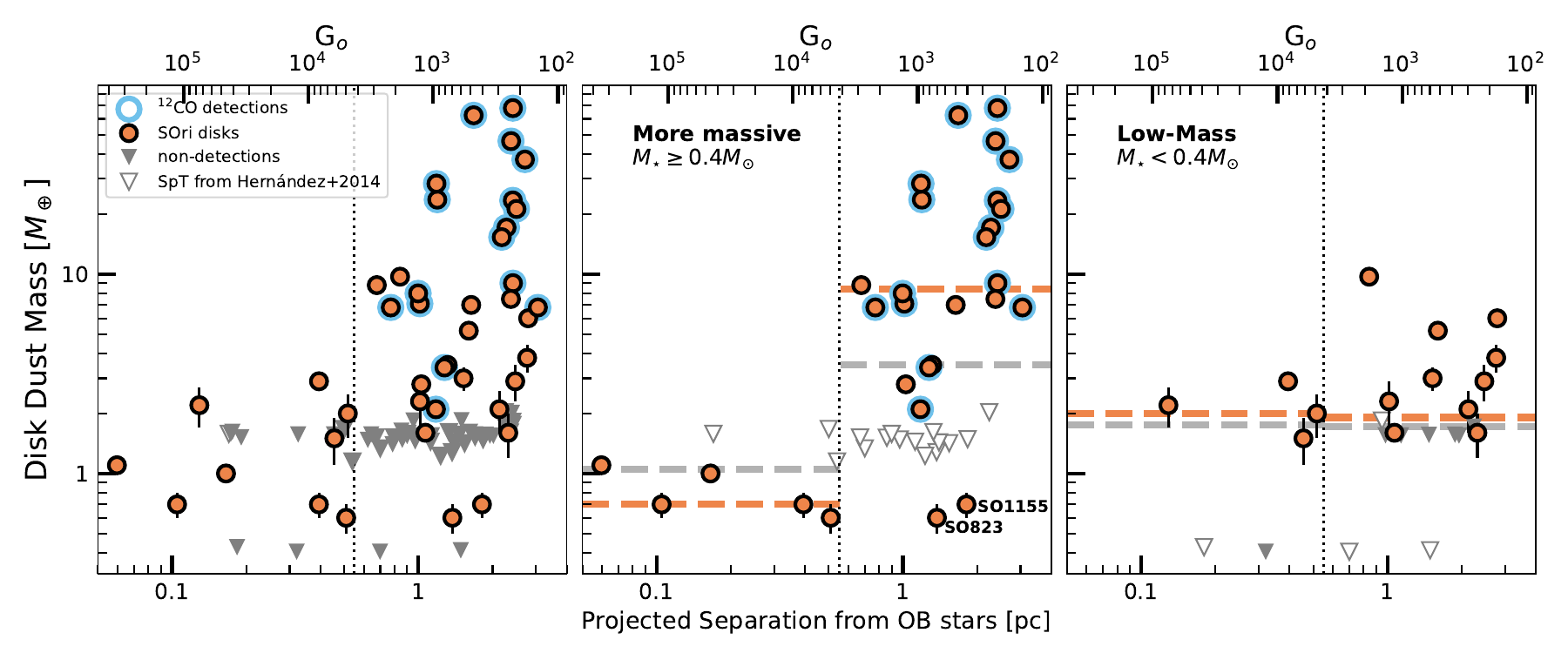}
\caption{Disk dust mass (\mdust) as a function of projected separation from \sori. \textit{Left:} Considering the whole sample of disks with ALMA observations.
\textit{Middle:} Considering the more massive (\mstar$\ge$ 0.4 $M_{\odot}$) stars in our $\sigma$ Orionis sample. \textit{Right:} Considering the less massive ones (\mstar$<$ 0.4 $M_{\odot}$). Dashed lines show the $M_{\rm dust}$ median inside and outside 0.5 pc for our X-Shooter sample (orange) and also including upper limits with reported SpT in \citet{Hernandez2014} (gray). Orange points are continuum detections, downward triangles are 3$\sigma$ upper limits and, $^{12}$CO detections (3$\sigma$) are indicated by an additional blue circle. The $^{12}$CO fluxes are listed in Table~\ref{tab:bestfit}.}\label{fig:mdisk_dp_sori}
\end{figure*}

\section{Discussion} \label{sec:discussion}

\subsection{Dependence of disk mass with projected separation (and UV flux) from \sori} \label{sec:masses_dp}

As discussed in \citet{Ansdell2017}, a dearth of massive ($M \rm _{dust} > 3 M_{\oplus}$) disks close ($<$ 0.5 pc) in projected distance to the central O9 star \sori\,was found in the \sorionis\, region, together with a shallow distance-dependent trend in disk dust mass. This result, similarly found in \citet{Mann2014,Mann2015} for other clusters in Orion, suggested that external photoevaporation may be a viable mechanism for disk depletion. 
In this work, we have included deeper ALMA data with 6 new detections (see Sect.~\ref{sec:obs_ALMA}). The updated \mdust\, distribution as a function of the projected separation from \sori\,is shown in Fig.~\ref{fig:mdisk_dp_sori}. We confirm the lack of any disk more massive than $\sim 3 M_{\oplus}$ in the inner $\sim$0.5 pc of the cluster, and again a shallow distance-dependent trend of \mdust. The new detections further reinforce the limit in the inner part of the cluster, with detections of disks as low mass as less than $1 M_{\oplus}$, and even more stringent upper limits. This strengthens the claim that many disks close to the ionizing star \sori\, have extremely low masses due to its irradiation. 

To further quantify the level at which \sori\,affects the stars, we calculate the FUV radiation field strength due to the central OB system (see Appendix~\ref{sec:UV_field} for details). This is dominated by the radiation of \sori\,alone. The top axis of Fig.~\ref{fig:mdisk_dp_sori} reports this FUV radiation strength expressed in terms of the Habing unit $G_0$ ($G_0 = \rm 1.6 \times 10^{-3}\, erg\, cm^{-2}\, s^{-1}$, \citealt{Habing1968}). The range of FUV values for this region is between $10^{2}$ and $10^{5}$ $G_0$, lower than what is usually observed in the Orion Nebula Cluster \citep[e.g.,][]{Winter2022}, but still significant.  
Indeed, previous findings suggested that even moderate FUV fields ($\ge$ 2$\times 10^3 G_0$) can drive significant disk mass loss \citep{Facchini2016,Kim2016,Haworth2018}, consistent with the observed trend. In particular, the radiation received by a disk at a projected separation of $\sim$0.5 pc from \sori\,is $\sim 10^4 G_0$, and in this range, the disks are found to have severely lower disk masses than at larger distances. However, the most massive disks (\mdust$\gtrsim 10 M_\oplus$) are found only at projected distances larger than $\sim$1 pc, corresponding to FUV fields of $\sim 10^3 G_0$. Moreover, the CO detections, reported in Fig.~\ref{fig:mdisk_dp_sori} as blue circles, are found only at projected distances larger than 0.5 pc, although in a much higher fraction than that reported by \citet{Ansdell2017}, mainly thanks to the deeper observations of C4 that were focused on the disks around higher-mass stars (\mstar\,$\geq\,0.5\,\rm M_{\odot}$) as they tend to have brighter millimeter emission. 

At this lower FUV field strength than the Orion Nebula Cluster or other massive regions, \sorionis\,is thus offering us the unique possibility to study external photoevaporation even at $\sim$3-5 Myr, where the effects are clearly detectable but the disks are not yet (all) dispersed. 
We note that the observed distance-dependent depletion of disks has been reproduced using external photoevaporative models \citep{Winter2020}, although with overestimated (by a factor of 2) disk dust masses. Although, according to \citet{Parker2021}, this could be coincidental, it is interesting to report on this new observational result to further constrain the models. Additional information to further support the external photoevaporation hypothesis is then discussed in the next sections. 

\subsection{Relations with stellar host mass} \label{sec:mstar_corr} 

Thanks to large surveys of young stars performed in various SFRs, global stellar and disk properties have been estimated revealing different relations between the various parameters. Among the well-established ones is that of the \macc\,vs \mstar, with a steeper-than-linear relation roughly as a power law with exponent $\sim$2 \citep[e.g.,][]{Hillenbrand1992,Muzerolle2003,Natta2006}, and reported spreads in \macc\,values of about 1-2 dex \citep[e.g.,][]{Alcala2014,Manara2016a,Manara2017a,Manara2023,venuti2014,venuti2019,Hartmann2016}. Recently, evidence of a double power-law fit of this relation has also been seen \citep{Alcala2017,Manara2017b}, with a very steep relation for the lowest-mass stars (\mstar$<0.2-0.3$ \msun) with slope $\sim$4.5 followed by a flatter relation (slope$\sim$1) at higher \mstar. 
The distribution of the measured \macc\, as a function of the \mstar\, for \sorionis\, sources are shown in the top panel of Fig.~\ref{fig:stellar_param_correlation}. These values reveal a great similarity with those found in other SFRs, like Lupus \citep{Alcala2017}, Chamaeleon I \citep{Manara2017a}, and even the older USco SFR \citep{Manara2020}. A flatter dependence of \macc\,on \mstar\,seems to be present at the highest \mstar\,even in our sample, suggesting that the broken power-law could be a better fit to the data, in line with previous studies. 

The similar range of \macc\,as in other typically younger SFRs is at odds with the usually assumed decline of \macc\,with age, a prediction of viscous evolution \citep[e.g.,][]{Hartmann1998}. This is however nowadays observed in several regions, from Orion OB1 \citep{Ingleby2014,Manara2021,Pittman2022}, to TWA \citep{venuti2019}, $\eta$-Cha \citep{Rugel2018}, or even in the 30 Dor region \citep{DeMarchi2017}. 
The reason why disks can have such a high accretion rate for a time not compatible with the amount of mass accreted over their lifetime and the total available mass in the disk, is still the subject of discussion \citep[e.g.,][]{Hartmann2006} and, it is possibly related with episodic accretion or other mechanisms \citep{Manara2020} but, in our specific case, it could be a selection effect due to a combination of enhanced accretion due to the effects of external photoevaporation \citep{Rosotti2017}, and the focus on just the disks that are not fully dispersed yet.
Similarly to other star-forming regions, a large scatter of \macc\, at any \mstar\, is observed for the  \sorionis\,sources. 
Such a spread has been demonstrated not to be due to accretion variability or other sources of uncertainty \citep[e.g.,][for a review]{Manara2023} and its origin remains an open question. 
As also shown in \citet{Rigliaco2012,Winter2020}, we find a positive correlation between \macc\,and \mstar\,and no correlation of \macc\,with proximity to \sori. 

The bottom panel of Fig.~\ref{fig:stellar_param_correlation} shows another correlation also well established empirically for individual regions, the \mdust\, vs \mstar\,relation. Several works surveying different SFRs have shown that \mdust\,directly depends on \mstar\,with a slope around 1.8-2.7 with the larger values describing the older Upper Scorpius region \citep{Ansdell2016,Ansdell2017,Barenfeld2016,Pascucci2016,Manara2023}, and holds down to the brown dwarf regime \citep[e.g.,][]{testi2016,sanchis2021,rilinger2021}. The steepening with age has been interpreted as faster evolution of dust around low-mass stars, whether as a result of more efficient conversion of millimeter grains into larger centimeter grains or more efficient radial drift. Interestingly, the dispersion around the relation is very similar in all the regions ($\sim$0.8 dex).
In the case of the \sorionis\,cluster, we find similar results as \citet{Ansdell2017} with sources populating a similar locus on this plane as in other SFRs. We also find a large scatter in \mdust\,for a given stellar mass, particularly large around the more massive stars (\mstar\,$\ge$ 0.4 \msun) in our sample.  
Since the dispersion is present for all regions, regardless of age and environment, it has been acknowledged as an inherent property of disk populations resulting from the range of disk initial conditions and has been explained theoretically, by invoking a mixture of both the initial conditions and the evolutionary process \citep{Pascucci2016,pinilla2020}. However, we think that the origin of this dispersion at high stellar masses (\mstar\,$\ge$ 0.4 \msun) is possibly related to the effects of the massive star \sori\,on the surrounding disks, as we discuss in the next subsection. 

\subsubsection{The effect of stellar mass on the disk mass depletion} \label{sec:photo_massvie}

In the middle and right panels of Fig.~\ref{fig:mdisk_dp_sori}, we show the distribution of \mdust\,as a function of projected separation from \sori\, for stars with $M_{\star} \ge 0.4\,M_{\odot}$ and $M_{\star} < 0.4\,M_{\odot}$, respectively. Dashed lines indicate the median values of \mdust\,for sources inside and outside a projected distance of 0.5 pc from the position of \sori. Since SpT estimates are available from \citet{Hernandez2014} for a sub-sample of stars with \mdust\,upper limits (gray triangles on the left panel) and without X-Shooter spectra (i.e., without stellar mass estimates), we have added them as white downward triangles on these panels. 
Our SpT estimates are in good agreement within the uncertainties with those reported in \citet{Hernandez2014}. The only three targets that deviate more than expected are two strong accretors (SO562, SO1075) and one highly extincted star (SO823).
We assigned the objects with SpT earlier than M2 in the higher mass panel, and for later SpT to the lower mass panel. The choice is motivated by the correspondence between SpT and \mstar\, found in the X-Shooter sample. 
The \mdust\,medians taking into account these additional values are shown with a gray dashed line, while those estimated from the X-Shooter sample alone are shown with an orange dashed line.   

Looking at Fig.~\ref{fig:mdisk_dp_sori} we note that, within the inner 0.5 pc from \sori, the more massive (\mstar\,$\ge$ 0.4 $M_{\odot}$) stars in \sorionis\,(middle panel), show \mdust\,about an order of magnitude lower than the more distant ones considering only the targets with measured \mstar\,(orange dashed lines), or about 4 times lower when including those where only the SpT is measured (gray dashed lines), although, in this case, the median inside 0.5 pc is more uncertain given the less stringent upper limits. By contrast, low-mass stars (\mstar\,$<$ 0.4 $M_{\odot}$) have an apparent constant distribution of \mdust, regardless of their distance from the ionizing stars (right panel). Even though this trend will still hold including the additional ($\sim$19) upper limits shown in the left panel for which no SpT nor \mstar\, estimate exists (as these upper limits are of the same order as our detections), this apparent flatness in the \mdust\, distribution for the low-mass stars in our sample is surely affected by the low numbers statistics in this stellar mass range, mainly due to the distance of the cluster (d = 401 pc) which makes it harder to survey low-mass stars with respect to closer star-forming regions. 
It could be, therefore, that there are
more low-mass stars inside 0.5 pc that were not targeted in the ALMA surveys because they were not part of the initial \textit{Spitzer} catalogs. If these are fainter at mm-wavelengths than our targets, the few low-mass objects that are detected in close proximity to \sori\,could represent the high upper tail of the low-mass distribution. Deeper ALMA observations on these additional targets along with spectroscopic follow-up are needed in order to probe the apparent flatness of the \mdust\,distribution of the low-mass stars in \sorionis.
At the same time, the lower median \mdust\,for the low-mass stars compared to the more massive ones in the outer part of the cluster (beyond 0.5 pc), is due to the known steep dependence of \mdust\,with \mstar\,just discussed. It is possible, therefore, to ascribe the differences in the outer part of the cluster to other (internal) effects related to the evolution of disks as well \citep{Pascucci2016,pinilla2020}.

The large difference between the median \mdust\,for the more massive stars inside and outside projected distances of 0.5 pc from \sori\, points, instead, to environmental factors, like external photoevaporation, affecting the closest disks to \sori, decreasing significantly their \mdust, as discussed in Sect.~\ref{sec:masses_dp}. We note that this discrepancy holds even considering the additional upper limits for targets without \mstar\, estimates from the spectroscopy presented in this work (gray dashed lines). Note as well that this discrepancy can be even larger if the two outliers (SO823 and SO1155, see Sect.~\ref{sec:sample}) are not taken into account. Although the disks in the low-mass sample are in general less massive, as expected due to their faster dust evolution, the median \mdust\,within the innermost region of the cluster is still lower for the high-mass star sample than for the low-mass star regime (see Fig.~\ref{fig:mdisk_dp_sori}). It is worth discussing, therefore, why such an effect is observed.

A possible solution to this puzzling result could be that the effects of external photoevaporation depend on the stellar mass of the host star in a more complex fashion than what is typically assumed. Indeed, for the fact that the gravitational potential is stronger for higher-mass stars, it is usually assumed that photoevaporation is more effective around lower-mass stars. This however is a very simplistic assumption, since it is known that the disk radii depend on the stellar mass as well, albeit indirectly through the already mentioned dependence of the continuum flux with the disk radii, and the fact that the disk masses are measured from the continuum flux. If the relation between the disk radii and the stellar mass is not linear, then external photoevaporation should affect the disks in a different way depending on the (unperturbed) disk radius. 

External photoevaporation would result in a lower disk mass obtained as a result of eroding the disk in the outer regions, at disk radii ($R_{\rm disk}$) larger than the gravitational radius, defined as $R_{\rm grav} = (G M_\star)/c_s^2$ in an isothermal system, where $c_s$ is the sound speed \citep{Winter2022}, or even down to 0.15 $\cdot ~R_{\rm grav}$ \citep{Adams2004}, although with lower mass-loss rates. If disks are eroded by this process, we expect the disk radii to be typically smaller than $R_{\rm grav}$. Unfortunately, the spatial resolution of our observations ($\sim$0.2''$\sim$80 au, see Sect.~\ref{sec:obs_ALMA}) is not sufficient to properly resolve the disks. 
However, we obtain indirect estimates of the disk radii using the measured continuum flux, known to correlate with the disk dust radii \citep{Tripathi2017,Andrews2018,Long2022}, and the measured $^{12}$CO fluxes, which can be related to the disk gas sizes under the assumption that the emission is optically thick \citep[e.g.,][]{Zagaria2023,toci2023,trapman2023}. In the cases where the $^{12}$CO is not detected, it is possible to extrapolate the gas radii from the dust radii assuming a ratio of 3, found here for the targets with both continuum and $^{12}$CO detections, and typically found in other star-forming regions \citep[e.g.,][]{Ansdell2018}. 
We note that this procedure is based on several assumptions, and, in particular, the latter is most probably not valid in the case of external photoevaporation, which mainly affects the gaseous component of the disk, where we expect a lower gas-to-dust radii ratio. 

Assuming $c_s=1$ km/s (which gives $\sim$120 K in 1000 $G_{0}$ environment) as representative for our sample, we can compare the inferred disk radii with the inferred gravitational radii ($R_{\rm disk}$/$R_{\rm grav}$) for each target. Although with many caveats, this analysis results in disk radii that are always smaller than the gravitational radii for all the stars in the cluster, with the lowest ratios ($R_{\rm disk}$/$R_{\rm grav}$ $<$ 0.1) for disks around stars \mstar$>$0.4 \msun\, and with projected separation from \sori\,smaller than 0.5 pc, whereas they are larger in the outer part of the cluster.
This is in line with the expectations of the imprint of external photoevaporation, with a stronger effect on the inner regions of the cluster. As shown in \citet{Adams2004}, the mass-loss rate due to externally irradiated disks can still be significant even for disk radii much smaller than the gravitational radius, in particular for $R_{\rm disk}$/$R_{\rm grav} > $ 0.15, as we found in the outer part of the cluster.  This reinforces the claim that even at intermediate FUV radiation fields (1-1000 $G_0$) the effects of this process can have a significant impact on the evolution of protoplanetary disks \citep{vanTerwisga2023}. However, we note that the dependence of \mdisk\, with the distance from \sori\, is not as steep as it would be expected from the results of  \citet{vanTerwisga2023}. 
The disks around the lowest mass stars, however, seem to have a constant ratio $R_{\rm disk}$/$R_{\rm grav}\sim$0.4 regardless of the distance to \sori, which is a consequence of the flat distribution of fluxes (and disk dust masses) with projected distance from \sori\,shown in Fig.~\ref{fig:mdisk_dp_sori}. 

With all the several assumptions of our approach, namely the dependence of continuum and gas emission with the disk radii, the ratio between gas and dust disk radii, and the sensitivity of the $R_{\rm disk}$/$R_{\rm grav}$ ratio to the value assumed for $c_s$, our approach points to a different dependence of the effect of external photoevaporation with stellar host mass. This is particularly evident in the dependence of \mdisk\, with the projected distance from \sori\, (Fig.~\ref{fig:mdisk_dp_sori}). 

Our findings suggest that the large spread in the \mdisk-\mstar\, relation observed for disks around stars with \mstar$> 0.4 M_\odot$ is an effect of the environment in the \sorionis\, cluster. 
If confirmed, this would shed new light on the evolution of the \mdisk-\mstar\, relation with age, which is mainly driven by the large scatter \citep[e.g.,][]{Manara2023}, leading to an interpretation where, at least for the mid-aged \sorionis\, region, the steepening of the relation is an effect of external photoevaporation. 
Work should be done in trying to properly measure disk radii in these systems, particularly around low-mass stars, to confirm whether they are less affected by external photoevaporation, or whether the different behavior with respect to the disks around higher-mass stars is due to other processes. 

\begin{figure}
\centering \includegraphics[width = 0.47\textwidth]{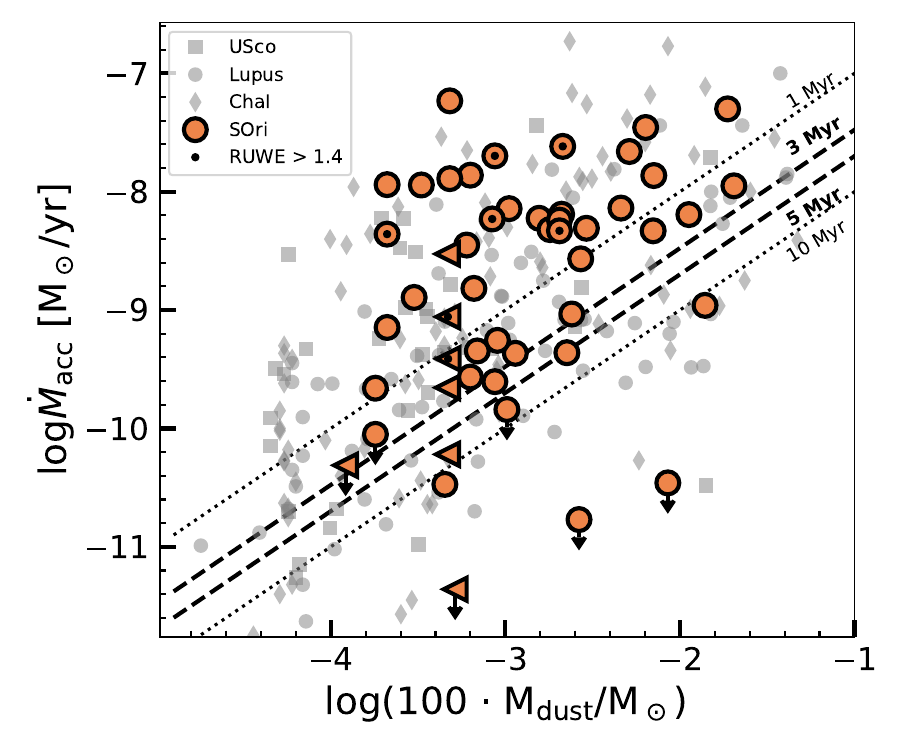}
\caption{Distribution of the \macc--\mdisk\,in \sorionis. The triangles indicate the upper limit on \mdisk, while the vertical arrows correspond to the upper limits on the \macc. The dotted and dashed lines are the isochrones at some relevant ages. The ones in bold are related to the estimated age of the cluster, i.e. 3-5 Myr \citep{Oliveira2004}. Values for other star-forming regions \citep{Manara2023} are shown as gray symbols for comparison.}\label{fig:macc_mdisk_sori}
\end{figure}

\begin{figure}
\centering \includegraphics[width = 0.47\textwidth]{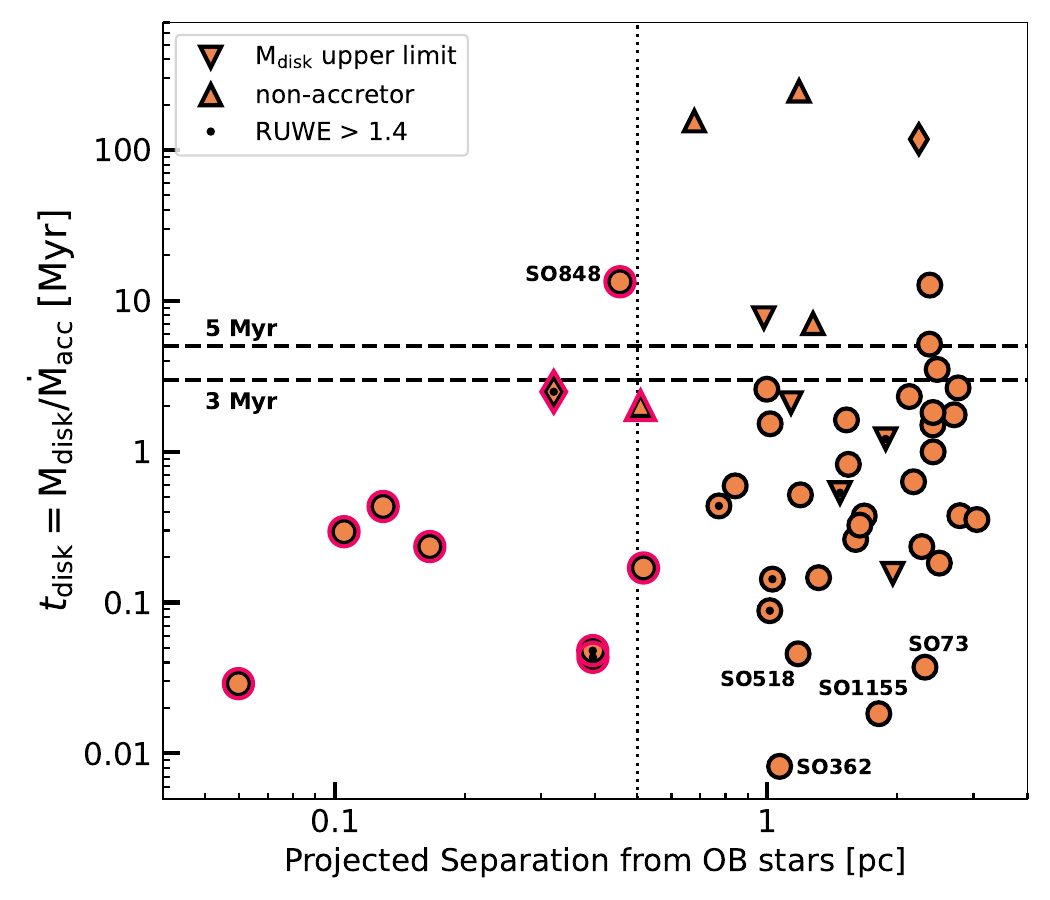}
\caption{Distribution of $t_{\rm disk}$ as a function of projected separation from \sori. The triangles indicate the upper/lower limits on \macc\, and \mdisk. The vertical dotted line is located at 0.5 pc from \sori. Targets within 0.5 pc are highlighted with an additional red outline. The dashed lines are related to the estimated age of the cluster, i.e. 3-5 Myr \citep{Oliveira2004}. }\label{fig:tdisk_dp_sori}
\end{figure}

\subsection{\macc--\mdisk\, plane as a proxy of Disk Evolution}

According to the disk viscous evolution framework,  \macc\,should directly correlate with \mdisk\,\citep[e.g.,][]{Hartmann1998,Rosotti2017,Lodato2017,Mulders2017,Manara2023}. The viscous quasi-steady state is characterized by the condition \mdisk $\sim$ \macc$\tau$, with $\tau$ as the viscous time-scale at the outer radius of the disk \citep{Rosotti2017}. One property of this paradigm is that $\tau$ is of the order of the system age independent of the initial conditions and the assumptions on disk viscosity \citep{Jones2012,Lodato2017}. Therefore, the ratio \mdisk/\macc, the so-called "disk lifetime" ($t_{\rm disk}$) can be used as a proxy of disk evolution \citep{Manara2016b,Manara2023,Rosotti2017}. The dependence between \mdisk\, and \macc\, has been explored extensively in the literature and found to be almost linear, albeit with a very large scatter \citep[e.g.,][]{Manara2016b,Manara2020,Manara2023,Mulders2017}.

The origin of the observed scatter at all ages is still unclear, although it points toward particular conditions in the viscous framework \citep{Lodato2017}, or to the necessity to include other mechanisms to explain the observations. 
Both \citet{Rosotti2017} and \citet{Zagaria2022} suggest that external disturbances, such as external photoevaporation or multiplicity, lead to shorter disk lifetimes, that is higher \macc\, than the value expected by viscous evolution corresponding to the measured disk mass. 
\citet{Zagaria2022} found that multiplicity can explain the high accretors found in the Upper Scorpius region \citep{Manara2020}. 

The data presented in this work allows us, for the first time, to test whether the \macc-\mdisk\, relation can be used to confirm the effect of external photoevaporation on disks close to a massive star. 
Fig.~\ref{fig:macc_mdisk_sori} shows the \macc--\mdisk\, plane for our $\sigma$-Orionis disk sample. 
We highlight the $t_{\rm disk}$ = 3 Myr and, 5 Myr (dashed lines), representative of the age of the cluster \citep{Oliveira2004, Hernandez2014}, for reference. 
We observe that the majority of the targets are located at shorter disk lifetimes than the age of the region, in line with the expectations from external photoevaporation models \citep{Rosotti2017}.
In particular, 28 targets ($\sim54\%$) lay above the 1 Myr line, 17 targets ($\sim34\%$) are between the 1 Myr and 10 Myr lines, while the remaining five targets ($\sim10\%$) are below the 10 Myr line, and they are mainly non-accreting objects. 
This points toward confirming the effect of external photoevaporation on the evolution of these disks. 

We note, however, that the distribution of data on the \macc-\mdisk\,plane is similar to what is observed in other SFRs. 
According to \citet{Zagaria2022}, most of the stars in Lupus, Chameleon I, and USCo SFRs that have higher \macc\,given their \mdisk\,can be explained by multiplicity (tidally truncated disks), with the bulk of the binary population being clustered around \mdisk/\macc\,= 0.1 Myr. Unfortunately, we do not have multiplicity information for our \sorionis\,sample to further test this scenario, but we have indicated in the plots the stars with RUWE values greater than 1.4, which may point to possible binaries in the cluster. Interestingly, most of the targets with high RUWE have also short disk lifetimes, suggesting that binarity might play a role also in the \sorionis\, cluster in the observed spread in the \macc-\mdisk\, relation. 

To further check whether the short disk lifetimes could be instead related to the presence of the massive \sori\, star, we show in Fig.~\ref{fig:tdisk_dp_sori} how $t_{\rm disk}$ depends on the projected distance to the massive \sori\,system. 
As shown, \textit{all} objects within 0.5 pc from \sori\,(red circles) have $t_{\rm disk} <$ 0.5 Myr, while disks further out can reach higher values. Outliers, having $t_{\rm disk} <$ 0.05 Myr at 1 pc or beyond, correspond to objects whose distances deviate in more than $\sim$40 pc to the median (SO73, SO848), strong accretors (SO1155, SO362) and/or edge-on disks candidates (SO518). 
The low disk lifetimes of the disks closest to the OB stars along with the distance-dependent trend in disk dust mass shown in Fig.~\ref{fig:mdisk_dp_sori}, robustly evidence the fact that, at least within 0.5 pc from the center, the disks are actively being externally photoevaporated. 

The dependence of the disk lifetime with the projected separation from \sori\, further suggests that, despite the similar distribution on the \macc-\mdisk\, plane as in other regions, the large spread observed in our \sorionis\,sample also supports the outside-in depletion of these disks. As stated in \citet{Lodato2017}, from disk population synthesis models, a tighter \mdisk-\macc\,correlation is expected at longer ages, so the fact that these sources show a similar spread, even at these intermediate ages, to other younger SFRs implies a more significant deviation of these stars from purely viscous evolution.   
Enlarging the sample on the low disk mass side, by adding additional disk detections to the available spectroscopic data, would constrain quantitatively how many disks are consistent with the effects of external photoevaporation, or whether other effects must be considered in order to explain the observations, such as the effects of dust evolution \citep[e.g.,][]{Sellek2020a} or binarity \citep[e.g.,][]{Zagaria2022}. 


\section{Conclusions} \label{sec:conclusions}

We conducted the first large-scale survey with both, UV-to-NIR spectroscopy with X-Shooter, and mm-interferometry with ALMA, for disk-bearing stars in the mid-age \sorionis\,cluster. We have derived the stellar and accretion properties of 50 targets, and shown new ALMA detections to complement the data presented by \citet{Ansdell2017}.
This has allowed us to test the effect of external photoevaporation from the massive star \sori\, on the surrounding population of disks. 

Our main conclusions are:

\begin{itemize}

    \item The disks in the \sorionis\,cluster show similar values and spread in the \macc\,--\mstar\, and \mdisk\,--\mstar\,relations as those in surveys of protoplanetary disks in other young SFRs. No correlation of \macc\,with proximity to \sori\,was found, in agreement with previous works. 

    \item We confirm the trend of decreasing \mdust\, at shorter distances from the massive star \sori, as expected from external photoevaporation.
    Disks around more massive stars show a more pronounced reduction in their masses if they are located in the inner 0.5 pc of the cluster than disks in the outer regions. They were also found to have the smallest $R_{\rm disk}$/$R_{\rm grav}$ at these separations, which corresponds to a value of FUV radiation of $\sim 10^4 G_0$. This effect is less pronounced in the lowest mass stars, either due to a stellar mass-dependent effect of external photoevaporation or to observational biases. Due to the low number statistics, the conclusions for the low-mass regime are still to be firmly established.
    Our results stress the need to develop a deeper understanding of disk evolution around very low-mass stars in clustered environments. 

    \item Half of the sample lies in the expected region for externally irradiated disks on the \macc\,vs \mdisk\, plane, showing disk lifetime ($t_{\rm disk}$) lower than expected given the age of the system. This implies that external photoevaporation may be a viable mechanism for disk depletion in the cluster.
    
    \item We found a tentative increasing trend of $t_{\rm disk}$ with projected separation from the massive OB stars. Within the first 0.5 pc, sources have very low $t_{\rm disk}$ ($\le 0.5$ Myrs). This strengthens the claim that outside-in depletion plays an important role in the evolution of disks, particularly those that are in close proximity ($<$ 0.5 pc) to the central OB system \sori.  
    
\end{itemize}

While this work has shown the power of combining information on disk properties with measurements of stellar and accretion parameters as a function of projected separation from the massive OB-system \sori, the final tell-tale test of external photoevaporation in this region is to detect the photoevaporating winds in these targets. 
A detailed study on wind tracers and mass-loss diagnostic (e.g., optical forbidden emission lines) of these sources using X-Shooter and high-resolution spectra can potentially confirm the above result and put better constraints on disk dispersal mechanisms in clustered environments \citep[e.g.,][]{Yasuhiro2022}. This has been attempted in a limited number of targets \citep{Rigliaco2009,Gangi2023}, and it will be assessed in a future paper (Maucó et al., in prep).

\begin{acknowledgements}

We thank the anonymous referee for the critical review of our work that improved the presented study.
Funded by the European Union (ERC, WANDA, 101039452). Views and opinions expressed are however those of the author(s) only and do not necessarily reflect those of the European Union or the European Research Council Executive Agency. Neither the European Union nor the granting authority can be held responsible for them.
This work benefited from discussions with the ODYSSEUS team (HST AR-16129), \url{https://sites.bu.edu/odysseus/}.
This research received financial support from the project PRIN-INAF 2019 "Spectroscopically Tracing the Disk Dispersal Evolution" (STRADE) and from the Large Grant INAF 2022 YODA (YSOs Outflows, Disks and Accretion: towards a global framework for the evolution of planet-forming systems). TJH is funded by a Royal Society Dorothy Hodgkin Fellowship. S.F. is funded by the European Union under the European Union's Horizon Europe Research \& Innovation Programme 101076613 (UNVEIL). Views and opinions expressed are however those of the author(s) only and do not necessarily reflect those of the European Union or the European Research Council. Neither the European Union nor the granting authority can be held responsible for them.
This paper makes use of the following ALMA data: ADS/JAO.ALMA\#2016.1.00447.S. ALMA is a partnership of ESO (representing its member states), NSF (USA) and NINS (Japan), together with NRC (Canada), MOST and ASIAA (Taiwan), and KASI (Republic of Korea), in cooperation with the Republic of Chile. The Joint ALMA Observatory is operated by ESO, AUI/NRAO and NAOJ.
This project has received funding from the European Union's Horizon 2020 research and innovation program under the Marie Sklodowska-Curie grant agreement No 823823 (DUSTBUSTERS).
This work was partly supported by the Deutsche Forschungs-Gemeinschaft (DFG, German Research Foundation) - Ref no. FOR 2634/1 TE 1024/1-1. 

This work has made use of data from the European Space Agency (ESA) mission
{\it Gaia} (\url{https://www.cosmos.esa.int/gaia}), processed by the {\it Gaia}
Data Processing and Analysis Consortium (DPAC,
\url{https://www.cosmos.esa.int/web/gaia/dpac/consortium}). Funding for the DPAC
has been provided by national institutions, in particular, the institutions
participating in the {\it Gaia} Multilateral Agreement.

\end{acknowledgements}

\bibliography{bibliography}

\begin{thebibliography}{123}
\expandafter\ifx\csname natexlab\endcsname\relax\def\natexlab#1{#1}\fi

\bibitem[{{Abergel} {et~al.}(2003){Abergel}, {Teyssier}, {Bernard},
  {Boulanger}, {Coulais}, {Fosse}, {Falgarone}, {Gerin}, {Perault}, {Puget},
  {Nordh}, {Olofsson}, {Huldtgren}, {Kaas}, {Andr{\'e}}, {Bontemps}, {Casali},
  {Cesarsky}, {Copet}, {Davies}, {Montmerle}, {Persi}, \&
  {Sibille}}]{abergel2003}
{Abergel}, A., {Teyssier}, D., {Bernard}, J.~P., {et~al.} 2003, \aap, 410, 577

\bibitem[{{Adams}(2010)}]{Adams2010}
{Adams}, F.~C. 2010, \araa, 48, 47

\bibitem[{{Adams} {et~al.}(2004){Adams}, {Hollenbach}, {Laughlin}, \&
  {Gorti}}]{Adams2004}
{Adams}, F.~C., {Hollenbach}, D., {Laughlin}, G., \& {Gorti}, U. 2004, \apj,
  611, 360

\bibitem[{{Alcal{\'a}} {et~al.}(2017){Alcal{\'a}}, {Manara}, {Natta}, {Frasca},
  {Testi}, {Nisini}, {Stelzer}, {Williams}, {Antoniucci}, {Biazzo}, {Covino},
  {Esposito}, {Getman}, \& {Rigliaco}}]{Alcala2017}
{Alcal{\'a}}, J.~M., {Manara}, C.~F., {Natta}, A., {et~al.} 2017, {\aap}, 600,
  A20

\bibitem[{{Alcal{\'a}} {et~al.}(2014){Alcal{\'a}}, {Natta}, {Manara}, {Spezzi},
  {Stelzer}, {Frasca}, {Biazzo}, {Covino}, {Randich}, {Rigliaco}, {Testi},
  {Comer{\'o}n}, {Cupani}, \& {D'Elia}}]{Alcala2014}
{Alcal{\'a}}, J.~M., {Natta}, A., {Manara}, C.~F., {et~al.} 2014, {\aap}, 561,
  A2

\bibitem[{{Anderson} {et~al.}(2013){Anderson}, {Adams}, \&
  {Calvet}}]{Anderson2013}
{Anderson}, K.~R., {Adams}, F.~C., \& {Calvet}, N. 2013, \apj, 774, 9

\bibitem[{{Andrews} {et~al.}(2018){Andrews}, {Huang}, {P{\'e}rez}, {Isella},
  {Dullemond}, {Kurtovic}, {Guzm{\'a}n}, {Carpenter}, {Wilner}, {Zhang}, {Zhu},
  {Birnstiel}, {Bai}, {Benisty}, {Hughes}, {{\"O}berg}, \&
  {Ricci}}]{Andrews2018}
{Andrews}, S.~M., {Huang}, J., {P{\'e}rez}, L.~M., {et~al.} 2018, The
  Messenger, 174, 19

\bibitem[{{Andrews} {et~al.}(2013){Andrews}, {Rosenfeld}, {Kraus}, \&
  {Wilner}}]{Andrews2013}
{Andrews}, S.~M., {Rosenfeld}, K.~A., {Kraus}, A.~L., \& {Wilner}, D.~J. 2013,
  \apj, 771, 129

\bibitem[{{Ansdell} {et~al.}(2017){Ansdell}, {Williams}, {Manara}, {Miotello},
  {Facchini}, {van der Marel}, {Testi}, \& {van Dishoeck}}]{Ansdell2017}
{Ansdell}, M., {Williams}, J.~P., {Manara}, C.~F., {et~al.} 2017, The \aj, 153,
  240

\bibitem[{{Ansdell} {et~al.}(2018){Ansdell}, {Williams}, {Trapman}, {van
  Terwisga}, {Facchini}, {Manara}, {van der Marel}, {Miotello}, {Tazzari},
  {Hogerheijde}, {Guidi}, {Testi}, \& {van Dishoeck}}]{Ansdell2018}
{Ansdell}, M., {Williams}, J.~P., {Trapman}, L., {et~al.} 2018, \apj, 859, 21

\bibitem[{{Ansdell} {et~al.}(2016){Ansdell}, {Williams}, {van der Marel},
  {Carpenter}, {Guidi}, {Hogerheijde}, {Mathews}, {Manara}, {Miotello},
  {Natta}, {Oliveira}, {Tazzari}, {Testi}, {van Dishoeck}, \& {van
  Terwisga}}]{Ansdell2016}
{Ansdell}, M., {Williams}, J.~P., {van der Marel}, N., {et~al.} 2016, \apj,
  828, 46

\bibitem[{{Antoniucci} {et~al.}(2014){Antoniucci}, {Garc{\'\i}a L{\'o}pez},
  {Nisini}, {Caratti o Garatti}, {Giannini}, \& {Lorenzetti}}]{Antoniucci2014}
{Antoniucci}, S., {Garc{\'\i}a L{\'o}pez}, R., {Nisini}, B., {et~al.} 2014,
  \aap, 572, A62

\bibitem[{{Bacciotti} {et~al.}(2011){Bacciotti}, {Whelan}, {Alcal{\'a}},
  {Nisini}, {Podio}, {Randich}, {Stelzer}, \& {Cupani}}]{Bacciotti2011}
{Bacciotti}, F., {Whelan}, E.~T., {Alcal{\'a}}, J.~M., {et~al.} 2011, \apjl,
  737, L26

\bibitem[{{Ballabio} {et~al.}(2023){Ballabio}, {Haworth}, \&
  {Henney}}]{ballabio2023}
{Ballabio}, G., {Haworth}, T.~J., \& {Henney}, W.~J. 2023, \mnras, 518, 5563

\bibitem[{{Baraffe} {et~al.}(2015){Baraffe}, {Homeier}, {Allard}, \&
  {Chabrier}}]{Baraffe2015}
{Baraffe}, I., {Homeier}, D., {Allard}, F., \& {Chabrier}, G. 2015, {\aap},
  577, A42

\bibitem[{{Barenfeld} {et~al.}(2016){Barenfeld}, {Carpenter}, {Ricci}, \&
  {Isella}}]{Barenfeld2016}
{Barenfeld}, S.~A., {Carpenter}, J.~M., {Ricci}, L., \& {Isella}, A. 2016,
  \apj, 827, 142

\bibitem[{{Beckwith} {et~al.}(1990){Beckwith}, {Sargent}, {Chini}, \&
  {Guesten}}]{Beckwith1990}
{Beckwith}, S. V.~W., {Sargent}, A.~I., {Chini}, R.~S., \& {Guesten}, R. 1990,
  \aj, 99, 924

\bibitem[{{B{\'e}jar} {et~al.}(1999){B{\'e}jar}, {Zapatero Osorio}, \&
  {Rebolo}}]{Bejar1999}
{B{\'e}jar}, V.~J.~S., {Zapatero Osorio}, M.~R., \& {Rebolo}, R. 1999, \apj,
  521, 671

\bibitem[{{Brice{\~n}o} {et~al.}(2019){Brice{\~n}o}, {Calvet}, {Hern{\'a}ndez},
  {Vivas}, {Mateu}, {Downes}, {Loerincs}, {P{\'e}rez-Blanco}, {Berlind},
  {Espaillat}, {Allen}, {Hartmann}, {Mateo}, \& {Bailey}}]{briceno2019}
{Brice{\~n}o}, C., {Calvet}, N., {Hern{\'a}ndez}, J., {et~al.} 2019, \aj, 157,
  85

\bibitem[{{Brown} {et~al.}(1994){Brown}, {de Geus}, \& {de Zeeuw}}]{Brown1994}
{Brown}, A.~G.~A., {de Geus}, E.~J., \& {de Zeeuw}, P.~T. 1994, \aap, 289, 101

\bibitem[{{Caballero}(2007)}]{caballero2007}
{Caballero}, J.~A. 2007, \aap, 466, 917

\bibitem[{{Calvet} \& {Gullbring}(1998)}]{Calvet1998}
{Calvet}, N. \& {Gullbring}, E. 1998, \apj, 509, 802

\bibitem[{{Cardelli} {et~al.}(1989){Cardelli}, {Clayton}, \&
  {Mathis}}]{Cardelli1989}
{Cardelli}, J.~A., {Clayton}, G.~C., \& {Mathis}, J.~S. 1989, {\apj}, 345, 245

\bibitem[{{Clarke}(2007)}]{Clarke2007}
{Clarke}, C.~J. 2007, Monthly Notices of the Royal Astronomical Society, 376,
  1350

\bibitem[{{Coleman} \& {Haworth}(2022)}]{coleman2022}
{Coleman}, G. A.~L. \& {Haworth}, T.~J. 2022, \mnras, 514, 2315

\bibitem[{{Cuello} {et~al.}(2023){Cuello}, {M{\'e}nard}, \&
  {Price}}]{cuello2023}
{Cuello}, N., {M{\'e}nard}, F., \& {Price}, D.~J. 2023, European Physical
  Journal Plus, 138, 11

\bibitem[{{Da Rio} {et~al.}(2014){Da Rio}, {Jeffries}, {Manara}, \&
  {Robberto}}]{DaRio2014}
{Da Rio}, N., {Jeffries}, R.~D., {Manara}, C.~F., \& {Robberto}, M. 2014,
  \mnras, 439, 3308

\bibitem[{{Damian} {et~al.}(2023{\natexlab{a}}){Damian}, {Jose}, {Biller},
  {Herczeg}, {Albert}, {Allers}, {Zhang}, {Liu}, {Dubber}, {Paul}, {Chen},
  {Lalchand}, {Sharma}, \& {Oasa}}]{damian2023a}
{Damian}, B., {Jose}, J., {Biller}, B., {et~al.} 2023{\natexlab{a}}, arXiv
  e-prints, arXiv:2303.17424

\bibitem[{{Damian} {et~al.}(2023{\natexlab{b}}){Damian}, {Jose}, {Biller}, \&
  {Paul}}]{damian2023b}
{Damian}, B., {Jose}, J., {Biller}, B., \& {Paul}, K. 2023{\natexlab{b}}, arXiv
  e-prints, arXiv:2305.18147

\bibitem[{{De Marchi} {et~al.}(2017){De Marchi}, {Panagia}, \&
  {Beccari}}]{DeMarchi2017}
{De Marchi}, G., {Panagia}, N., \& {Beccari}, G. 2017, \apj, 846, 110

\bibitem[{{Ercolano} \& {Pascucci}(2017)}]{Ercolano2017}
{Ercolano}, B. \& {Pascucci}, I. 2017, Royal Society Open Science, 4, 170114

\bibitem[{{Facchini} {et~al.}(2016){Facchini}, {Clarke}, \&
  {Bisbas}}]{Facchini2016}
{Facchini}, S., {Clarke}, C.~J., \& {Bisbas}, T.~G. 2016, Monthly Notices of
  the Royal Astronomical Society, 457, 3593

\bibitem[{{Fatuzzo} \& {Adams}(2008)}]{fatuzzo2008}
{Fatuzzo}, M. \& {Adams}, F.~C. 2008, \apj, 675, 1361

\bibitem[{{Feiden}(2016)}]{Feiden2016}
{Feiden}, G.~A. 2016, \aap, 593, A99

\bibitem[{{Frank} {et~al.}(2014){Frank}, {Ray}, {Cabrit}, {Hartigan}, {Arce},
  {Bacciotti}, {Bally}, {Benisty}, {Eisl{\"o}ffel}, {G{\"u}del}, {Lebedev},
  {Nisini}, \& {Raga}}]{frank2014}
{Frank}, A., {Ray}, T.~P., {Cabrit}, S., {et~al.} 2014, in Protostars and
  Planets VI, ed. H.~{Beuther}, R.~S. {Klessen}, C.~P. {Dullemond}, \&
  T.~{Henning}, 451--474

\bibitem[{{Frasca} {et~al.}(2017){Frasca}, {Biazzo}, {Alcal{\'a}}, {Manara},
  {Stelzer}, {Covino}, \& {Antoniucci}}]{Frasca2017}
{Frasca}, A., {Biazzo}, K., {Alcal{\'a}}, J.~M., {et~al.} 2017, \aap, 602, A33

\bibitem[{{Freudling} {et~al.}(2013){Freudling}, {Romaniello}, {Bramich},
  {Ballester}, {Forchi}, {Garc{\'\i}a-Dabl{\'o}}, {Moehler}, \&
  {Neeser}}]{Freuding2013}
{Freudling}, W., {Romaniello}, M., {Bramich}, D.~M., {et~al.} 2013, \aap, 559,
  A96

\bibitem[{{Gaia Collaboration} {et~al.}(2021){Gaia Collaboration}, {Brown},
  {Vallenari}, {Prusti}, {de Bruijne}, {Babusiaux}, {Biermann}, {Creevey},
  {Evans}, {Eyer}, {Hutton}, {Jansen}, {Jordi}, {Klioner}, {Lammers},
  {Lindegren}, {Luri}, {Mignard}, {Panem}, {Pourbaix}, {Randich}, {Sartoretti},
  {Soubiran}, {Walton}, {Arenou}, {Bailer-Jones}, {Bastian}, {Cropper},
  {Drimmel}, {Katz}, {Lattanzi}, {van Leeuwen}, {Bakker}, {Cacciari},
  {Casta{\~n}eda}, {De Angeli}, {Ducourant}, {Fabricius}, {Fouesneau},
  {Fr{\'e}mat}, {Guerra}, {Guerrier}, {Guiraud}, {Jean-Antoine Piccolo},
  {Masana}, {Messineo}, {Mowlavi}, {Nicolas}, {Nienartowicz}, {Pailler},
  {Panuzzo}, {Riclet}, {Roux}, {Seabroke}, {Sordo}, {Tanga}, {Th{\'e}venin},
  {Gracia-Abril}, {Portell}, {Teyssier}, {Altmann}, {Andrae}, {Bellas-Velidis},
  {Benson}, {Berthier}, {Blomme}, {Brugaletta}, {Burgess}, {Busso}, {Carry},
  {Cellino}, {Cheek}, {Clementini}, {Damerdji}, {Davidson}, {Delchambre},
  {Dell'Oro}, {Fern{\'a}ndez-Hern{\'a}ndez}, {Galluccio}, {Garc{\'\i}a-Lario},
  {Garcia-Reinaldos}, {Gonz{\'a}lez-N{\'u}{\~n}ez}, {Gosset}, {Haigron},
  {Halbwachs}, {Hambly}, {Harrison}, {Hatzidimitriou}, {Heiter},
  {Hern{\'a}ndez}, {Hestroffer}, {Hodgkin}, {Holl}, {Jan{\ss}en}, {Jevardat de
  Fombelle}, {Jordan}, {Krone-Martins}, {Lanzafame}, {L{\"o}ffler}, {Lorca},
  {Manteiga}, {Marchal}, {Marrese}, {Moitinho}, {Mora}, {Muinonen}, {Osborne},
  {Pancino}, {Pauwels}, {Petit}, {Recio-Blanco}, {Richards}, {Riello},
  {Rimoldini}, {Robin}, {Roegiers}, {Rybizki}, {Sarro}, {Siopis}, {Smith},
  {Sozzetti}, {Ulla}, {Utrilla}, {van Leeuwen}, {van Reeven}, {Abbas}, {Abreu
  Aramburu}, {Accart}, {Aerts}, {Aguado}, {Ajaj}, {Altavilla}, {{\'A}lvarez},
  {{\'A}lvarez Cid-Fuentes}, {Alves}, {Anderson}, {Anglada Varela}, {Antoja},
  {Audard}, {Baines}, {Baker}, {Balaguer-N{\'u}{\~n}ez}, {Balbinot}, {Balog},
  {Barache}, {Barbato}, {Barros}, {Barstow}, {Bartolom{\'e}}, {Bassilana},
  {Bauchet}, {Baudesson-Stella}, {Becciani}, {Bellazzini}, {Bernet}, {Bertone},
  {Bianchi}, {Blanco-Cuaresma}, {Boch}, {Bombrun}, {Bossini}, {Bouquillon},
  {Bragaglia}, {Bramante}, {Breedt}, {Bressan}, {Brouillet}, {Bucciarelli},
  {Burlacu}, {Busonero}, {Butkevich}, {Buzzi}, {Caffau}, {Cancelliere},
  {C{\'a}novas}, {Cantat-Gaudin}, {Carballo}, {Carlucci}, {Carnerero},
  {Carrasco}, {Casamiquela}, {Castellani}, {Castro-Ginard}, {Castro Sampol},
  {Chaoul}, {Charlot}, {Chemin}, {Chiavassa}, {Cioni}, {Comoretto}, {Cooper},
  {Cornez}, {Cowell}, {Crifo}, {Crosta}, {Crowley}, {Dafonte}, {Dapergolas},
  {David}, {David}, {de Laverny}, {De Luise}, {De March}, {De Ridder}, {de
  Souza}, {de Teodoro}, {de Torres}, {del Peloso}, {del Pozo}, {Delbo},
  {Delgado}, {Delgado}, {Delisle}, {Di Matteo}, {Diakite}, {Diener},
  {Distefano}, {Dolding}, {Eappachen}, {Edvardsson}, {Enke}, {Esquej}, {Fabre},
  {Fabrizio}, {Faigler}, {Fedorets}, {Fernique}, {Fienga}, {Figueras},
  {Fouron}, {Fragkoudi}, {Fraile}, {Franke}, {Gai}, {Garabato},
  {Garcia-Gutierrez}, {Garc{\'\i}a-Torres}, {Garofalo}, {Gavras}, {Gerlach},
  {Geyer}, {Giacobbe}, {Gilmore}, {Girona}, {Giuffrida}, {Gomel}, {Gomez},
  {Gonzalez-Santamaria}, {Gonz{\'a}lez-Vidal}, {Granvik},
  {Guti{\'e}rrez-S{\'a}nchez}, {Guy}, {Hauser}, {Haywood}, {Helmi}, {Hidalgo},
  {Hilger}, {H{\l}adczuk}, {Hobbs}, {Holland}, {Huckle}, {Jasniewicz},
  {Jonker}, {Juaristi Campillo}, {Julbe}, {Karbevska}, {Kervella}, {Khanna},
  {Kochoska}, {Kontizas}, {Kordopatis}, {Korn}, {Kostrzewa-Rutkowska},
  {Kruszy{\'n}ska}, {Lambert}, {Lanza}, {Lasne}, {Le Campion}, {Le Fustec},
  {Lebreton}, {Lebzelter}, {Leccia}, {Leclerc}, {Lecoeur-Taibi}, {Liao},
  {Licata}, {Lindstr{\o}m}, {Lister}, {Livanou}, {Lobel}, {Madrero Pardo},
  {Managau}, {Mann}, {Marchant}, {Marconi}, {Marcos Santos}, {Marinoni},
  {Marocco}, {Marshall}, {Martin Polo}, {Mart{\'\i}n-Fleitas}, {Masip},
  {Massari}, {Mastrobuono-Battisti}, {Mazeh}, {McMillan}, {Messina},
  {Michalik}, {Millar}, {Mints}, {Molina}, {Molinaro}, {Moln{\'a}r},
  {Montegriffo}, {Mor}, {Morbidelli}, {Morel}, {Morris}, {Mulone}, {Munoz},
  {Muraveva}, {Murphy}, {Musella}, {Noval}, {Ord{\'e}novic}, {Orr{\`u}},
  {Osinde}, {Pagani}, {Pagano}, {Palaversa}, {Palicio}, {Panahi}, {Pawlak},
  {Pe{\~n}alosa Esteller}, {Penttil{\"a}}, {Piersimoni}, {Pineau}, {Plachy},
  {Plum}, {Poggio}, {Poretti}, {Poujoulet}, {Pr{\v{s}}a}, {Pulone}, {Racero},
  {Ragaini}, {Rainer}, {Raiteri}, {Rambaux}, {Ramos}, {Ramos-Lerate}, {Re
  Fiorentin}, {Regibo}, {Reyl{\'e}}, {Ripepi}, {Riva}, {Rixon}, {Robichon},
  {Robin}, {Roelens}, {Rohrbasser}, {Romero-G{\'o}mez}, {Rowell}, {Royer},
  {Rybicki}, {Sadowski}, {Sagrist{\`a} Sell{\'e}s}, {Sahlmann}, {Salgado},
  {Salguero}, {Samaras}, {Sanchez Gimenez}, {Sanna}, {Santove{\~n}a},
  {Sarasso}, {Schultheis}, {Sciacca}, {Segol}, {Segovia}, {S{\'e}gransan},
  {Semeux}, {Shahaf}, {Siddiqui}, {Siebert}, {Siltala}, {Slezak}, {Smart},
  {Solano}, {Solitro}, {Souami}, {Souchay}, {Spagna}, {Spoto}, {Steele},
  {Steidelm{\"u}ller}, {Stephenson}, {S{\"u}veges}, {Szabados}, {Szegedi-Elek},
  {Taris}, {Tauran}, {Taylor}, {Teixeira}, {Thuillot}, {Tonello}, {Torra},
  {Torra}, {Turon}, {Unger}, {Vaillant}, {van Dillen}, {Vanel}, {Vecchiato},
  {Viala}, {Vicente}, {Voutsinas}, {Weiler}, {Wevers}, {Wyrzykowski}, {Yoldas},
  {Yvard}, {Zhao}, {Zorec}, {Zucker}, {Zurbach}, \& {Zwitter}}]{Gaia2021}
{Gaia Collaboration}, {Brown}, A.~G.~A., {Vallenari}, A., {et~al.} 2021, \aap,
  649, A1

\bibitem[{{Gangi} {et~al.}(2023){Gangi}, {Nisini}, {Manara}, {France},
  {Antoniucci}, {Biazzo}, {Giannini}, {Herczeg}, {Alcal{\'a}}, {Frasca},
  {Mauc{\'o}}, {Campbell-White}, {Siwak}, {Venuti}, {Schneider},
  {K{\'o}sp{\'a}l}, {Garatti}, {Fiorellino}, {Rigliaco}, \&
  {Yadav}}]{Gangi2023}
{Gangi}, M., {Nisini}, B., {Manara}, C.~F., {et~al.} 2023, arXiv e-prints,
  arXiv:2305.18940

\bibitem[{{Garrison}(1967)}]{Garrison1967}
{Garrison}, R.~F. 1967, \pasp, 79, 433

\bibitem[{{Grant} {et~al.}(2021){Grant}, {Espaillat}, {Wendeborn}, {Tobin},
  {Mac{\'\i}as}, {Rilinger}, {Ribas}, {Megeath}, {Fischer}, {Calvet}, \& {Hee
  Kim}}]{Grant2021}
{Grant}, S.~L., {Espaillat}, C.~C., {Wendeborn}, J., {et~al.} 2021, \apj, 913,
  123

\bibitem[{{Habing}(1968)}]{Habing1968}
{Habing}, H.~J. 1968, Bulletin of the Astronomical Institutes of the
  Netherland, 19, 421

\bibitem[{{Hartmann} {et~al.}(1998){Hartmann}, {Calvet}, {Gullbring}, \&
  {D'Alessio}}]{Hartmann1998}
{Hartmann}, L., {Calvet}, N., {Gullbring}, E., \& {D'Alessio}, P. 1998, {\apj},
  495, 385

\bibitem[{{Hartmann} {et~al.}(2006){Hartmann}, {D'Alessio}, {Calvet}, \&
  {Muzerolle}}]{Hartmann2006}
{Hartmann}, L., {D'Alessio}, P., {Calvet}, N., \& {Muzerolle}, J. 2006, \apj,
  648, 484

\bibitem[{Hartmann {et~al.}(2016)Hartmann, Herczeg, \& Calvet}]{Hartmann2016}
Hartmann, L., Herczeg, G., \& Calvet, N. 2016, Annual Review of \aap, 54, 135

\bibitem[{{Hasegawa} {et~al.}(2022){Hasegawa}, {Haworth}, {Hoadley}, {Kim},
  {Goto}, {Juzikenaite}, {Turner}, {Pascucci}, \& {Hamden}}]{Yasuhiro2022}
{Hasegawa}, Y., {Haworth}, T.~J., {Hoadley}, K., {et~al.} 2022, \apjl, 926, L23

\bibitem[{{Haworth} {et~al.}(2018){Haworth}, {Clarke}, {Rahman}, {Winter}, \&
  {Facchini}}]{Haworth2018}
{Haworth}, T.~J., {Clarke}, C.~J., {Rahman}, W., {Winter}, A.~J., \&
  {Facchini}, S. 2018, \mnras, 481, 452

\bibitem[{{Herczeg} \& {Hillenbrand}(2014)}]{Herczeg2014}
{Herczeg}, G.~J. \& {Hillenbrand}, L.~A. 2014, \apj, 786, 97

\bibitem[{{Hern{\'a}ndez} {et~al.}(2014){Hern{\'a}ndez}, {Calvet}, {Perez},
  {Brice{\~n}o}, {Olguin}, {Contreras}, {Hartmann}, {Allen}, {Espaillat}, \&
  {Hernan}}]{Hernandez2014}
{Hern{\'a}ndez}, J., {Calvet}, N., {Perez}, A., {et~al.} 2014, \apj, 794, 36

\bibitem[{{Hern{\'a}ndez} {et~al.}(2007){Hern{\'a}ndez}, {Hartmann}, {Megeath},
  {Gutermuth}, {Muzerolle}, {Calvet}, {Vivas}, {Brice{\~n}o}, {Allen},
  {Stauffer}, {Young}, \& {Fazio}}]{Hernandez2007}
{Hern{\'a}ndez}, J., {Hartmann}, L., {Megeath}, T., {et~al.} 2007, \apj, 662,
  1067

\bibitem[{{Hillenbrand} {et~al.}(1992){Hillenbrand}, {Strom}, {Vrba}, \&
  {Keene}}]{Hillenbrand1992}
{Hillenbrand}, L.~A., {Strom}, S.~E., {Vrba}, F.~J., \& {Keene}, J. 1992, \apj,
  397, 613

\bibitem[{{Ingleby} {et~al.}(2014){Ingleby}, {Calvet}, {Hern{\'a}ndez},
  {Hartmann}, {Briceno}, {Miller}, {Espaillat}, \& {McClure}}]{Ingleby2014}
{Ingleby}, L., {Calvet}, N., {Hern{\'a}ndez}, J., {et~al.} 2014, \apj, 790, 47

\bibitem[{{Johnstone} {et~al.}(1998){Johnstone}, {Hollenbach}, \&
  {Bally}}]{Johnstone1998}
{Johnstone}, D., {Hollenbach}, D., \& {Bally}, J. 1998, \apj, 499, 758

\bibitem[{{Jones} {et~al.}(2012){Jones}, {Pringle}, \& {Alexander}}]{Jones2012}
{Jones}, M.~G., {Pringle}, J.~E., \& {Alexander}, R.~D. 2012, \mnras, 419, 925

\bibitem[{{Kim} {et~al.}(2016){Kim}, {Clarke}, {Fang}, \& {Facchini}}]{Kim2016}
{Kim}, J.~S., {Clarke}, C.~J., {Fang}, M., \& {Facchini}, S. 2016, \apj
  Letters, 826, L15

\bibitem[{{Lada} \& {Lada}(2003)}]{lada_lada2003}
{Lada}, C.~J. \& {Lada}, E.~A. 2003, \araa, 41, 57

\bibitem[{{Lodato} {et~al.}(2017){Lodato}, {Scardoni}, {Manara}, \&
  {Testi}}]{Lodato2017}
{Lodato}, G., {Scardoni}, C.~E., {Manara}, C.~F., \& {Testi}, L. 2017, Monthly
  Notices of the Royal Astronomical Society, 472, 4700

\bibitem[{{Long} {et~al.}(2022){Long}, {Andrews}, {Rosotti}, {Harsono},
  {Pinilla}, {Wilner}, {{\"O}berg}, {Teague}, {Trapman}, \&
  {Tabone}}]{Long2022}
{Long}, F., {Andrews}, S.~M., {Rosotti}, G., {et~al.} 2022, \apj, 931, 6

\bibitem[{{Luhman} {et~al.}(2008){Luhman}, {Hern{\'a}ndez}, {Downes},
  {Hartmann}, \& {Brice{\~n}o}}]{Luhman2008}
{Luhman}, K.~L., {Hern{\'a}ndez}, J., {Downes}, J.~J., {Hartmann}, L., \&
  {Brice{\~n}o}, C. 2008, \apj, 688, 362

\bibitem[{{Luhman} {et~al.}(2003){Luhman}, {Stauffer}, {Muench}, {Rieke},
  {Lada}, {Bouvier}, \& {Lada}}]{Luhman2003}
{Luhman}, K.~L., {Stauffer}, J.~R., {Muench}, A.~A., {et~al.} 2003, \apj, 593,
  1093

\bibitem[{{Lynden-Bell} \& {Pringle}(1974)}]{LyndenBell1974}
{Lynden-Bell}, D. \& {Pringle}, J.~E. 1974, Monthly Notices of the Royal
  Astronomical Society, 168, 603

\bibitem[{{Manara} {et~al.}(2023){Manara}, {Ansdell}, {Rosotti}, {Hughes},
  {Armitage}, {Lodato}, \& {Williams}}]{Manara2023}
{Manara}, C.~F., {Ansdell}, M., {Rosotti}, G.~P., {et~al.} 2023, in
  Astronomical Society of the Pacific Conference Series, Vol. 534, Protostars
  and Planets VII, ed. S.~{Inutsuka}, Y.~{Aikawa}, T.~{Muto}, K.~{Tomida}, \&
  M.~{Tamura}, 539

\bibitem[{{Manara} {et~al.}(2013{\natexlab{a}}){Manara}, {Beccari}, {Da Rio},
  {De Marchi}, {Natta}, {Ricci}, {Robberto}, \& {Testi}}]{Manara2013b}
{Manara}, C.~F., {Beccari}, G., {Da Rio}, N., {et~al.} 2013{\natexlab{a}},
  \aap, 558, A114

\bibitem[{{Manara} {et~al.}(2016{\natexlab{a}}){Manara}, {Fedele}, {Herczeg},
  \& {Teixeira}}]{Manara2016a}
{Manara}, C.~F., {Fedele}, D., {Herczeg}, G.~J., \& {Teixeira}, P.~S.
  2016{\natexlab{a}}, \aap, 585, A136

\bibitem[{{Manara} {et~al.}(2017{\natexlab{a}}){Manara}, {Frasca},
  {Alcal{\'a}}, {Natta}, {Stelzer}, \& {Testi}}]{Manara2017b}
{Manara}, C.~F., {Frasca}, A., {Alcal{\'a}}, J.~M., {et~al.}
  2017{\natexlab{a}}, {\aap}, 605, A86

\bibitem[{{Manara} {et~al.}(2021){Manara}, {Frasca}, {Venuti}, {Siwak},
  {Herczeg}, {Calvet}, {Hernandez}, {Tychoniec}, {Gangi}, {Alcal{\'a}},
  {Boffin}, {Nisini}, {Robberto}, {Briceno}, {Campbell-White},
  {Sicilia-Aguilar}, {McGinnis}, {Fedele}, {K{\'o}sp{\'a}l}, {{\'A}brah{\'a}m},
  {Alonso-Santiago}, {Antoniucci}, {Arulanantham}, {Bacciotti}, {Banzatti},
  {Beccari}, {Benisty}, {Biazzo}, {Bouvier}, {Cabrit}, {Caratti o Garatti},
  {Coffey}, {Covino}, {Dougados}, {Eisl{\"o}ffel}, {Ercolano}, {Espaillat},
  {Erkal}, {Facchini}, {Fang}, {Fiorellino}, {Fischer}, {France}, {Gameiro},
  {Garcia Lopez}, {Giannini}, {Ginski}, {Grankin}, {G{\"u}nther}, {Hartmann},
  {Hillenbrand}, {Hussain}, {James}, {Koutoulaki}, {Lodato}, {Mauc{\'o}},
  {Mendigut{\'\i}a}, {Mentel}, {Miotello}, {Oudmaijer}, {Rigliaco}, {Rosotti},
  {Sanchis}, {Schneider}, {Spina}, {Stelzer}, {Testi}, {Thanathibodee}, {Vink},
  {Walter}, {Williams}, \& {Zsidi}}]{Manara2021}
{Manara}, C.~F., {Frasca}, A., {Venuti}, L., {et~al.} 2021, \aap, 650, A196

\bibitem[{{Manara} {et~al.}(2020){Manara}, {Natta}, {Rosotti}, {Alcal{\'a}},
  {Nisini}, {Lodato}, {Testi}, {Pascucci}, {Hillenbrand}, {Carpenter},
  {Scholz}, {Fedele}, {Frasca}, {Mulders}, {Rigliaco}, {Scardoni}, \&
  {Zari}}]{Manara2020}
{Manara}, C.~F., {Natta}, A., {Rosotti}, G.~P., {et~al.} 2020, \aap, 639, A58

\bibitem[{{Manara} {et~al.}(2016{\natexlab{b}}){Manara}, {Rosotti}, {Testi},
  {Natta}, {Alcal{\'a}}, {Williams}, {Ansdell}, {Miotello}, {van der Marel},
  {Tazzari}, {Carpenter}, {Guidi}, {Mathews}, {Oliveira}, {Prusti}, \& {van
  Dishoeck}}]{Manara2016b}
{Manara}, C.~F., {Rosotti}, G., {Testi}, L., {et~al.} 2016{\natexlab{b}}, \aap,
  591, L3

\bibitem[{{Manara} {et~al.}(2017{\natexlab{b}}){Manara}, {Testi}, {Herczeg},
  {Pascucci}, {Alcal{\'a}}, {Natta}, {Antoniucci}, {Fedele}, {Mulders},
  {Henning}, {Mohanty}, {Prusti}, \& {Rigliaco}}]{Manara2017a}
{Manara}, C.~F., {Testi}, L., {Herczeg}, G.~J., {et~al.} 2017{\natexlab{b}},
  {\aap}, 604, A127

\bibitem[{{Manara} {et~al.}(2013{\natexlab{b}}){Manara}, {Testi}, {Rigliaco},
  {Alcal{\'a}}, {Natta}, {Stelzer}, {Biazzo}, {Covino}, {Covino}, {Cupani},
  {D'Elia}, \& {Randich}}]{Manara2013a}
{Manara}, C.~F., {Testi}, L., {Rigliaco}, E., {et~al.} 2013{\natexlab{b}},
  {\aap}, 551, A107

\bibitem[{{Mann} {et~al.}(2015){Mann}, {Andrews}, {Eisner}, {Williams},
  {Meyer}, {Di Francesco}, {Carpenter}, \& {Johnstone}}]{Mann2015}
{Mann}, R.~K., {Andrews}, S.~M., {Eisner}, J.~A., {et~al.} 2015, \apj, 802, 77

\bibitem[{{Mann} {et~al.}(2014){Mann}, {Di Francesco}, {Johnstone}, {Andrews},
  {Williams}, {Bally}, {Ricci}, {Hughes}, \& {Matthews}}]{Mann2014}
{Mann}, R.~K., {Di Francesco}, J., {Johnstone}, D., {et~al.} 2014, \apj, 784,
  82

\bibitem[{{Manzo-Mart{\'\i}nez} {et~al.}(2020){Manzo-Mart{\'\i}nez}, {Calvet},
  {Hern{\'a}ndez}, {Lizano}, {Hern{\'a}ndez}, {Miller}, {Mauc{\'o}},
  {Brice{\~n}o}, \& {D'Alessio}}]{Manzo2020}
{Manzo-Mart{\'\i}nez}, E., {Calvet}, N., {Hern{\'a}ndez}, J., {et~al.} 2020,
  \apj, 893, 56

\bibitem[{{Mauc{\'o}} {et~al.}(2016){Mauc{\'o}}, {Hern{\'a}ndez}, {Calvet},
  {Ballesteros-Paredes}, {Brice{\~n}o}, {McClure}, {D'Alessio}, {Anderson}, \&
  {Ali}}]{Mauco2016}
{Mauc{\'o}}, K., {Hern{\'a}ndez}, J., {Calvet}, N., {et~al.} 2016, \apj, 829,
  38

\bibitem[{{Miotello} {et~al.}(2023){Miotello}, {Kamp}, {Birnstiel}, {Cleeves},
  \& {Kataoka}}]{Miotello2023}
{Miotello}, A., {Kamp}, I., {Birnstiel}, T., {Cleeves}, L.~C., \& {Kataoka}, A.
  2023, in Astronomical Society of the Pacific Conference Series, Vol. 534,
  Protostars and Planets VII, ed. S.~{Inutsuka}, Y.~{Aikawa}, T.~{Muto},
  K.~{Tomida}, \& M.~{Tamura}, 501

\bibitem[{{Modigliani} {et~al.}(2010){Modigliani}, {Goldoni}, {Royer},
  {Haigron}, {Guglielmi}, {Fran{\c{c}}ois}, {Horrobin}, {Bristow}, {Vernet},
  {Moehler}, {Kerber}, {Ballester}, {Mason}, \& {Christensen}}]{Modigliani2010}
{Modigliani}, A., {Goldoni}, P., {Royer}, F., {et~al.} 2010, in Society of
  Photo-Optical Instrumentation Engineers (SPIE) Conference Series, Vol. 7737,
  Observatory Operations: Strategies, Processes, and Systems III, ed. D.~R.
  {Silva}, A.~B. {Peck}, \& B.~T. {Soifer}, 773728

\bibitem[{{Mulders} {et~al.}(2017){Mulders}, {Pascucci}, {Manara}, {Testi},
  {Herczeg}, {Henning}, {Mohanty}, \& {Lodato}}]{Mulders2017}
{Mulders}, G.~D., {Pascucci}, I., {Manara}, C.~F., {et~al.} 2017, \apj, 847, 31

\bibitem[{{Muzerolle} {et~al.}(2003){Muzerolle}, {Hillenbrand}, {Calvet},
  {Brice{\~n}o}, \& {Hartmann}}]{Muzerolle2003}
{Muzerolle}, J., {Hillenbrand}, L., {Calvet}, N., {Brice{\~n}o}, C., \&
  {Hartmann}, L. 2003, \apj, 592, 266

\bibitem[{{Natta} {et~al.}(2006){Natta}, {Testi}, \& {Randich}}]{Natta2006}
{Natta}, A., {Testi}, L., \& {Randich}, S. 2006, \aap, 452, 245

\bibitem[{{O'dell} {et~al.}(1993){O'dell}, {Wen}, \& {Hu}}]{Odell1993}
{O'dell}, C.~R., {Wen}, Z., \& {Hu}, X. 1993, \apj, 410, 696

\bibitem[{{Oliveira} {et~al.}(2004){Oliveira}, {Jeffries}, \& {van
  Loon}}]{Oliveira2004}
{Oliveira}, J.~M., {Jeffries}, R.~D., \& {van Loon}, J.~T. 2004, Monthly
  Notices of the Royal Astronomical Society, 347, 1327

\bibitem[{{Parker} {et~al.}(2021){Parker}, {Alcock}, {Nicholson}, {Pani{\'c}},
  \& {Goodwin}}]{Parker2021}
{Parker}, R.~J., {Alcock}, H.~L., {Nicholson}, R.~B., {Pani{\'c}}, O., \&
  {Goodwin}, S.~P. 2021, \apj, 913, 95

\bibitem[{{Pascucci} {et~al.}(2022){Pascucci}, {Cabrit}, {Edwards}, {Gorti},
  {Gressel}, \& {Suzuki}}]{pascucci2022}
{Pascucci}, I., {Cabrit}, S., {Edwards}, S., {et~al.} 2022, arXiv e-prints,
  arXiv:2203.10068

\bibitem[{{Pascucci} {et~al.}(2016){Pascucci}, {Testi}, {Herczeg}, {Long},
  {Manara}, {Hendler}, {Mulders}, {Krijt}, {Ciesla}, {Henning}, {Mohanty},
  {Drabek-Maunder}, {Apai}, {Sz{\H{u}}cs}, {Sacco}, \&
  {Olofsson}}]{Pascucci2016}
{Pascucci}, I., {Testi}, L., {Herczeg}, G.~J., {et~al.} 2016, \apj, 831, 125

\bibitem[{{Pecaut} \& {Mamajek}(2016)}]{Pecaut2016}
{Pecaut}, M.~J. \& {Mamajek}, E.~E. 2016, \mnras, 461, 794

\bibitem[{{P{\'e}rez-Blanco} {et~al.}(2018){P{\'e}rez-Blanco}, {Mauc{\'o}},
  {Hern{\'a}ndez}, {Calvet}, {Espaillat}, {McClure}, {Brice{\~n}o}, {Robinson},
  {Feldman}, {Villarreal}, \& {D'Alessio}}]{Perez2018}
{P{\'e}rez-Blanco}, A., {Mauc{\'o}}, K., {Hern{\'a}ndez}, J., {et~al.} 2018,
  \apj, 867, 116

\bibitem[{{Pety} {et~al.}(2005){Pety}, {Teyssier}, {Foss{\'e}}, {Gerin},
  {Roueff}, {Abergel}, {Habart}, \& {Cernicharo}}]{Pety2005}
{Pety}, J., {Teyssier}, D., {Foss{\'e}}, D., {et~al.} 2005, \aap, 435, 885

\bibitem[{{Pinilla} {et~al.}(2020){Pinilla}, {Pascucci}, \&
  {Marino}}]{pinilla2020}
{Pinilla}, P., {Pascucci}, I., \& {Marino}, S. 2020, \aap, 635, A105

\bibitem[{{Pittman} {et~al.}(2022){Pittman}, {Espaillat}, {Robinson},
  {Thanathibodee}, {Calvet}, {Wendeborn}, {Hern{\'a}ndez}, {Manara}, {Walter},
  {{\'A}brah{\'a}m}, {Alcal{\'a}}, {Alencar}, {Arulanantham}, {Cabrit},
  {Eisl{\"o}ffel}, {Fiorellino}, {France}, {Gangi}, {Grankin}, {Herczeg},
  {K{\'o}sp{\'a}l}, {Mendigut{\'\i}a}, {Serna}, \& {Venuti}}]{Pittman2022}
{Pittman}, C.~V., {Espaillat}, C.~C., {Robinson}, C.~E., {et~al.} 2022, \aj,
  164, 201

\bibitem[{{Reipurth}(2008)}]{Reipurth2008}
{Reipurth}, B. 2008, {Handbook of Star Forming Regions, Volume I: The Northern
  Sky}, Vol.~4

\bibitem[{{Reiter} \& {Parker}(2022)}]{reiter2022}
{Reiter}, M. \& {Parker}, R.~J. 2022, European Physical Journal Plus, 137, 1071

\bibitem[{{Rigliaco} {et~al.}(2009){Rigliaco}, {Natta}, {Randich}, \&
  {Sacco}}]{Rigliaco2009}
{Rigliaco}, E., {Natta}, A., {Randich}, S., \& {Sacco}, G. 2009, \aap, 495, L13

\bibitem[{{Rigliaco} {et~al.}(2011){Rigliaco}, {Natta}, {Randich}, {Testi}, \&
  {Biazzo}}]{Rigliaco2011}
{Rigliaco}, E., {Natta}, A., {Randich}, S., {Testi}, L., \& {Biazzo}, K. 2011,
  \aap, 525, A47

\bibitem[{{Rigliaco} {et~al.}(2012){Rigliaco}, {Natta}, {Testi}, {Randich},
  {Alcal{\`a}}, {Covino}, \& {Stelzer}}]{Rigliaco2012}
{Rigliaco}, E., {Natta}, A., {Testi}, L., {et~al.} 2012, \aap, 548, A56

\bibitem[{{Rilinger} \& {Espaillat}(2021)}]{rilinger2021}
{Rilinger}, A.~M. \& {Espaillat}, C.~C. 2021, \apj, 921, 182

\bibitem[{{Rosotti} {et~al.}(2017){Rosotti}, {Clarke}, {Manara}, \&
  {Facchini}}]{Rosotti2017}
{Rosotti}, G.~P., {Clarke}, C.~J., {Manara}, C.~F., \& {Facchini}, S. 2017,
  Monthly Notices of the Royal Astronomical Society, 468, 1631

\bibitem[{{Rugel} {et~al.}(2018){Rugel}, {Fedele}, \& {Herczeg}}]{Rugel2018}
{Rugel}, M., {Fedele}, D., \& {Herczeg}, G. 2018, \aap, 609, A70

\bibitem[{{Sanchis} {et~al.}(2021){Sanchis}, {Testi}, {Natta}, {Facchini},
  {Manara}, {Miotello}, {Ercolano}, {Henning}, {Preibisch}, {Carpenter}, {de
  Gregorio-Monsalvo}, {Jayawardhana}, {Lopez}, {Mu{\v{z}}i{\'c}}, {Pascucci},
  {Santamar{\'\i}a-Miranda}, {van Terwisga}, \& {Williams}}]{sanchis2021}
{Sanchis}, E., {Testi}, L., {Natta}, A., {et~al.} 2021, \aap, 649, A19

\bibitem[{{Sellek} {et~al.}(2020{\natexlab{a}}){Sellek}, {Booth}, \&
  {Clarke}}]{Sellek2020b}
{Sellek}, A.~D., {Booth}, R.~A., \& {Clarke}, C.~J. 2020{\natexlab{a}}, Monthly
  Notices of the Royal Astronomical Society, 498, 2845

\bibitem[{{Sellek} {et~al.}(2020{\natexlab{b}}){Sellek}, {Booth}, \&
  {Clarke}}]{Sellek2020a}
{Sellek}, A.~D., {Booth}, R.~A., \& {Clarke}, C.~J. 2020{\natexlab{b}}, Monthly
  Notices of the Royal Astronomical Society, 492, 1279

\bibitem[{{Sherry} {et~al.}(2008){Sherry}, {Walter}, {Wolk}, \&
  {Adams}}]{Sherry2008}
{Sherry}, W.~H., {Walter}, F.~M., {Wolk}, S.~J., \& {Adams}, N.~R. 2008, The
  \aj, 135, 1616

\bibitem[{{Sicilia-Aguilar} {et~al.}(2010){Sicilia-Aguilar}, {Henning}, \&
  {Hartmann}}]{Sicilia2010}
{Sicilia-Aguilar}, A., {Henning}, T., \& {Hartmann}, L.~W. 2010, \apj, 710, 597

\bibitem[{{Siess} {et~al.}(2000){Siess}, {Dufour}, \& {Forestini}}]{Siess2000}
{Siess}, L., {Dufour}, E., \& {Forestini}, M. 2000, {\aap}, 358, 593

\bibitem[{{Sim{\'o}n-D{\'\i}az} {et~al.}(2015){Sim{\'o}n-D{\'\i}az},
  {Caballero}, {Lorenzo}, {Ma{\'\i}z Apell{\'a}niz}, {Schneider}, {Negueruela},
  {Barb{\'a}}, {Dorda}, {Marco}, {Montes}, {Pellerin}, {Sanchez-Bermudez},
  {S{\'o}dor}, \& {Sota}}]{Simon2015}
{Sim{\'o}n-D{\'\i}az}, S., {Caballero}, J.~A., {Lorenzo}, J., {et~al.} 2015,
  \apj, 799, 169

\bibitem[{{Smette} {et~al.}(2015){Smette}, {Sana}, {Noll}, {Horst}, {Kausch},
  {Kimeswenger}, {Barden}, {Szyszka}, {Jones}, {Gallenne}, {Vinther},
  {Ballester}, \& {Taylor}}]{Smette2015}
{Smette}, A., {Sana}, H., {Noll}, S., {et~al.} 2015, \aap, 576, A77

\bibitem[{{Soderblom} {et~al.}(2014){Soderblom}, {Hillenbrand}, {Jeffries},
  {Mamajek}, \& {Naylor}}]{Soderblom2014}
{Soderblom}, D.~R., {Hillenbrand}, L.~A., {Jeffries}, R.~D., {Mamajek}, E.~E.,
  \& {Naylor}, T. 2014, in Protostars and Planets VI, ed. H.~{Beuther}, R.~S.
  {Klessen}, C.~P. {Dullemond}, \& T.~{Henning}, 219--241

\bibitem[{{Somigliana} {et~al.}(2022){Somigliana}, {Toci}, {Rosotti}, {Lodato},
  {Tazzari}, {Manara}, {Testi}, \& {Lepri}}]{Somigliana22}
{Somigliana}, A., {Toci}, C., {Rosotti}, G., {et~al.} 2022, \mnras, 514, 5927

\bibitem[{{Tabone} {et~al.}(2022){Tabone}, {Rosotti}, {Lodato}, {Armitage},
  {Cridland}, \& {van Dishoeck}}]{Tabone22}
{Tabone}, B., {Rosotti}, G.~P., {Lodato}, G., {et~al.} 2022, \mnras, 512, L74

\bibitem[{{Tazzari} {et~al.}(2021){Tazzari}, {Testi}, {Natta}, {Williams},
  {Ansdell}, {Carpenter}, {Facchini}, {Guidi}, {Hogherheijde}, {Manara},
  {Miotello}, \& {van der Marel}}]{tazzari2021}
{Tazzari}, M., {Testi}, L., {Natta}, A., {et~al.} 2021, \mnras, 506, 5117

\bibitem[{{Testi} {et~al.}(2022){Testi}, {Natta}, {Manara}, {de Gregorio
  Monsalvo}, {Lodato}, {Lopez}, {Muzic}, {Pascucci}, {Sanchis}, {Miranda},
  {Scholz}, {De Simone}, \& {Williams}}]{Testi2022}
{Testi}, L., {Natta}, A., {Manara}, C.~F., {et~al.} 2022, \aap, 663, A98

\bibitem[{{Testi} {et~al.}(2016){Testi}, {Natta}, {Scholz}, {Tazzari}, {Ricci},
  \& {de Gregorio Monsalvo}}]{testi2016}
{Testi}, L., {Natta}, A., {Scholz}, A., {et~al.} 2016, \aap, 593, A111

\bibitem[{{Toci} {et~al.}(2023){Toci}, {Lodato}, {Livio}, {Rosotti}, \&
  {Trapman}}]{toci2023}
{Toci}, C., {Lodato}, G., {Livio}, F.~G., {Rosotti}, G., \& {Trapman}, L. 2023,
  \mnras, 518, L69

\bibitem[{{Trapman} {et~al.}(2023){Trapman}, {Rosotti}, {Zhang}, \&
  {Tabone}}]{trapman2023}
{Trapman}, L., {Rosotti}, G., {Zhang}, K., \& {Tabone}, B. 2023, arXiv
  e-prints, arXiv:2307.07600

\bibitem[{{Tripathi} {et~al.}(2017){Tripathi}, {Andrews}, {Birnstiel}, \&
  {Wilner}}]{Tripathi2017}
{Tripathi}, A., {Andrews}, S.~M., {Birnstiel}, T., \& {Wilner}, D.~J. 2017,
  \apj, 845, 44

\bibitem[{{van Terwisga} \& {Hacar}(2023)}]{vanTerwisga2023}
{van Terwisga}, S.~E. \& {Hacar}, A. 2023, \aap, 673, L2

\bibitem[{{Venuti} {et~al.}(2014){Venuti}, {Bouvier}, {Flaccomio}, {Alencar},
  {Irwin}, {Stauffer}, {Cody}, {Teixeira}, {Sousa}, {Micela}, {Cuillandre}, \&
  {Peres}}]{venuti2014}
{Venuti}, L., {Bouvier}, J., {Flaccomio}, E., {et~al.} 2014, \aap, 570, A82

\bibitem[{{Venuti} {et~al.}(2019){Venuti}, {Stelzer}, {Alcal{\'a}}, {Manara},
  {Frasca}, {Jayawardhana}, {Antoniucci}, {Argiroffi}, {Natta}, {Nisini},
  {Randich}, \& {Scholz}}]{venuti2019}
{Venuti}, L., {Stelzer}, B., {Alcal{\'a}}, J.~M., {et~al.} 2019, \aap, 632, A46

\bibitem[{{Vernet} {et~al.}(2011){Vernet}, {Dekker}, {D'Odorico}, {Kaper},
  {Kjaergaard}, {Hammer}, {Randich}, {Zerbi}, {Groot}, {Hjorth}, {Guinouard},
  {Navarro}, {Adolfse}, {Albers}, {Amans}, {Andersen}, {Andersen}, {Binetruy},
  {Bristow}, {Castillo}, {Chemla}, {Christensen}, {Conconi}, {Conzelmann},
  {Dam}, {de Caprio}, {de Ugarte Postigo}, {Delabre}, {di Marcantonio},
  {Downing}, {Elswijk}, {Finger}, {Fischer}, {Flores}, {Fran{\c{c}}ois},
  {Goldoni}, {Guglielmi}, {Haigron}, {Hanenburg}, {Hendriks}, {Horrobin},
  {Horville}, {Jessen}, {Kerber}, {Kern}, {Kiekebusch}, {Kleszcz}, {Klougart},
  {Kragt}, {Larsen}, {Lizon}, {Lucuix}, {Mainieri}, {Manuputy}, {Martayan},
  {Mason}, {Mazzoleni}, {Michaelsen}, {Modigliani}, {Moehler}, {M{\o}ller},
  {Norup S{\o}rensen}, {N{\o}rregaard}, {P{\'e}roux}, {Patat}, {Pena}, {Pragt},
  {Reinero}, {Rigal}, {Riva}, {Roelfsema}, {Royer}, {Sacco}, {Santin},
  {Schoenmaker}, {Spano}, {Sweers}, {Ter Horst}, {Tintori}, {Tromp}, {van
  Dael}, {van der Vliet}, {Venema}, {Vidali}, {Vinther}, {Vola}, {Winters},
  {Wistisen}, {Wulterkens}, \& {Zacchei}}]{Vernet2011}
{Vernet}, J., {Dekker}, H., {D'Odorico}, S., {et~al.} 2011, \aap, 536, A105

\bibitem[{{Winter} {et~al.}(2020){Winter}, {Ansdell}, {Haworth}, \&
  {Kruijssen}}]{Winter2020}
{Winter}, A.~J., {Ansdell}, M., {Haworth}, T.~J., \& {Kruijssen}, J.~M.~D.
  2020, Monthly Notices of the Royal Astronomical Society, 497, L40

\bibitem[{{Winter} {et~al.}(2018){Winter}, {Clarke}, {Rosotti}, {Ih},
  {Facchini}, \& {Haworth}}]{winter2018}
{Winter}, A.~J., {Clarke}, C.~J., {Rosotti}, G., {et~al.} 2018, \mnras, 478,
  2700

\bibitem[{{Winter} \& {Haworth}(2022)}]{Winter2022}
{Winter}, A.~J. \& {Haworth}, T.~J. 2022, European Physical Journal Plus, 137,
  1132

\bibitem[{{Zagaria} {et~al.}(2022){Zagaria}, {Clarke}, {Rosotti}, \&
  {Manara}}]{Zagaria2022}
{Zagaria}, F., {Clarke}, C.~J., {Rosotti}, G.~P., \& {Manara}, C.~F. 2022,
  \mnras, 512, 3538

\bibitem[{{Zagaria} {et~al.}(2023){Zagaria}, {Facchini}, {Miotello}, {Manara},
  {Toci}, \& {Clarke}}]{Zagaria2023}
{Zagaria}, F., {Facchini}, S., {Miotello}, A., {et~al.} 2023, \aap, 672, L15

\end{thebibliography}

%
%

\appendix

\section{UV radiation field strength} \label{sec:UV_field}

We estimated the FUV radiation field strength due to the central OB system \sori. For this, we followed the same approach as \citet{Winter2022}. We considered the three most massive stars in the hierarchical system \sori: $\sigma$ Ori Aa (\mstar = 20 \msun), $\sigma$ Ori Ab (\mstar = 14.6 \msun), and $\sigma$ Ori B (\mstar = 13.6 \msun) from \citet{Simon2015}, and using Fig. 14 of \citet{Winter2022} we estimated the FUV luminosity (L$_{\rm FUV}$) of the system. The FUV field strength due to the massive OB stars was then obtained in terms of the dimensionless parameter $G_0$ as:

\begin{equation}
    G_0 = \frac{1}{F_0} \frac{L_{\rm FUV}}{4 \pi d_{p}^2},
\end{equation}
\noindent
where $F_0$ is the typical interstellar flux level of $1.6 \times 10^{-3}\, \rm erg\, cm^{-2}\, s^{-1}$ \citep{Habing1968}, and $d_{\rm p}$ represents the distance from the photoevaporative source \sori\, in parsecs (the projected distance).

\begin{figure}[ht]
\centering 
\includegraphics[width = 0.47\textwidth]{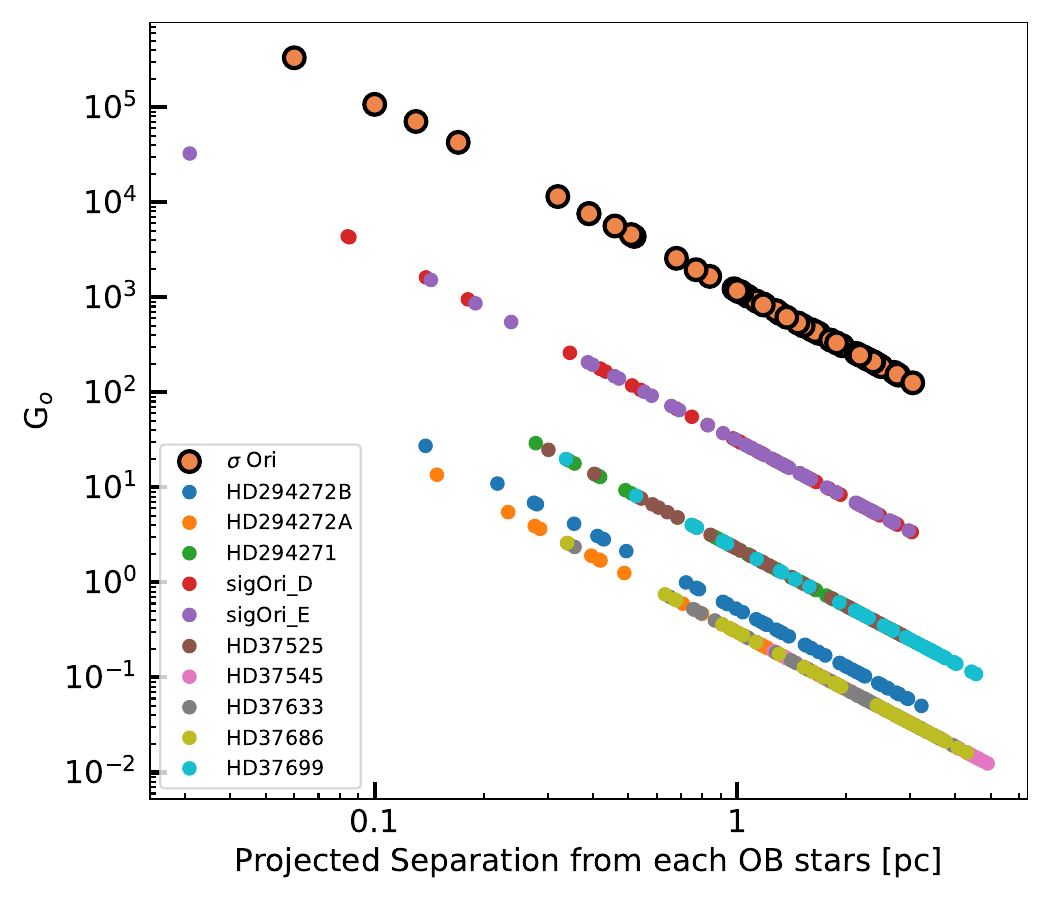}
\caption{G$_0$ of O and B stars in the $\sigma$ Orionis cluster. The FUV radiation is dominated by the central system \sori.}\label{fig:Go_massive_stars}
\end{figure}

We found values of FUV flux between $10^{2}$ and $10^{5}$ $G_0$ (see Fig.~\ref{fig:mdisk_dp_sori}, top axis). We also estimated the FUV field strength produced by the other B-stars in the cluster and found that the FUV field is completely dominated by the multiple system \sori. In Fig.~\ref{fig:spatial_distribution_stars} we showed the spatial distribution of disks in the cluster (circles) along with the massive O and B stars (gray stars). The color bar indicates the \textit{total} $G_0$ values (\sori\,+ B-stars) which as shown in Fig.~\ref{fig:Go_massive_stars} mainly corresponds to the FUV field produced by \sori.

\section{Stellar and accretion parameters from previous works}

We show on Table~\ref{tab:comparison} 
the stellar and accretion parameters of $\sigma$-Orionis sources from \citet{Manara2021}. 
Overall, we found a good agreement (given our uncertainties) between our best-fit model and the results from \citet{Manara2021}. SO518 and SO1153 are peculiar sources. SO518 seems to be an almost edge-on disk and, therefore, the estimate of its stellar parameters, particularly \macc, is more uncertain \citep{Alcala2014}. SO1153, on the other hand, is a strong accretor where the best-fit slab spectrum is much brighter than the photospheric template. Such strong accretors are problematic to fit, since the photospheric emission is veiled by the accretion emission, leading to large uncertainties in the best fit SpT \citep[e.g.,][]{Calvet1998}, and many photospheric absorption lines are seen in emission. To obtain the final results for this source and for SO518 we had to constrain $A_V$ to low values. Considering these caveats, we were able to obtain best-fit results compatible with the ones found by \citet{Manara2021}.  

\begin{table*}
\begin{center}
\caption{ \label{tab:comparison}Comparison between our best fit results and the ones of \cite{Manara2021} for the targets SO518, SO583, and SO1153. }
\begin{tabular}{lcccccc}
\hline\hline
Name & SpT & $A_V$ & $L_\star$ & $\log L_\mathrm{acc}$ & $M_\star$ & $\log\dot{M}_\mathrm{acc}$ \\
\hline
SO518 & K6 & 1.6 & 0.48 & -0.69 & 0.80 & -7.86 \\[0ex]
SO518 \cite{Manara2021} & K7 & 1.0 & 0.24 & -1.22 & 0.81 & -8.53 \\[1ex]
SO583 & K4 & 1.0 & 4.06 & -0.69 & 1.18 & -7.62 \\[0ex]
SO583 \cite{Manara2021}  & K5 & 0.4 & 3.61 & -0.30 & 1.09 & -7.21 \\[1ex]
SO1153 & K5 & 1.5 & 0.33 & 0.02 & 0.90 & -7.30 \\[0ex]
SO1153 \cite{Manara2021}  & K7 & 0.1 & 0.17 & -0.88 & 0.76 & -8.24 \\[1ex]
\hline
\end{tabular}
\end{center}
\end{table*}

\section{Additional information on targets}

\subsection{ALMA observations} \label{sec:appendix_ALMA}

\begin{figure}[ht]
\centering \includegraphics[width = 0.42\textwidth]{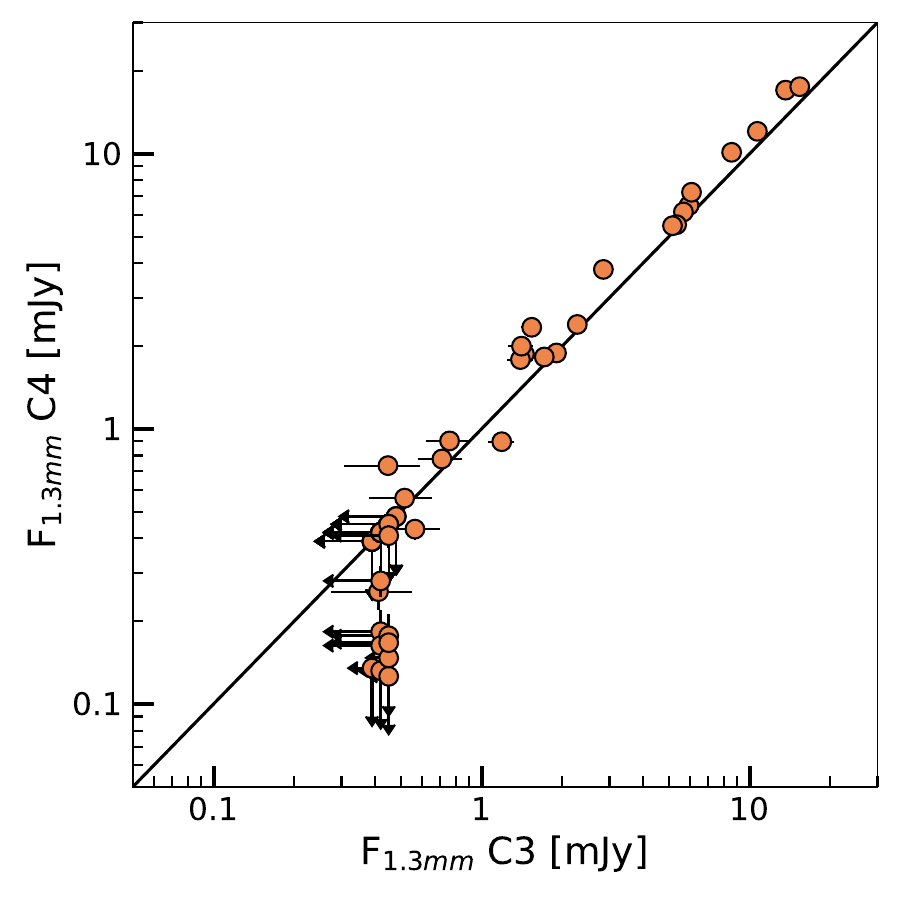}
\caption{Comparison between ALMA fluxes from C3 and C4 for \sorionis\,sources. The arrows show 3$\sigma$ upper limits.
}\label{fig:ALMA_fluxes_comparison}
\end{figure}

\begin{figure*}[ht]
\centering \includegraphics[width = 0.90\textwidth]{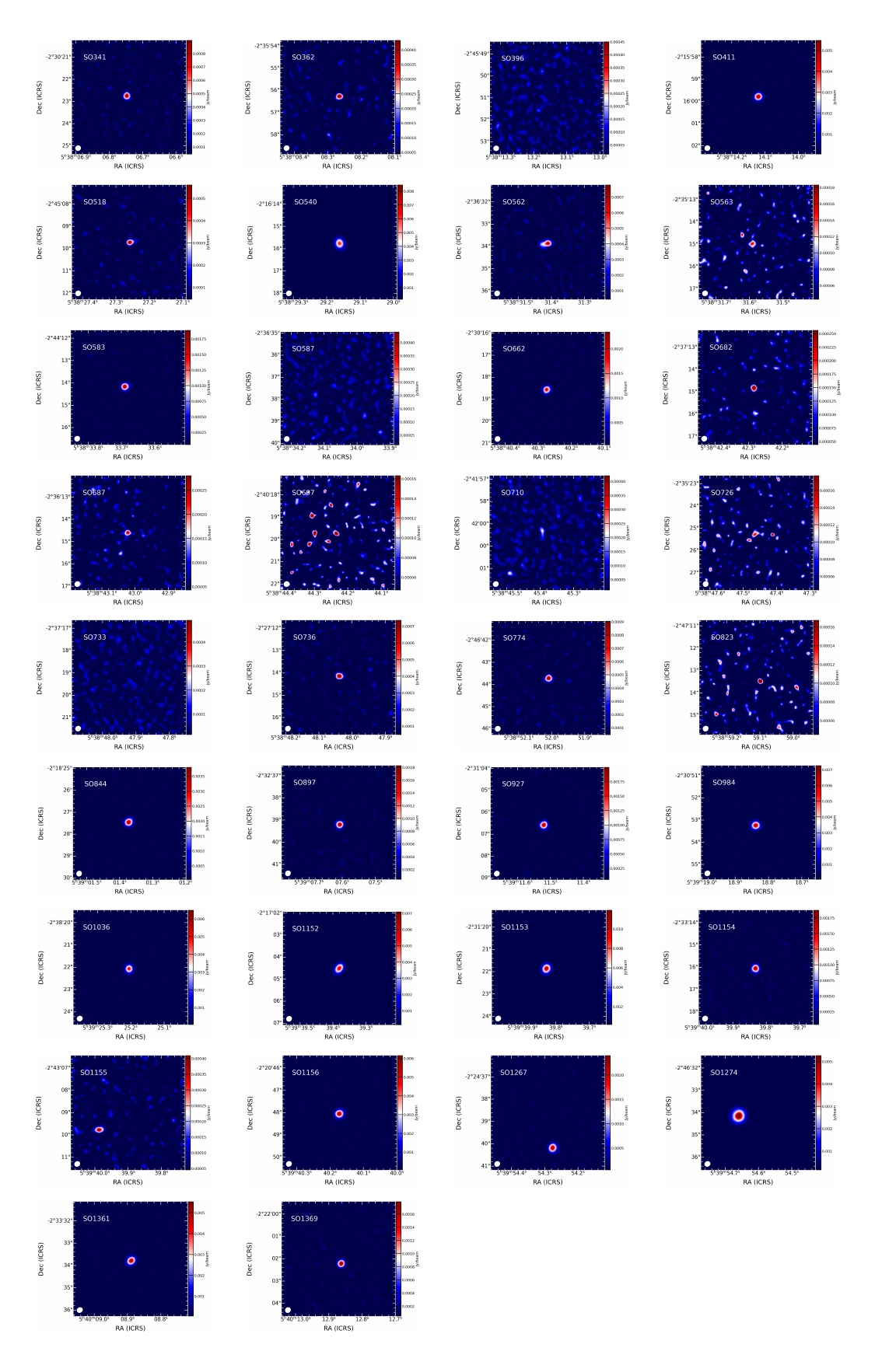}
\caption{Continuum images at 1.33 mm of the 34 disks sampled by ALMA in Cycle 4, ordered by source name.}\label{fig:cont_images}
\end{figure*}

\begin{figure*}[ht]
\centering \includegraphics[width = 0.90\textwidth]{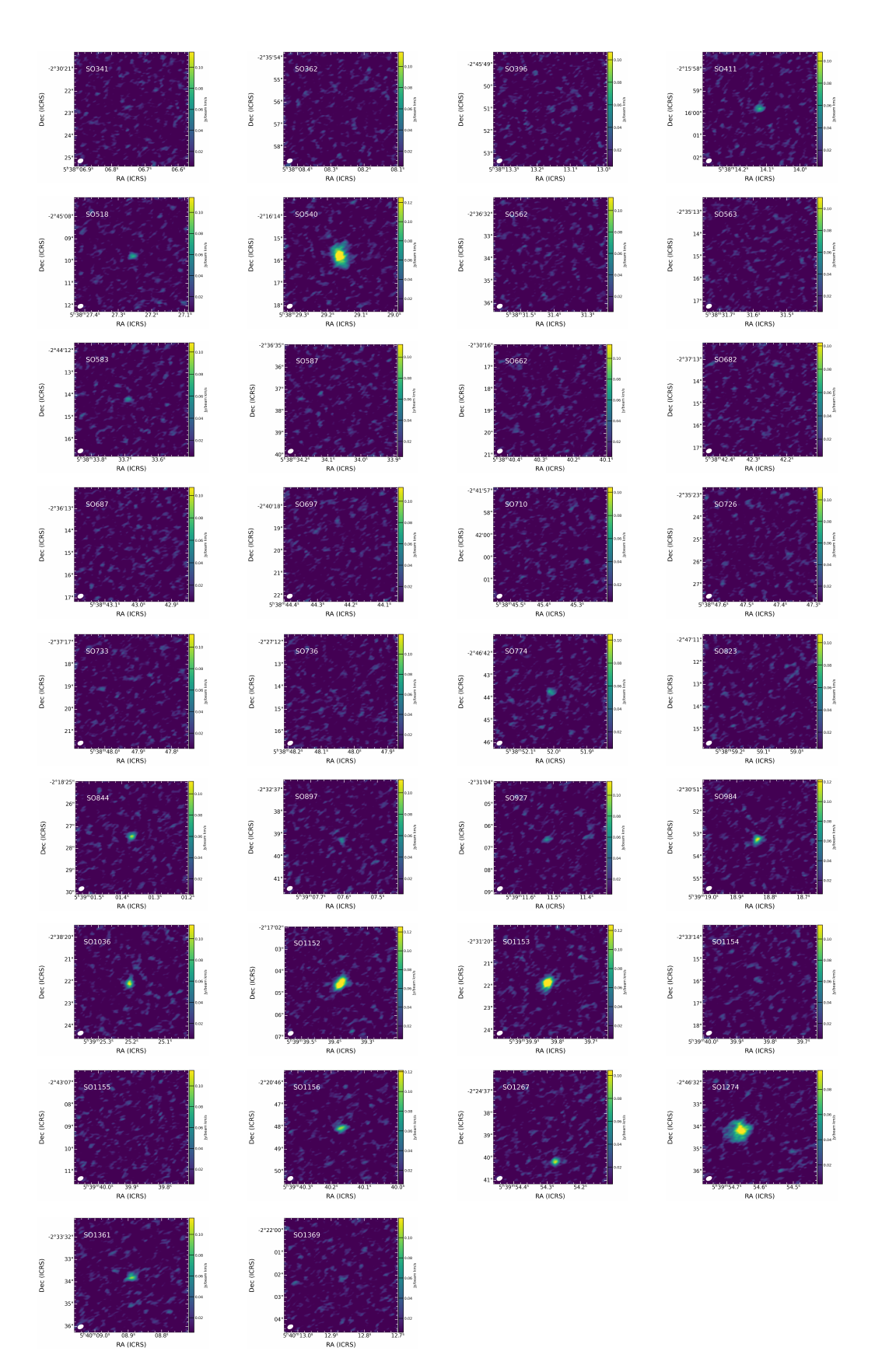}
\caption{$^{12}$CO Moment 0 maps of the 34 disks sampled by ALMA in Cycle 4, ordered by source name.}\label{fig:12COmom0_images}
\end{figure*}

We obtained deeper ALMA Cycle 4 (C4) observations for disks in $\sigma$-Orionis. Fig.~\ref{fig:ALMA_fluxes_comparison} shows the comparison between ALMA observations from \citet[][C3]{Ansdell2017} and our new C4 observations. ALMA observations for our X-Shooter sample are reported in Table~\ref{tab:bestfit}. The list of ALMA non-detections is reported here (Table~\ref{tab:ALMA_nondetections}). 
The data between both cycles agree very well. In general, the C4 fluxes are higher than in C3 by a factor $\sim$1.13 and the new detections have fluxes just at the level of the C3 upper limits (3$\sigma$). The C4 observations allow the detection of 6 new sources in the continuum. Furthermore, in an effort to establish more robust population statistics, especially for the gas, the higher sensitivity C4 data resulted in 13 new detections in $^{12}$CO. The $^{12}$CO fluxes are also reported in Table~\ref{tab:bestfit}. Continuum and $^{12}$CO images are shown in Fig.~\ref{fig:cont_images}, and ~\ref{fig:12COmom0_images}, respectively.

\onecolumn
\begin{table*}
\begin{center}
\caption{\label{tab:ALMA_nondetections} ALMA non-detections}	
\begin{tabular}{lccccccc}
\hline\hline 
Name & RA$_{2000}$ & Dec$_{2000}$ & Distance & $d_{\rm p}$ & Log $G_o$ & Disk type & $M_\mathrm{dust}$\\
     & hh:mm:ss.s  & dd:mm:ss.s   &  [pc]    &  [pc]  &   &  & $[\mT{}]$ \\  
\hline
 SO247 & 05:37:54.86 & -02:41:09.2 & $392.2_{- 9.5}^{+10.0}$ &  1.54 &  2.69 & II    & <1.42 \\ [0.5ex]
 SO254 & 05:37:55.60 & -02:33:05.3 & $373.4_{-33.8}^{+41.3}$ &  1.37 &  2.79 & II    & <1.28 \\ [0.5ex]
 SO271 & 05:37:57.46 & -02:38:44.4 & $396.0_{-18.4}^{+20.2}$ &   1.4 &  2.78 & II    & <1.43 \\ [0.5ex]
 SO300 & 05:38:01.07 & -02:45:38.0 & $401.0_{-15.0}^{+16.7}$ &   1.7 &  2.61 & --    & <1.5  \\ [0.5ex]
 SO327 & 05:38:05.52 & -02:35:57.1 & $392.3_{-45.6}^{+59.3}$ &  1.12 &  2.97 & II    & <1.44 \\ [0.5ex]
 SO396 & 05:38:13.16 & -02:45:51.0 & $404.6_{- 3.8}^{+ 3.9}$ &  1.49 &  2.73 & II    & <0.41 \\ [0.5ex]
 SO435 & 05:38:17.78 & -02:40:50.1 & $401.0$                 &  0.97 &   3.1 & II    & <1.48 \\ [0.5ex]
 SO451 & 05:38:18.86 & -02:51:38.8 & $409.2_{- 5.9}^{+ 6.1}$ &  2.02 &  2.46 & --    & <1.55 \\ [0.5ex]
 SO462 & 05:38:20.50 & -02:34:09.0 & $379.7_{-11.0}^{+11.7}$ &   0.7 &  3.38 & II    & <1.34 \\ [0.5ex]
 SO482 & 05:38:23.08 & -02:36:49.4 & $403.3_{-20.1}^{+22.3}$ &  0.64 &  3.45 & --    & <1.57 \\ [0.5ex]
 SO485 & 05:38:23.33 & -02:25:34.6 & $360.7_{- 9.6}^{+10.2}$ &  1.23 &  2.89 & II    & <1.23 \\ [0.5ex]
 SO514 & 05:38:26.84 & -02:38:46.1 & $351.5_{-14.6}^{+15.9}$ &  0.54 &  3.61 & II    & <1.15 \\ [0.5ex]
 SO537 & 05:38:28.97 & -02:48:47.3 & $416.7_{-32.4}^{+38.3}$ &  1.62 &  2.65 & --    & <1.62 \\ [0.5ex]
 SO598 & 05:38:34.60 & -02:41:08.8 & $400.4_{- 7.3}^{+ 7.6}$ &  0.67 &  3.42 & II    & <1.51 \\ [0.5ex]
 SO657 & 05:38:39.76 & -02:32:20.3 & $402.8_{-57.8}^{+81.1}$ &  0.45 &  3.76 & --    & <1.57 \\ [0.5ex]
 SO663 & 05:38:40.54 & -02:33:27.6 & $405.5_{- 8.7}^{+ 9.0}$ &  0.32 &  4.05 & --    & <1.57 \\ [0.5ex]
 SO674 & 05:38:41.60 & -02:30:28.9 & $383.8_{- 6.4}^{+ 6.6}$ &  0.62 &  3.48 & --    & <1.47 \\ [0.5ex]
 SO707 & 05:38:45.28 & -02:37:29.3 & $405.5_{-16.7}^{+18.2}$ &  0.18 &  4.58 & --    & <1.62 \\ [0.5ex]
 SO710 & 05:38:45.38 & -02:41:59.4 & $401.0_{-16.8}^{+19.4}$ &   0.7 &  3.38 & II    & <0.4  \\ [0.5ex]
 SO723 & 05:38:47.19 & -02:34:36.8 & $392.0_{- 8.8}^{+ 9.2}$ &  0.17 &  4.59 & II    & <1.58 \\ [0.5ex]
 SO733 & 05:38:47.92 & -02:37:19.2 & $411.4_{- 6.5}^{+ 6.7}$ &  0.18 &  4.54 & II    & <0.43 \\ [0.5ex]
 SO738 & 05:38:48.10 & -02:28:53.6 & $377.2_{-21.3}^{+24.0}$ &  0.79 &  3.28 & --    & <1.4  \\ [0.5ex]
 SO750 & 05:38:49.29 & -02:23:57.6 & $425.0_{-27.0}^{+31.0}$ &   1.5 &  2.72 & --    & <1.85 \\ [0.5ex]
 SO754 & 05:38:49.70 & -02:34:52.6 & $392.6_{-11.8}^{+12.6}$ &  0.19 &  4.51 & --    & <1.52 \\ [0.5ex]
 SO762 & 05:38:50.61 & -02:42:42.9 & $394.0_{-14.4}^{+15.5}$ &  0.79 &  3.28 & --    & <1.53 \\ [0.5ex]
 SO827 & 05:38:59.23 & -02:33:51.4 & $410.7_{- 8.0}^{+ 8.4}$ &   0.5 &  3.67 & II    & <1.67 \\ [0.5ex]
 SO865 & 05:39:03.57 & -02:46:27.0 & $399.5_{- 7.1}^{+ 7.3}$ &  1.33 &  2.82 & II    & <1.61 \\ [0.5ex]
 SO866 & 05:39:03.87 & -02:20:08.2 & $379.7_{-12.8}^{+13.7}$ &  1.83 &  2.54 & II    & <1.48 \\ [0.5ex]
 SO871 & 05:39:04.59 & -02:41:49.4 & $429.4_{-12.1}^{+12.8}$ &  0.95 &  3.11 & II    & <1.84 \\ [0.5ex]
 SO908 & 05:39:08.78 & -02:31:11.5 & $383.6_{- 7.8}^{+ 8.1}$ &  0.86 &   3.2 & II    & <1.51 \\ [0.5ex]
 SO936 & 05:39:13.08 & -02:37:50.9 & $401.0_{-63.3}^{+85.4}$ &  0.85 &  3.21 & --    & <1.64 \\ [0.5ex]
 SO967 & 05:39:15.83 & -02:36:50.7 & $396.0_{- 8.3}^{+ 8.6}$ &   0.9 &  3.16 & II    & <1.58 \\ [0.5ex]
SO1050 & 05:39:26.33 & -02:28:37.7 & $389.5_{-10.8}^{+11.5}$ &  1.44 &  2.75 & --    & <1.58 \\ [0.5ex]
SO1182 & 05:39:43.19 & -02:32:43.3 & $390.0_{- 9.1}^{+ 9.5}$ &   1.7 &  2.61 & --    & <1.57 \\ [0.5ex]
SO1193 & 05:39:44.51 & -02:24:43.2 & $439.8_{-21.2}^{+23.5}$ &   2.4 &  2.31 & --    & <2.01 \\ [0.5ex]
SO1230 & 05:39:49.45 & -02:23:45.9 & $413.6_{-10.3}^{+10.8}$ &  2.44 &  2.29 & --    & <1.77 \\ [0.5ex]
SO1268 & 05:39:54.33 & -02:37:18.9 & $441.6_{-37.7}^{+45.4}$ &  2.24 &  2.37 & TD    & <2.03 \\ [0.5ex]
SO1338 & 05:40:04.54 & -02:36:42.1 & $419.7_{-56.6}^{+77.5}$ &  2.44 &   2.3 & --    & <1.84 \\ [0.5ex]
SO1344 & 05:40:05.26 & -02:30:52.3 & $394.3_{-16.6}^{+18.1}$ &  2.38 &  2.32 & --    & <1.62 \\ [0.5ex]

\hline
\end{tabular}
\end{center}
\end{table*}
\twocolumn


\subsection{Gaia observations}
Gaia information for our X-Shooter sample is listed on Table~\ref{tab:parallax_s}. Information on parallaxes ($\varpi$), proper motions ($\mu_{\alpha}$, $\mu_{\beta}$), and RUWE values are provided for each source. 

 \begin{table*}
\footnotesize
\begin{center}
\caption{\label{tab:parallax_s}Information on distances from Gaia for our X-Shooter sample}
\begin{tabular}{lccccc}
\hline\hline
Name & $\varpi$ [mas] & $\sigma_\varpi$ [mas] & $\mu_{\alpha}$ [mas/yr] & $\mu_{\delta}$ [mas/yr] & RUWE  \\  
\hline 
SO73  &   2.7843 &   0.0333 &  1.757 $\pm$  0.034 & -1.313 $\pm$  0.027 &   1.14 \\ [0.5ex] 
 SO299 &   2.8126 &   0.0344 &  1.749 $\pm$  0.036 &  -1.41 $\pm$  0.028 &   0.99 \\ [0.5ex] 
 SO341 &   2.4451 &   0.0259 &  1.334 $\pm$  0.028 & -0.529 $\pm$  0.021 &    1.1 \\ [0.5ex] 
 SO362 &   2.4859 &   0.0291 &  1.057 $\pm$  0.034 & -0.615 $\pm$  0.026 &   1.08 \\ [0.5ex] 
 SO397 &   2.4075 &   0.2217 &  0.816 $\pm$  0.193 &  0.846 $\pm$  0.161 &   2.45 \\ [0.5ex] 
 SO411 &    2.736 &   0.0164 &  1.923 $\pm$  0.014 &  -1.37 $\pm$  0.013 &   0.83 \\ [0.5ex] 
 SO467 &   2.6087 &   0.0601 &  1.237 $\pm$  0.055 & -1.306 $\pm$  0.049 &   1.03 \\ [0.5ex] 
 SO490 &   3.8006 &   0.6852 & -0.999 $\pm$  0.671 & -3.064 $\pm$  0.571 &   7.94 \\ [0.5ex] 
 SO500 &   2.4437 &   0.2443 &  0.977 $\pm$  0.293 & -0.944 $\pm$  0.229 &    1.1 \\ [0.5ex] 
 SO518 &   2.5064 &    0.025 &  1.191 $\pm$  0.027 & -0.895 $\pm$  0.022 &    1.2 \\ [0.5ex] 
 SO520 &   2.4839 &   0.0397 &  1.409 $\pm$  0.039 & -0.571 $\pm$  0.032 &    1.0 \\ [0.5ex] 
 SO540 &   2.4629 &   0.0216 &  0.316 $\pm$  0.019 &  0.709 $\pm$  0.017 &   1.16 \\ [0.5ex] 
 SO563 &   7.7206 &   0.8106 &  3.868 $\pm$  0.767 &    5.7 $\pm$  0.633 &  35.66 \\ [0.5ex] 
 SO583 &   2.4699 &    0.223 &   1.07 $\pm$  0.238 & -0.485 $\pm$  0.191 &  19.86 \\ [0.5ex] 
 SO587 &   2.1903 &   0.2054 &  0.298 $\pm$  0.202 &  0.855 $\pm$  0.165 &   6.77 \\ [0.5ex] 
 SO646 &   2.4714 &   0.0409 &  1.243 $\pm$  0.044 &  -0.72 $\pm$  0.037 &   1.04 \\ [0.5ex] 
 SO662 &   2.4925 &   0.0208 &  0.871 $\pm$  0.018 &  -0.61 $\pm$  0.017 &   1.12 \\ [0.5ex] 
 SO682 &   2.4402 &   0.0284 &  1.072 $\pm$  0.027 & -0.349 $\pm$  0.022 &   1.22 \\ [0.5ex] 
 SO687 &   2.4226 &   0.0251 &  1.566 $\pm$  0.026 &  0.132 $\pm$  0.022 &   1.13 \\ [0.5ex] 
 SO694 &   2.5492 &   0.0612 &  1.733 $\pm$   0.06 & -1.114 $\pm$   0.05 &   1.12 \\ [0.5ex] 
 SO697 &   2.4723 &   0.0148 &  1.459 $\pm$  0.014 & -1.009 $\pm$  0.012 &   1.03 \\ [0.5ex] 
 SO726 &   2.4761 &   0.0425 &   2.21 $\pm$  0.045 &  0.097 $\pm$   0.04 &   1.14 \\ [0.5ex] 
 SO736 &   2.5267 &   0.0608 &  1.456 $\pm$  0.054 &  0.147 $\pm$   0.05 &   3.32 \\ [0.5ex] 
 SO739 &   2.3081 &   0.1132 &  1.368 $\pm$  0.125 & -0.946 $\pm$  0.095 &   0.98 \\ [0.5ex] 
 SO774 &   2.4796 &   0.0205 &   1.63 $\pm$  0.021 & -1.074 $\pm$  0.016 &   1.11 \\ [0.5ex] 
 SO818 &   2.4665 &   0.0253 &  0.746 $\pm$  0.024 &    0.4 $\pm$  0.021 &    1.1 \\ [0.5ex] 
 SO823 &   2.2249 &    0.058 &  0.845 $\pm$  0.062 &  0.104 $\pm$  0.053 &   1.46 \\ [0.5ex] 
 SO844 &   2.4067 &   0.0216 &  1.704 $\pm$  0.023 & -0.052 $\pm$   0.02 &    1.1 \\ [0.5ex] 
 SO848 &   2.8064 &   0.1347 &  1.689 $\pm$  0.128 & -0.225 $\pm$  0.119 &   1.08 \\ [0.5ex] 
 SO859 &   2.4517 &   0.0389 &  1.561 $\pm$  0.037 & -0.715 $\pm$  0.033 &   1.04 \\ [0.5ex] 
 SO897 &   2.6239 &   0.0517 &  1.588 $\pm$  0.049 & -0.922 $\pm$  0.044 &   3.28 \\ [0.5ex] 
 SO927 &   2.4178 &   0.0276 &  1.832 $\pm$  0.026 & -0.316 $\pm$  0.024 &   1.27 \\ [0.5ex] 
 SO984 &   2.4412 &   0.0188 &  1.808 $\pm$  0.021 & -0.654 $\pm$  0.017 &   1.09 \\ [0.5ex] 
SO1036 &   2.5318 &   0.0221 &  1.843 $\pm$  0.018 & -0.349 $\pm$  0.018 &   1.19 \\ [0.5ex] 
SO1075 &   2.5638 &   0.0554 &  1.701 $\pm$  0.053 & -0.122 $\pm$  0.047 &   1.18 \\ [0.5ex] 
SO1152 &   2.5091 &   0.0245 &  2.105 $\pm$  0.024 & -0.033 $\pm$   0.02 &    1.3 \\ [0.5ex] 
SO1153 &   2.5212 &   0.0268 &  1.798 $\pm$  0.023 & -0.094 $\pm$   0.02 &   1.23 \\ [0.5ex] 
SO1154 &   2.4985 &   0.0797 &  2.113 $\pm$  0.075 & -0.824 $\pm$  0.068 &    1.4 \\ [0.5ex] 
SO1156 &   2.4765 &   0.0159 &   2.44 $\pm$  0.017 & -0.255 $\pm$  0.014 &   1.08 \\ [0.5ex] 
SO1248 &     2.51 &   0.0486 &  2.221 $\pm$  0.049 & -0.183 $\pm$  0.042 &   1.09 \\ [0.5ex] 
SO1260 &   2.5888 &   0.0419 &  2.259 $\pm$  0.039 & -0.318 $\pm$  0.032 &   1.03 \\ [0.5ex] 
SO1266 &   2.5055 &   0.0673 &  2.004 $\pm$  0.062 & -0.496 $\pm$  0.053 &   0.98 \\ [0.5ex] 
SO1267 &   2.4967 &   0.0325 &  2.149 $\pm$   0.03 & -0.193 $\pm$  0.026 &   1.29 \\ [0.5ex] 
SO1274 &    2.455 &   0.0164 &  2.166 $\pm$  0.017 & -0.781 $\pm$  0.014 &   1.01 \\ [0.5ex] 
SO1327 &   2.5147 &   0.0363 &  2.336 $\pm$  0.037 &  0.018 $\pm$   0.03 &   1.01 \\ [0.5ex] 
SO1361 &   2.4632 &   0.0239 &  2.253 $\pm$  0.019 & -0.254 $\pm$  0.018 &   1.16 \\ [0.5ex] 
SO1362 &   2.5041 &   0.0656 &  2.228 $\pm$  0.058 &  -0.39 $\pm$  0.052 &   1.08 \\ [0.5ex] 
SO1369 &   2.4843 &   0.0153 & -2.476 $\pm$  0.014 &  -4.16 $\pm$  0.012 &   0.97 \\ [0.5ex] 
\hline
\end{tabular}
\end{center}
\end{table*}

\section{Plots of the Balmer jump fits}

Here, we present the best fit of the X-Shooter spectra of our sample
obtained following \citet{Manara2013b} and described in Sect.~\ref{sec:stellar_param}. We show the Balmer jump region of the spectra for each target.

\begin{figure*}
\centering 
\includegraphics[width = 0.96\textwidth]{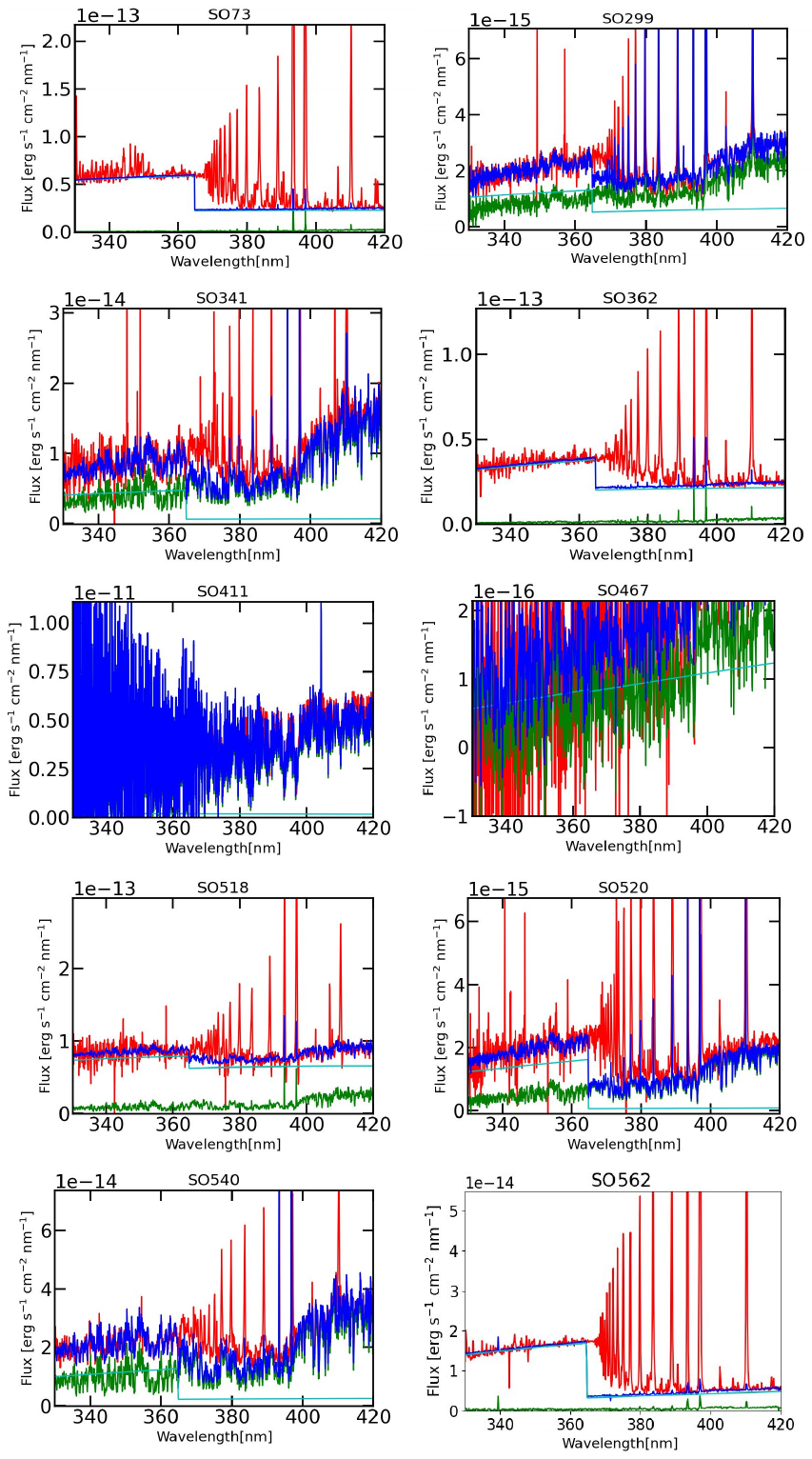}
\caption{Best fit for the Balmer continuum region for targets in the \sorionis\,cluster, ordered by source name.}\label{fig:best_fit_plots}
\end{figure*}

\begin{figure*}
\centering 
\includegraphics[width = 0.96\textwidth]{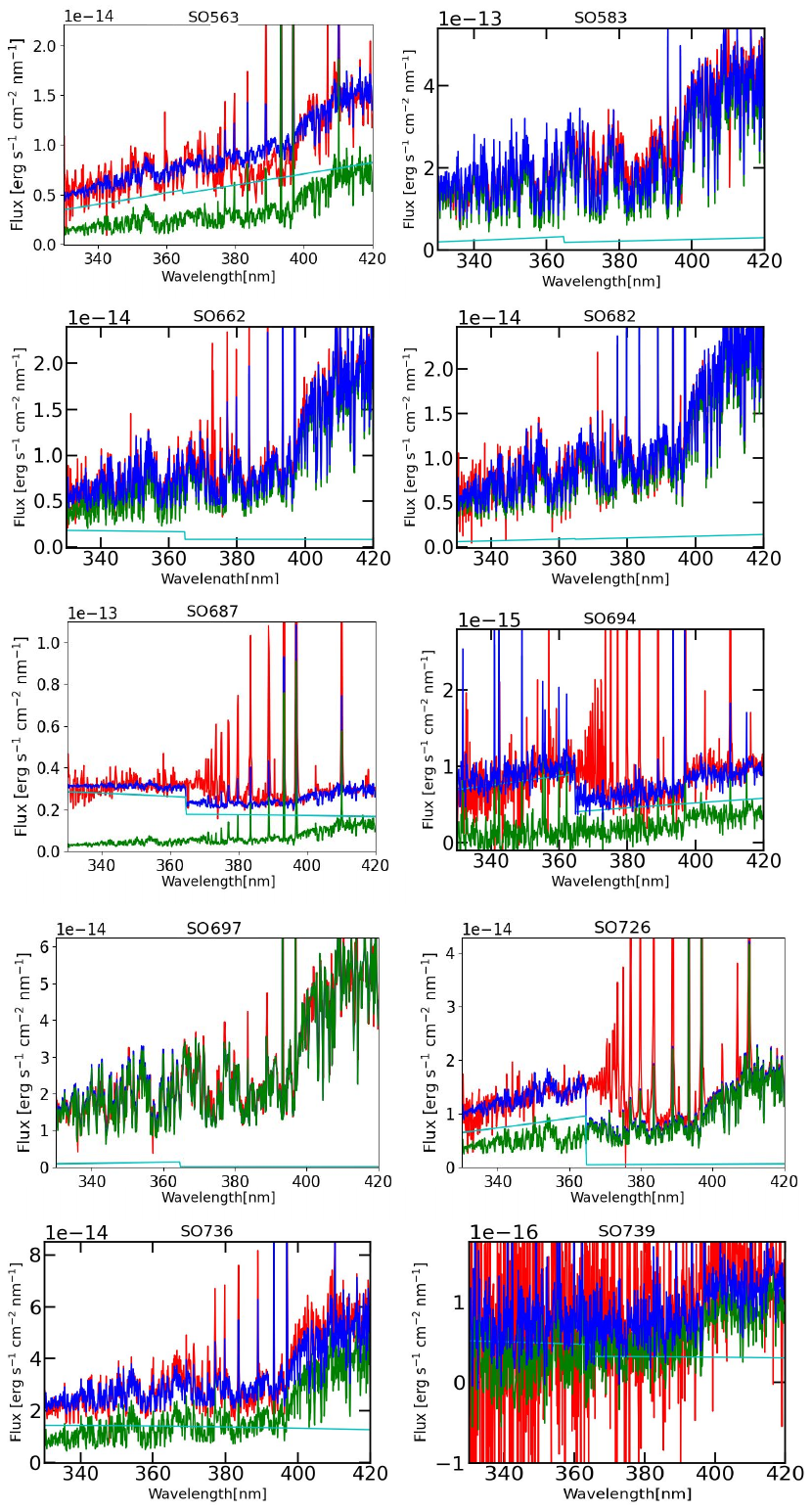}
\caption{Best fit for the Balmer continuum region for targets in the \sorionis\,cluster,  ordered by source name. - continued}\label{fig:best_fit_plots}
\end{figure*}

\begin{figure*}
\centering 
\includegraphics[width = 0.96\textwidth]{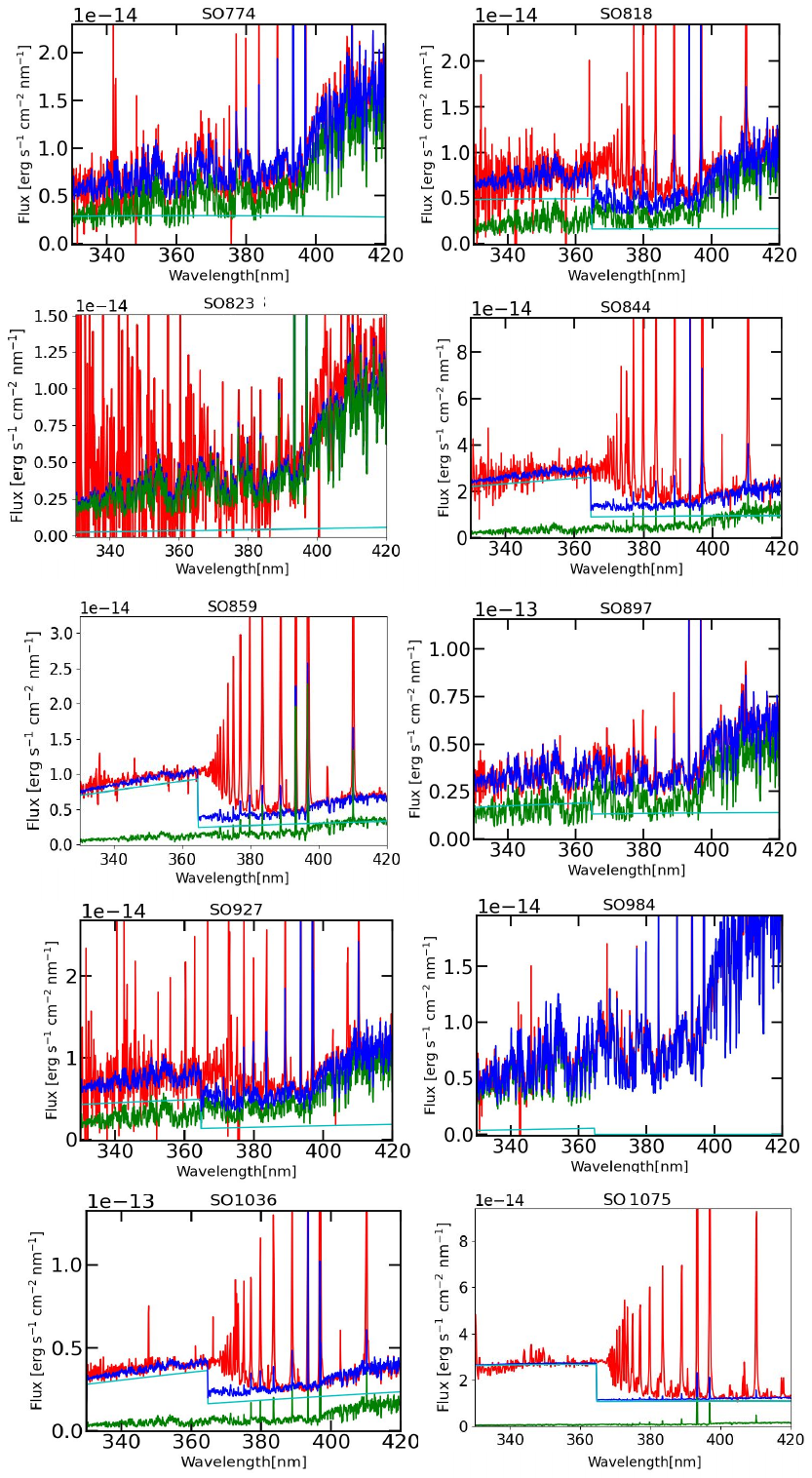}
\caption{Best fit for the Balmer continuum region for targets in the \sorionis\,cluster, ordered by source name. - continued}\label{fig:best_fit_plots}
\end{figure*}

\begin{figure*}
\centering 
\includegraphics[width = 0.96\textwidth]{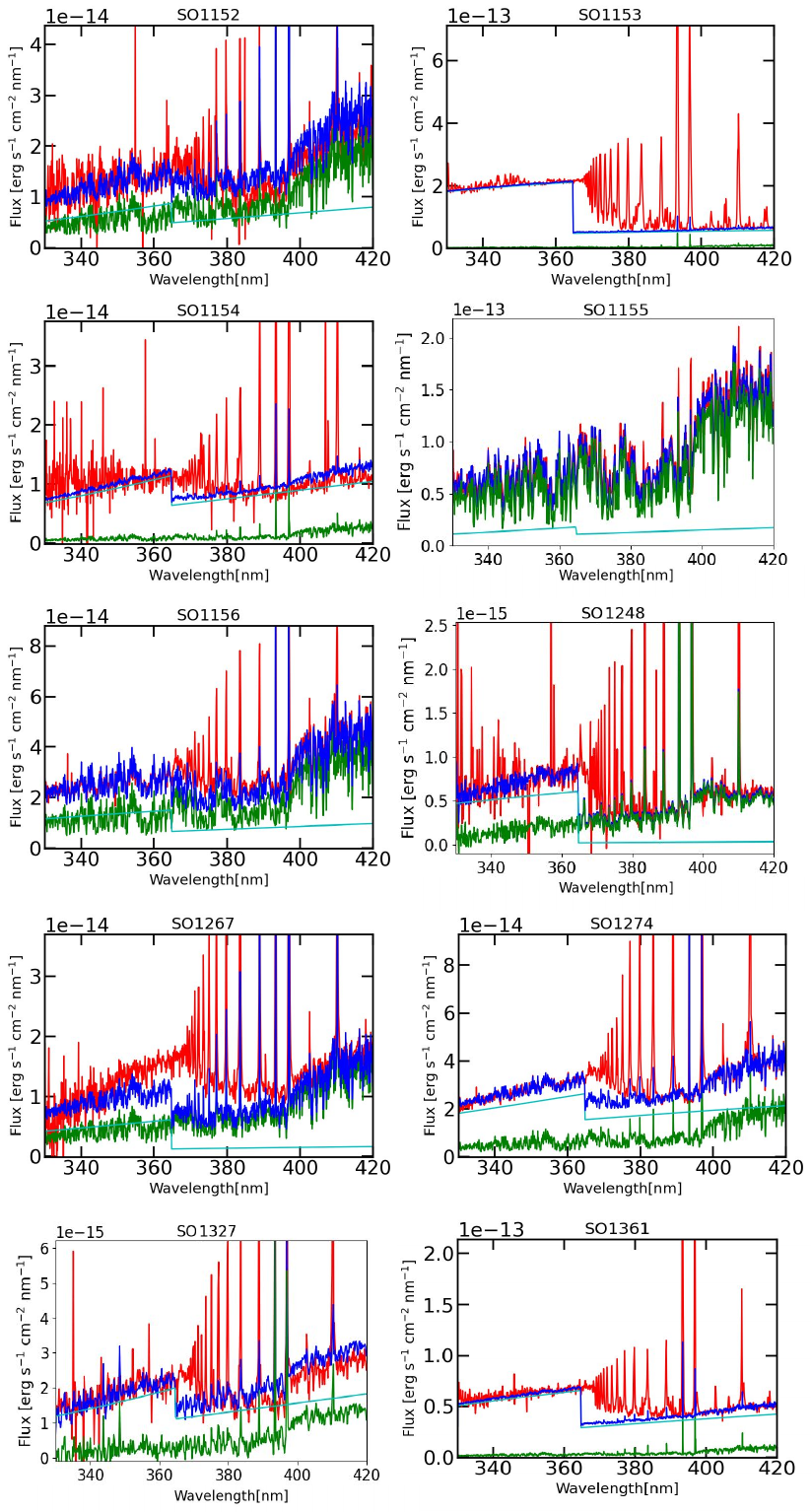}
\caption{Best fit for the Balmer continuum region for targets in the \sorionis\,cluster, ordered by source name. - continued}\label{fig:best_fit_plots}
\end{figure*}

\begin{figure*}
\centering 
\includegraphics[width = 0.65\textwidth]{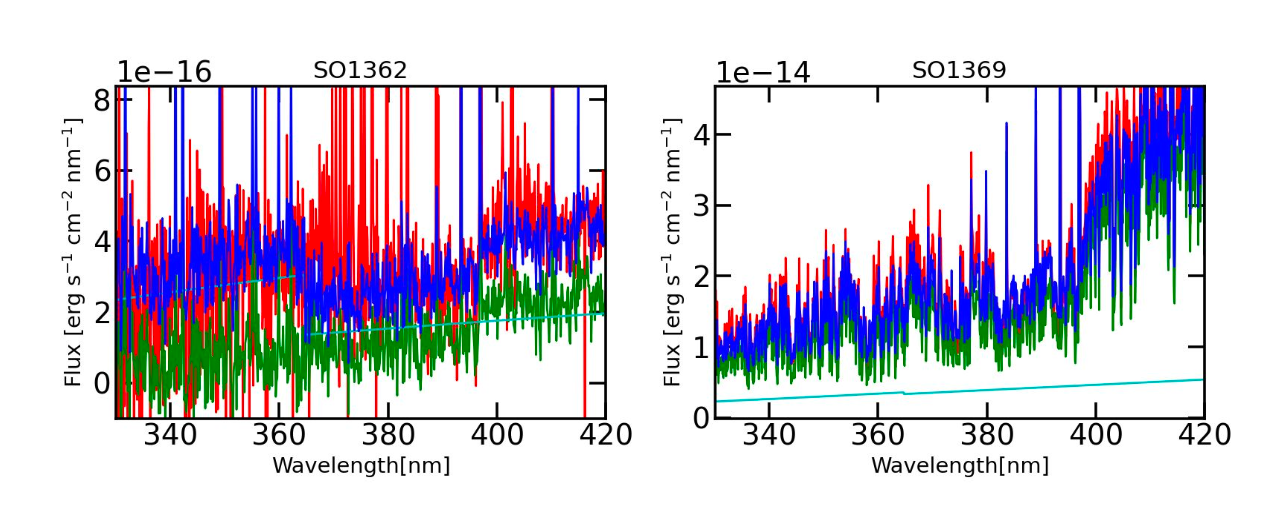}
\caption{Best fit for the Balmer continuum region for targets in the \sorionis\,cluster, ordered by source name. - continued}\label{fig:best_fit_plots}
\end{figure*}

\end{document}